\documentclass[12pt,a4paper,final]{iopart}
\usepackage{iopams}
\usepackage{graphicx}
\usepackage{amssymb,amsthm,color}
\usepackage{mathptmx}
\usepackage{amsfonts,bm}
\usepackage{lscape}

\expandafter\let\csname equation*\endcsname\relax

\expandafter\let\csname endequation*\endcsname\relax

\usepackage{amsmath}
\usepackage[breaklinks=true,colorlinks=true,linkcolor=blue,urlcolor=blue,citecolor=blue]{hyperref}

\usepackage{tikz}
\usetikzlibrary{matrix}
\usetikzlibrary{positioning}
\usetikzlibrary{intersections,calc}
\usetikzlibrary{arrows.meta}
\newcommand{\myGlobalTransformation}[2]

\makeatletter

\@addtoreset{figure}{section}
\makeatother

\makeatletter

\@addtoreset{equation}{section}
\makeatother

\def\eq{eq.\ }
\def\eqs{eqs.\ }
\def\dimcase{d}
\def\dimiso{m}

\newcommand{\D}{{\rm d}}

\newtheorem*{theorem}{Theorem}

\begin{document}

\title[]{Counting the number of Killing vectors in a 3D spacetime}

\author{Masato Nozawa}
\address{Center for Gravitational Physics, Yukawa Institute for Theoretical Physics, Kyoto University, Kyoto 606-8502, Japan}
\ead{masato.nozawa@yukawa.kyoto-u.ac.jp}

\author{Kentaro Tomoda}
\address{Advanced Mathematical Institute, Osaka City University, Osaka 558-8585 Japan}
\ead{k-tomoda@sci.osaka-cu.ac.jp}

\begin{abstract}
	We devise an algorithm which allows one to count the number of Killing vectors for a Lorentzian manifold of dimension $3$.
	Our algorithm relies on the principal traces of powers of the Ricci tensor and
	branches intricately according to the values of differential invariants arising from the compatibility conditions of the Killing equation.
	As illustrating examples, we classify the Lifshitz and pp-wave spacetimes into a hierarchy based on their level of symmetry.
	A complete classification of spacetimes admitting 4 Killing vectors is also presented.
\end{abstract}

\vspace{2pc}


\section{Introduction and summary of our work}
\label{sec:introduction}

Klein's Erlangen programme \cite{Klein:2008} provided an attempt to make a connection between geometry and symmetry.  
In modern parlance, this programme tried to classify the geometry according to the invariant properties under a certain (finite) group action. 
Whilst this paradigm has turned out to be too narrow to encompass the Riemannian geometry, it nevertheless has significant implications and influences to Riemannian geometry, as well as for theoretical physics. The advent of general relativity stimulated Klein to investigate the role of groups from relativistic standpoint. 
He had regarded special relativity as the theory of invariants of Minkowski spacetime under the Lorentz group action
and also contributed to the formulation of the conservation laws in general relativity, see \cite{Zuber:2013rha} for a review.

An isometry group acting on a Lorentzian manifold $(M, g_{ab})$ is of fundamental importance 
in understanding the geometrical and physical properties of a spacetime. 
For instance, the isometry group allows one to set the privileged local coordinate system of $M$ and 
also gives rise to first integrals of geodesic flow that are linear in momenta. 
In particular, the globally conserved quantities associated with isometries---such as the mass and the angular momentum---play 
a significant role in the analysis of black holes.
It is then natural to ask whether there exists a set of invariants in Klein's sense which is associated with the isometry group.

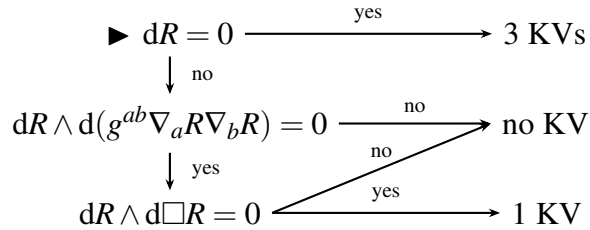
\begin{figure}[t]
	\begin{center}
		\begin{tikzpicture}
		[
		every node/.style={outer sep=0.15cm, inner sep=0cm},
		arrow/.style={-{Stealth[length=0.15cm]},thick},
		rblock/.style={rectangle, rounded corners,draw, minimum height = 0.5cm,
			minimum width=1.6cm, thick, outer sep = 0},
		point/.style={radius=2pt}
		]
		\node [] (Obst1){$\blacktriangleright \:\D R = 0$};
		\node [below=0.5 of Obst1] (Obst2){$\D R \wedge \D (g^{ab} \nabla_a R\nabla_b R) =0$};
		\node [below=0.5 of Obst2] (Obst3){$\D R \wedge \D \Box R = 0$};
		\node [right=3.25 of Obst1] (3KVs){3 KVs};
		\node [right=2 of Obst2] (noKV){no KV};
		\node [right=3 of Obst3] (1KV){1 KV};
		\draw[arrow] (Obst1) -- (3KVs) node[above,midway] {{\scriptsize yes}};
		\draw[arrow] (Obst1) -- (Obst2) node[right,midway] {{\scriptsize ~~no}};
		\draw[arrow] (Obst2) -- (noKV) node[above,midway] {{\scriptsize no}};
		\draw[arrow] (Obst2) -- (Obst3) node[right,midway] {{\scriptsize ~~yes}};
		\draw[arrow] (Obst3) -- (1KV) node[above,midway] {{\scriptsize yes}};
		\draw[arrow] (Obst3.east) -- (noKV.west) node[above,midway] {{\scriptsize no}};
		\end{tikzpicture}
		\caption{The Darboux algorithm for a (semi-)Riemannian space of dimension $2$.
			Here $R$ denotes the Ricci scalar; $\nabla_a$ denotes the Levi-Civita connection;
			a box denotes d'Alembertian, $\Box \equiv g^{ab}\nabla_a \nabla_b$;
			a triangle symbol $\blacktriangleright$ stands for a root of this algorithm.
		}
		\label{fig:2-dim}
	\end{center}
\end{figure}

In this paper, we focus on the local existence of infinitesimal isometries of $(M, g_{ab})$.
The problem of finding all of them is equivalent to that of finding all independent Killing vectors (KVs).
To give an exhaustive list of KVs for a given Riemannian/Lorentzian manifold has a long history in geometry, which dates back at least to Darboux \cite{Darboux:1894}.
For dimension $n=2$, he found a set of invariants for determining the existence and the number of KVs as shown in Figure \ref{fig:2-dim}.
See also \cite{Eisenhart:1909,Singer:1960,Nomizu:1960,PuferTricerriVanhecke:1996,ConsoleOlmos:2008} for pertinent results.

For determining the number of KVs in arbitrary dimension, 
Cartan's equivalence method or the Cartan--Karlhede algorithm \cite{Cartan:1946,KarlhedeMaccallum:1982} is known as an effective technique (see also
section 9.2 of \cite{Stephani:2003tm}). 
Though it was originally developed to solve  alias the {\it equivalence problem},
that is the problem of deciding whether given two (semi-)Riemannian manifolds are locally isometric,
this amounts to determining the dimension of their local isometry group, as well as the structure constants of the group.
Their machinery is explained as follows. 
Let $(M,g_{ab})$ be a (semi-)Riemannian manifold of dimension $n$ and
$\mathcal{C}^p$ be the set of Cartan scalars of order $p$.
Note that Cartan scalars of order $p$ are defined as the frame components of
the Riemann--Christoffel tensor and its first $p$ covariant derivatives.
As with Newman--Penrose scalars, Cartan scalars are functions on the tangent frame bundle $F(M)$ but not on $M$.
Cartan showed that the local geometry of $M$ is completely determined by $\mathcal{C}^{p_0}$ with $p_0 \leq \tfrac 12 n(n+1)$, where  $p_0$ is the smallest natural number such that
the elements of $\mathcal{C}^{p_0+1}$ are functionally dependent on those in $\mathcal{C}^{p_0}$. 
Subsequently, Karlhede demonstrated that in dimension $n=4$ at most $7$ differentiations suffice,
whereat $3,156$ functionally independent scalars are required.
Given the set $\mathcal{C}^{p_0}$ consisting of $q$ functionally independent scalars,
the manifold $M$ admits $\tfrac 12n(n+1) - q$ independent KVs. 
Unfortunately, the actual computations required to perform the Cartan--Karlhede algorithm remain formidable, 
even though it indeed ensures that the problem of finding all KVs is computable.

For a Riemannian space of dimension $n=3$, an important progress has been made 
in this problem rather recently in \cite{Kruglikov:2018qcn}.
The scheme exploits the compatibility condition of the Killing equation (see \eq \eqref{eq:comp_cond} below),
and aims exclusively at determining the dimension of the isometry group.
This provides a more efficient and effective algorithm to count the number of KVs,
allowing us to circumvent enormous amount of computational efforts.
We discuss in this paper its extension to a Lorentzian manifold $(M, g_{ab})$ of dimension $3$. 
An essential ingredient which operates 
the mechanism is as follows:
Recall that any vector $K^a$ is a KV on $(M, g_{ab})$ if and only if
the Killing equation is satisfied
\begin{align}
&\pounds_K g_{ab}~=2\nabla_{(a} K_{b)} ~=~0 \,. 
\label{eq:Killing}
\end{align}
 Here $\pounds_K$ is the Lie derivative along $K^a$,
$\nabla_a$ denotes the Levi-Civita connection and
indices are raised and lowered with $g_{ab}$ and its inverse.
The round brackets denote symmetrisation over the enclosed indices.
As the compatibility condition of \eq \eqref{eq:Killing}, one finds the curvature collineation \cite{Kerr:1962}
\begin{align}
\pounds_K R_{abc}{}^d ~=~ 0\:,
\label{eq:comp_cond}
\end{align}
where $R_{abc}{}^d$ is the Riemann--Christoffel tensor defined by $2\nabla_{[a}\nabla_{b]}V_c=R_{abc}{}^d V_d$. 
Here the square brackets over indices is used for skew-symmetrisation.
Any solution to the Killing equation \eqref{eq:Killing} automatically
solves the equation \eqref{eq:comp_cond}, but in general the converse is not true.
In dimension $3$, the following condition is an immediate corollary of (\ref{eq:comp_cond}):
\begin{align}
&\pounds_K R~=~0\:,
&&\pounds_K S^{(2)}~=~0\:,
&&\pounds_K S^{(3)}~=~0\:,
\end{align}
where $R_{ab} \equiv R_{acb}{}^c$ is the Ricci tensor, $R\equiv R^a{}_a=g^{ab}R_{ab}$ is the scalar curvature, and
$S^{(2)} \equiv S^a{}_bS^b{}_a$, $S^{(3)} \equiv S^a{}_b S^b{}_c S^c{}_a$ are the principal traces of powers of
the traceless Ricci tensor $S_{ab} \equiv R_{ab}-(R/3) g_{ab}$.
Thus, any solution to \eq \eqref{eq:Killing} must satisfy the following matrix equation\footnote{
The terms $S^{(m)}$ ($m\ge 4$) fail to contribute to the obstruction matrix, since 
the Cayley-Hamilton theorem allows one to express them as lower matrix powers. }
\begin{align}
&
{\boldsymbol R}_a\:K^a ~=~ 0\:,
&&
{\boldsymbol R}_a ~\equiv~
\begin{bmatrix}
\nabla_a R~~~\\
\nabla_a S^{(2)}\\
\nabla_a S^{(3)}
\end{bmatrix}\:.
\label{eq:matrix_eq}
\end{align}
We shall refer to the $3 \times 3$ matrix ${\boldsymbol R}_a$ as the {\it first obstruction matrix}.
This equation implies that any KV must be in $\ker {\boldsymbol R}_a$ and hence
$\det {\boldsymbol R}_a~=0$ for $\ker {\boldsymbol R}_a \neq \emptyset$, where 
the determinant of ${\boldsymbol R}_a$ is given by 
\begin{align}
\det {\boldsymbol R}_a~=~\D R \wedge \D S^{(2)} \wedge \D S^{(3)}\:. 
\end{align}
The first obstruction matrix ${\boldsymbol R}_a$ is classified by the dimension of its kernel,
\begin{align}
\dimcase~\equiv~\dim \ker {\boldsymbol R}_a ~=~ 3- \mathrm{rank} {\boldsymbol R}_a\:,
\end{align}
which can be determined according to the minors of ${\boldsymbol R}_a$:
\begin{subequations}
\begin{align}
&\D R \wedge \D  S^{(2)}\:,
&&\D S^{(2)} \wedge \D S^{(3)}\:,
&&\D S^{(3)} \wedge \D  R\:,
\label{eq:ann_caseI}\\
&\D  R\:,
&&\D  S^{(2)} \:,
&&\D  S^{(3)} \:.
\label{eq:ann_caseII}
\end{align}
\label{eq:ann_cases}
\end{subequations}
It follows that $(M,g_{ab})$ enjoys a local isometry group of dimension $d_{\mathrm{iso.}} \leq \tfrac 12\dimcase(\dimcase+1)$
with an isotropy subgroup of dimension $d_{\mathrm{sub.}} \leq \tfrac 12 \dimcase(\dimcase-1)$,
acting on orbits of dimension $d_{\mathrm{orb.}} \leq \dimcase$.

In any of these cases, a general solution of \eq \eqref{eq:matrix_eq} can be written in the form
\begin{align}
&K^a ~=~ \sum_{\alpha} \omega_\alpha\:u_\alpha^a\:,
&&\left( \alpha=1,\:\ldots,\:\dimcase \right)
\label{eq:ansatz_g}
\end{align}
where $\{u_\alpha^a\}$ are linearly independent vectors that span $\ker {\boldsymbol R}_a$ and the coefficients $\{\omega_\alpha\}$ are left arbitrary.
In what follows, we refer to the case in which \eq \eqref{eq:ansatz_g} holds true as {\it class $\dimcase$}.
Substituting the form \eqref{eq:ansatz_g} into \eq \eqref{eq:Killing}, we obtain a PDE system of the form
\begin{align}
&\nabla_a {\boldsymbol \omega} ~=~ {\boldsymbol \Omega}_a\:{\boldsymbol \omega}\:,
&&{\boldsymbol \omega} ~\equiv~ [\omega_\alpha ~~\varpi_\beta]^T\:,
&&\left(\beta = 1,\:\ldots,\:\dimiso~\equiv~\frac 12\dimcase(\dimcase-1)\right)
\label{eq:basic_PDE}
\end{align}
where $\{\varpi_\beta\}$ are the 1-jet variables and the connection ${\boldsymbol \Omega}_a$ is expressed in terms of  the Ricci rotation coefficients and their ratio. Since eq.~\eqref{eq:basic_PDE} is the first-order system, 
its compatibility gives rise to algebraic constraints on ${\boldsymbol \omega}$, 
\begin{align}
(\nabla_{[a} {\boldsymbol \Omega}_{b]} - {\boldsymbol \Omega}_{[a} {\boldsymbol \Omega}_{b]})\:{\boldsymbol \omega}~=~0\:. 
\label{eq:comp_cond_PDE}
\end{align}
Equivalently, \eq \eqref{eq:comp_cond_PDE} can be written in component form of curvature of the bundle
\begin{align}
{\boldsymbol R}_{\mathrm{cls. }\dimcase}\:{\boldsymbol \omega} ~=~0\:.
\label{eq:matrix_eq_k}
\end{align}
We henceforth call ${\boldsymbol R}_{\mathrm{cls. }\dimcase}$ as the {\it second obstruction matrix of class $\dimcase$}. 
Since the matrix equation \eqref{eq:matrix_eq} is a necessary condition for \eq \eqref{eq:Killing},
the second obstruction matrix yields obstructions to the existence of KVs as differential invariants.
A noteworthy asset of this method is that the obstruction is measured by a purely algebraic fashion. 

Let us now outline our strategy to be carried out. 
In each class $\dimcase$, we solve \eq \eqref{eq:comp_cond_PDE} and then update the form \eqref{eq:ansatz_g}.
When this achives a decrease in the number $\dimcase$ in \eq \eqref{eq:ansatz_g}
or in the number $\dimiso$ in \eq \eqref{eq:basic_PDE},
which are initially given by $\dimcase = \dim \ker {\boldsymbol R}_a$ and $\dimiso = \tfrac 12 \dimcase (\dimcase-1)$ respectively,
we write out \eqs \eqref{eq:basic_PDE}--\eqref{eq:matrix_eq_k} with the updated form,
that is the form \eqref{eq:ansatz_g} with ${\boldsymbol \omega}$ being the solution of \eq \eqref{eq:comp_cond_PDE}.
Once again, we solve \eq\eqref{eq:comp_cond_PDE} and re-update the form \eqref{eq:ansatz_g}. 
We iterate this procedure until the latest compatibility is met trivially, or until the number $\dimcase$ vanishes as a consequence thereof.
This procedure is amendable to a follow-up study. This is a prime advantage of our formulation over the treatment of Cartan--Karlhede.

In this paper, we classify the number of local isometry group for a Lorentzian manifold of dimension $3$, by presenting the explicit forms of the second obstruction matrix ${\boldsymbol R}_{\mathrm{cls. }\dimcase}$ in all classes.
This survey is essentially based on the procedure developed in \cite{Kruglikov:2018qcn},
but differs from it in that: In Lorentzian signature, there appear null KVs and
the Ricci tensor is not always diagonalisable.
It is this aspect that prohibits the direct application of the previous analysis of Riemannian case \cite{Kruglikov:2018qcn} and requires the separate study, complicating attempts to pin it down discursively.
The strategy employed here is similar in spirit to the Erlangen programme, since the symmetry is classified in terms of differential invariants. On top of the intrinsic interest in 3 spacetime dimensions,
the method developed here can be applicable also for the 3 dimensional induced metrics as well as for quotient metrics.
For $n\ge 4$ dimensions, a considerable number of loose ends are left over and the study of counting KVs remains open. See e.g. \cite{Houri:2014hma} for the analysis giving rise to an upper bound of KVs. We hope to address the issue for $n\ge 4$ in the future.

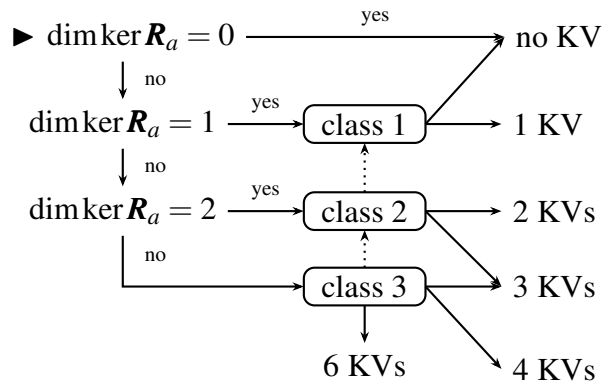
\begin{figure}[h]
	\begin{center}
				\begin{tikzpicture}
				[
						every node/.style={outer sep=0.15cm, inner sep=0cm},
						arrow/.style={-{Stealth[length=0.15cm]},thick},
						rblock/.style={rectangle, rounded corners,draw, minimum height = 0.5cm,
							minimum width=1.6cm, thick, outer sep = 0},
						point/.style={radius=2pt}
				]
				\node [] (ker0){$\blacktriangleright \:\dim \ker {\boldsymbol R}_a = 0$};
				\node [below=0.5 of ker0] (ker1){$\dim \ker {\boldsymbol R}_a = 1$};
				\node [below=0.5 of ker1] (ker2){$\dim \ker {\boldsymbol R}_a = 2$};
				\node [rblock,right=1 of ker1] (caseI){class 1};
				\node [rblock,right=1 of ker2] (caseII){class 2};
				\node [rblock,below=0.5 of caseII] (caseIII){class 3};
				\node [right=1 of caseI] (1KV){1 KV};
				\node [right=3.4 of ker0] (0KV){no KV};
				\node [right=1 of caseII] (2KVs){2 KVs};
				\node [right=1 of caseIII] (3KVs){3 KVs};
				\node [below=0.5 of 3KVs] (4KVs){4 KVs};
				\node [below=0.5 of caseIII] (6KVs){6 KVs};
				\draw[arrow] (ker0) -- (ker1) node[right,midway] {{\scriptsize ~~no}};
				\draw[arrow] (ker1) -- (ker2) node[right,midway] {{\scriptsize ~~no}};
				\draw[arrow] (ker2) |- (caseIII) node[right,pos=0.2] {{\scriptsize ~~no}};
				\draw[arrow] (ker0) -- (0KV) node[above,midway] {{\scriptsize yes}};
				\draw[arrow] (ker1) -- (caseI) node[above,midway] {{\scriptsize yes}};
				\draw[arrow] (ker2) -- (caseII) node[above,midway] {{\scriptsize yes}};
				\draw[arrow] (caseI.east) -- (0KV.west);
				\draw[arrow] (caseI.east) -- (1KV.west);
				\draw[arrow] (caseII.east) -- (2KVs.west);
				\draw[arrow] (caseII.east) -- (3KVs.west);
				\draw[arrow] (caseIII.east) -- (3KVs.west);
				\draw[arrow] (caseIII.east) -- (4KVs.west);
				\draw[arrow] (caseIII.south) -- (6KVs.north);
				\draw[arrow,dotted] (caseII.north) -- (caseI.south);
				\draw[arrow,dotted] (caseIII.north) -- (caseII.south);
				\end{tikzpicture}
		\caption{Main algorithm to determine the number of linearly independent KVs for a 3-dimensional spacetime.
			It consists of the three criteria [see \eqs \eqref{eq:ann_cases}] and the 3-round blocks.
			The blocks include sub-algorithms which are respectively given in Figures \ref{fig:caseI}--\ref{fig:Segre}.
			It has the nest structures indicated by dotted lines.
			By connecting a path which is possibly a combination of solid and dotted lines, the number of KVs is exactly counted.
			Note that the case admitting $5$ KVs is rigid, in the sense that it inevitably generates 6th KV (see e.g. \cite{Kruglikov11} and references therein).
		}
		\label{fig:main}
	\end{center}
\end{figure}

Our main results in this paper can be summarised as follows:
\begin{theorem}
	Let $(M, g_{ab})$ be a 3-dimensional Lorentzian manifold.
	The number of linearly independent Killing vectors is counted by an algorithm described in Figure \ref{fig:main}.
	It includes sub-algorithms given in Figures \ref{fig:caseI}--\ref{fig:Segre}.
\end{theorem}

It is noteworthy that the algorithm shown in Figure \ref{fig:main} has the {\it nest structures}:
The sub-algorithm for the class 2 contains that for the class 1 as sub-sub-algorithms;
Similarly, the sub-algorithm for the class 3 contains not just that for the class 1 but also that for the class 2 as sub-sub-algorithms.
As we will see in Sections \ref{sec:caseII} and \ref{sec:caseIII}, such structures serve
to avoid unnecessary repetition and to simplify the whole algorithm.

The rest of this paper is organised as follows.
In Sections \ref{sec:caseI}--\ref{sec:caseIII},
we analyse the Killing equation in classes  1 to 3 in sequence.
The corresponding obstruction matrices and sub-algorithm are given explicitly.
We enlighten readers about the power of our algorithm with some instructive examples in Section \ref{sec:application}. 
We first inspect the Lifshitz spacetime admitting a single parameter $z$, the value of which controls the 
number of local isometry. After this simple exercise, a special attention is payed to the plane-fronted wave with parallel rays (pp-wave), which admits a 
covariantly constant null vector and vanishing scalar invariants. This metric epitomises the Lorentzian signature 
and is specified by a single function. We provide the complete classification of local isometry which turns out to be controlled by the profile of this function.
We close this paper with some comments in Section \ref{sec:conclusion}.
Technical formulae are summarised in \ref{app:rels}. An exhaustive classification of spacetimes admitting 4 KVs is given in \ref{app:4KVs}.  
This also serves as an insightful guide to confirm the vindication of the present paper.

Remark that we shall  use the same symbol for different sections and subsections recurrently
in order to minimise the number of symbols and to lighten the notation.
We caution the readers not to be confused by this abuse of notation.

\section{Analysis of class 1}\label{sec:caseI}

Let us begin our analysis with the class 1, in which any KV can be written as
\begin{align}
K^a ~\propto~ u^a\:,
\label{eq:caseI}
\end{align}
where $u^a$ is a vector that annihilates ${\boldsymbol R}_a$.
The annihilator $u^a$ must be specified beforehand, but the results in this section does not depend on the explicit form of $u^a$.

For $\dim \ker {\boldsymbol R}_a =1$, there is at least one non-vanishing 2-form in \eq \eqref{eq:ann_caseI}, which allows us to 
take $u^a$ to be the Hodge dual of it.
For instance, $u^a \equiv \epsilon^{abc} \nabla_b R \nabla_c S^{(2)}$ for $\D  R \wedge \D  S^{(2)} \neq 0$,
where $\epsilon^{abc}$ is the Levi-Civita tensor.

Our discussion has two offshoots according to 
whether $u^a$ is timelike or spacelike (Subsection \ref{subsec:caseI_nonnull}), whilst
$u^a$ is null (Subsection \ref{subsec:caseI_null}).
Subsection \ref{sec:summary_I} gives short summary of this section.

\subsection{Non-null case}\label{subsec:caseI_nonnull}

When $u^a$ is timelike or spacelike, we can normalise an annihilator of ${\boldsymbol R}_a$ to unity by setting
\begin{align}
e^a~\equiv~ \frac{u^a}{\sqrt{\iota\:g_{bc} u^b u^c}} \:, \qquad 
g_{ab}e^a e^b= \iota \,, 
\label{eq:annihilator_I}
\end{align}
where $\iota \equiv \mathrm{sgn} (g_{ab}u^a u^b)$.
We also introduce the  tensor
\begin{align}
h_{ab}(e) ~\equiv~ g_{ab} -\iota\:e_a e_b\:,
\label{eq:projection}
\end{align}
that is endowed with a projection property and an orthogonality
\begin{align}
&h^a{}_c h^c{}_b ~=~ h^a{}_b\:,
&&h_{ab}\:e^b ~=~ 0\:.
\end{align}
In this case any KV takes the form
\begin{align}
K^a = \omega \:e^a\:,
\label{eq:ansatz_I}
\end{align}
where $\omega$ is an unknown scalar.
By using the projection tensor \eqref{eq:projection} and the form \eqref{eq:ansatz_I},
one can boil down each component of \eq \eqref{eq:Killing} to
\begin{subequations}
	\begin{align}
	\label{eq:Killing_case1nonnull_ee}
	0~=~ &e^a e^b \nabla_{(a} K_{b)} ~=~\iota\: \pounds_e \omega\:,
	\\
		\label{eq:Killing_case1nonnull_eh}
	0~=~ &e^a h^b{}_c \nabla_{(a} K_{b)} ~=~
	\tfrac{1}{2}(\iota \nabla_c \omega -e_c \pounds_e \omega -\iota \Omega_c^\iota \omega)\:,\\
	0~=~ &h^a{}_c h^b{}_d \nabla_{(a} K_{b)}
	~=~\omega\:\kappa_{cd}\:,
	\end{align}
\end{subequations}
where
\begin{align}
&\Omega_a^\iota (e) ~\equiv~ -\iota\:e^b \nabla_b e_a\:,
&& \kappa_{ab}(e) ~\equiv~
h^c{}_a h^d{}_b \nabla_{(c} e_{d)}\:.
\label{eq:obsts_I}
\end{align}
It follows that the Killing equation \eqref{eq:Killing} amounts to 
\begin{align}
&\kappa_{ab} ~=~0\:,
&&\nabla_a \omega ~=~ \Omega_a^\iota\:\omega\:.
\label{eq:Killing_caseI}
\end{align}
It is noted that the condition (\ref{eq:Killing_case1nonnull_ee}) follows from 
the second equation in (\ref{eq:Killing_caseI}). 
The compatibility condition of the latter equation reads $\nabla_{[a} \Omega_{b]}^\iota = 0$.

As a result, the necessary and sufficient conditions for the local solvability of \eq \eqref{eq:Killing}
are aggregated into an algebraic equation
\begin{align}
&({\boldsymbol R}_{\mathrm{cls. 1}}^\iota)_{ab}~=~0\:,
&&({\boldsymbol R}_{\mathrm{cls. 1}}^\iota)_{ab} ~\equiv~
\begin{bmatrix}
\kappa_{ab} \\
\nabla_{[a} \Omega_{b]}^\iota
\end{bmatrix}
\:,
\label{eq:compatibility_I_nonnull}
\end{align}
yielding tests for $e^a$.
If the equation \eqref{eq:compatibility_I_nonnull} is satisfied, 
there are no extra obstructions for the existence of the Killing vector.
This means  that precisely one KV exists. On the other hand, the failure of \eqref{eq:compatibility_I_nonnull} 
means that there exist no KVs.

\subsection{Null case}
\label{subsec:caseI_null}

In this case, we directly write the KV as
\begin{align}
K^a = \omega \:u^a\:,
\label{eq:ansatz_In}
\end{align}
where $\omega$ is an unknown scalar, keeping the same notation in Subsection \ref{subsec:caseI_nonnull}.
We also define the projection tensor
\begin{align}
q_{ab}(u,v) ~\equiv~ g_{ab} -\:u_a v_b - v_a u_b\:,
\label{eq:projection_n}
\end{align}
where $v^a$ is a vector filed satisfying $g_{ab} v^a v^b=0$ and $g_{ab} v^a u^b =1$\footnote{Remark that the two conditions do not determine $v^a$ uniquely.
We need to make a particular choice of $v^a$ in $M$ [a section of a frame bundle $F(M)$]. 
In spite of this ambiguity, the final outcomes  are insensitive to the choice of $v^a$.}.
The tensor \eqref{eq:projection_n} is projective and orthogonal 
\begin{align}
&q^a{}_c q^c{}_b ~=~ q^a{}_b\:,
&&q^a{}_b\:u^b ~=~ 0\:,
&&q^a{}_b\:v^b ~=~ 0\:.
\end{align}

With the help of \eqs \eqref{eq:ansatz_In} and \eqref{eq:projection_n}, 
the components of \eq \eqref{eq:Killing} can be written as follows:
\begin{subequations}
\label{eq:Killing_case1null}
	\begin{align}
	0~=~ &v^a v^b \nabla_{(a} K_{b)} ~=~ \omega\:v^a v^b \nabla_a u_b + \pounds_v \omega\:,
	\\
	0~=~ &u^a v^b \nabla_{(a} K_{b)} ~=~ \tfrac{1}{2} (\omega\:u^a v^b \nabla_a u_b + \pounds_u \omega )\:,
	\\
	0~=~ &u^a q^b{}_c \nabla_{(a} K_{b)}
	~=~\tfrac{1}{2} \omega\: \theta_c\:,\\
	0~=~ &v^a q^b{}_c \nabla_{(a} K_{b)}
	~=~q^b{}_c( \omega\:v^a \nabla_{(a} u_{b)}+\tfrac{1}{2} \nabla_b \omega)\:,\\
	0~=~ &q^a{}_c q^b{}_d \nabla_{(a} K_{b)}
	~=~\omega\:\kappa\:q_{cd}\:,
	\label{eq:Killing_case1null_qq}
	\end{align}
\end{subequations}
where we remark that the $uu$-component is satisfied automatically and 
the ``shear term'' in (\ref{eq:Killing_case1null_qq}) identically vanishes since $q_{ab}$ admits only a single nonvanishing component. 
Here we have introduced
\begin{align}
&\kappa(u,v)~\equiv~ q^{ab}\:\nabla_a u_b\:,
&&\theta_a(u,v)~\equiv~ u^b \nabla_b u_a - (u^b v^c \nabla_b u_c) u_a\:.
\label{eq:obsts_I_n}
\end{align}
From above equations, it follows that the satisfaction of the Killing equation 
is tantamount to
\begin{align}
&\kappa~=~0\:,
&&\theta_a ~=~ 0\:,
&&\nabla_a \omega~=~\Omega_a\:\omega\:,
\label{eq:Killing_caseI_null}
\end{align}
where
\begin{align}
\Omega_a (u,v)~\equiv~-2v^b\nabla_{(a} u_{b)} +u_a v^b v^c \nabla_{b} u_c\:.
\label{eq:obsts_I_n2}
\end{align}
The compatibility condition of the third equation in (\ref{eq:Killing_caseI_null}) reads $\nabla_{[a} \Omega_{b]} = 0$.

As a result, the necessary and sufficient conditions for the local solvability of \eq \eqref{eq:Killing}
are aggregated into an algebraic equation
\begin{align}
&({\boldsymbol R}_{\mathrm{cls. 1}})_{ab}~=~0\:,
&&({\boldsymbol R}_{\mathrm{cls. 1}})_{ab}~\equiv~
\begin{bmatrix}
\kappa\:q_{ab} \\
u_{(a}\theta_{b)}\\
\nabla_{[a} \Omega_{b]}
\end{bmatrix}\:,
\label{eq:compatibility_I_null}
\end{align}
yielding tests for $u^a$ and $v^a$.
If the equation \eqref{eq:compatibility_I_null} is satisfied, there are no extra conditions to be satisfied.
Hence, one null KV exists.

\subsection{Short summary of class 1}\label{sec:summary_I}

\begin{figure}[h]
	\begin{center}
						\begin{tikzpicture}
						[
						every node/.style={outer sep=0.15cm, inner sep=0cm},
						arrow/.style={-{Stealth[length=0.15cm]},thick},
						rblock/.style={rectangle, rounded corners,draw, minimum height = 0.5cm,
							minimum width=1.6cm, thick, outer sep = 0},
						point/.style={radius=2pt}
						]
						\node [] (NullorNot){$\blacktriangleright \:g_{ab}u^a u^b = 0$};
						\node [rblock,above=0.25 of NullorNot, xshift=-1.0cm] (title){class 1};
						\node [right=1 of NullorNot] (ObstNull){$({\boldsymbol R}_{\mathrm{cls. 1}})_{ab}= 0$};
						\node [below=0.5 of ObstNull] (Obst){$({\boldsymbol R}_{\mathrm{cls. 1}}^\iota)_{ab} = 0$};
						\node [right=1.5 of ObstNull] (0KV){no KV};
						\node [right=1.5 of Obst] (1KV){1 KV};
						\draw[arrow] (NullorNot) -- (ObstNull) node[above,midway] {{\scriptsize yes}};
						\draw[arrow] (NullorNot) |- (Obst) node[right,pos=0.3] {{\scriptsize ~~no}};
						\draw[name path=cross1, arrow] (ObstNull.east) -- (1KV.west);
						\node[left=0.1 of 1KV, yshift=0.3cm, xshift=0.45cm] {{\scriptsize yes}};
						\draw[name path=cross2, arrow] (Obst.east) -- (0KV.west);
						\node[left=0.1 of 0KV, yshift=0.3cm, xshift=0.4cm] {{\scriptsize no}};
						\path[name intersections={of=cross1 and cross2,by={cross}}];
						\fill [point] (cross) circle;
						\end{tikzpicture}
		\caption{The sub-algorithm for the class 1, $K^a \propto u^a$.
			See \eqs \eqref{eq:compatibility_I_nonnull} and
			\eqref{eq:compatibility_I_null} for notations.}
		\label{fig:caseI}
	\end{center}
\end{figure}
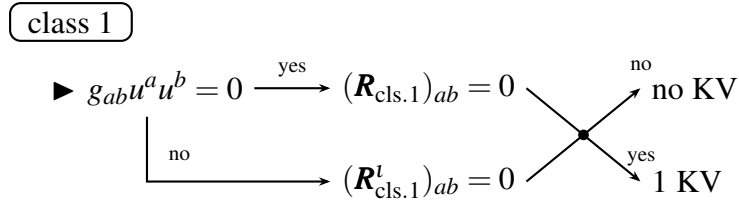

We summarise the results here in Figure \ref{fig:caseI}.

It deserves to emphasise that the sub-algorithm in Figure \ref{fig:caseI} is  applicable also for some cases in class $2$ and $3$ as explained in Section \ref{sec:introduction}:
It might be seemingly appreciated that the prescription in class 2 and 3 is not be adaptive to 
class 1, since KVs in either case are expressed as a linear combination of
two (or three) annihilators of ${\boldsymbol R}_a$ with $\dim \ker {\boldsymbol R}_a > 1$. 
Nevertheless, an essential {\it terminus a quo} for the argument in class 1 is 
the assumption  $K^a \propto u^a$ in \eqref{eq:caseI}, 
rather than $\dim \ker {\boldsymbol R}_a=1$.  Indeed, the situation we shall encounter in class 2 and 3 is that the KVs in several branches are proportional to an annihilator of ${\boldsymbol R}_a$, whereas $\dim \ker {\boldsymbol R}_a > 1$. 
Since $K^a \propto u^a$ conforms to the applicability of class 1, 
no harm is caused in pretending that the recipe in this section is reusable also for such branches.
The recyclability of the analysis significantly reduces the total amount of calculations.

\section{Analysis of class 2}\label{sec:caseII}

Our focus in this section is centred on the class 2, in which any KV can be written as
\begin{align}
K^a ~=~ \sum_{\alpha=1}^2 \omega_\alpha \: u^a_\alpha\:,
\label{eq:caseII}
\end{align}
where $\{u^a_\alpha\}$ are vectors that annihilates ${\boldsymbol R}_a$.
The annihilators must be specified beforehand, but the results in this section do not depend on their explicit form.

For $\dim \ker {\boldsymbol R}_a =2$, there is only one linearly independent 1-form in \eq \eqref{eq:ann_caseII}, say $u_a$
\footnote{
To reduce the number of symbols needed, we have employed the same symbol as that in Section \ref{sec:caseI}.
The reader is cautioned not to confuse it with \eq \eqref{eq:caseI}.
}.
It turns out that  any vector orthogonal to $u^a=g^{ab}u_b$ is an annihilator of ${\boldsymbol R}_a$.
In particular, $u^a$ itself is the annihilator if it is null, for which a special handling is required.

We are proceeding along two cases
where Subsection \ref{subsec:caseII_nonnull} treats the case where the two annihilators are both non-null, whilst
Subsection \ref{subsec:caseI_null} discusses either of them is null. 
We do not try to examine the case in which two annihilators are null and non-parallel, since 
we can bring this case to the non-null case by a suitable change of basis.\footnote{
In an arbitrary dimensional Lorentzian manifold, two null vectors orthogonal to each other must be proportional. This fact allows us to exclude this possibility.}
Subsection \ref{sec:summary_II} gives short summary of this section.

\subsection{Non-null case}
\label{subsec:caseII_nonnull}

In this case, it is assumed that $\{e_i^a, i=1,2,3\}$ forms an orthonormal basis of $T(M)$,
\begin{align}
g^{ab} ~=~ \iota\:e_1^a e_1^b + e_2^a e_2^b -\iota\: e_3^a e_3^b\:,
\label{eq:frame_caseII}
\end{align}
where $\iota \equiv \mathrm{sgn} (g_{ab}e_1^a e_1^b)$, and
two vectors $\{e_2^a, e_3^a\}$ are two annihilators of ${\boldsymbol R}_a$.
Remark that $e_2^a$ is spacelike, whereas $e_3^a$ is either spacelike for $\iota=-1$ or timelike for $\iota=+1$.

For $\dim \ker {\boldsymbol R}_a =2$, the basis $\{e_i^a\}$ is taken as follows:
One can choose $e_1^a$ in such a way that it is proportional to $u^a$, i.e.
\begin{align}
e_1^a ~\equiv~ \frac{u^a}{\sqrt{\iota_u\:g_{bc}u^b u^c}}\:,
\label{eq:1vec_II}
\end{align}
where $\iota_u \equiv \mathrm{sgn} (g_{ab}u^a u^b)$.
So $e_2^a$ can be taken as a vector such that $g_{ab}e_1^a e_2^b =0$ and $g_{ab} e_2^a e_2^b =1$,\footnote{
	Similar to $v^a$ in Subsection \ref{subsec:caseI_null} 
	the two conditions do not determine $e_2^a$ uniquely.
	We need to make a particular choice of $e_2^a$ in $M$.
	Again, the results here do not depend on the choice of $e_2^a$.} 
in terms of which one can specify $e_3^a$ to be $e_3^a \equiv \epsilon^{abc}e_{1b}e_{2c}$.

Given these orthogonal frame $\{e_i^a\}$, any KV can be written in the form
\begin{align}
K^a ~=~\omega_2\:e_2^a +\omega_3\:e_3^a\:,
\label{eq:ansantz_II}
\end{align}
where scalars $\{ \omega_2, \omega_3\}$ are yet indeterminate.

Instead of writing down the components of \eq \eqref{eq:Killing} with \eq \eqref{eq:ansantz_II} by the coordinate basis, it is much more convenient to work with the connection components. 
For this purpose, let us introduce the Ricci rotation coefficients as
\begin{subequations}
	\begin{align}
	e_1^b \nabla_b
	\left[\begin{matrix}
	e_1^a \\
	e_2^a \\
	e_3^a
	\end{matrix}\right]
	&~=~
	\left[
	\begin{matrix}
	0 	& \kappa_1 & - \iota  \eta_1 \\
	-\iota \kappa_1	& 0 & -\iota \tau_1 \\
	-\iota \eta_1	& -\tau_1 & 0
	\end{matrix}\right]
	\left[\begin{matrix}
	e_1^a \\
	e_2^a \\
	e_3^a
	\end{matrix}\right]\:,\\
	e_2^b \nabla_b
	\left[\begin{matrix}
	e_1^a \\
	e_2^a \\
	e_3^a
	\end{matrix}\right]
	&~=~
	\left[\begin{matrix}
	0& -\kappa_2 & -\iota \tau_2 \\
	\iota \kappa_2	& 0 & -\iota \eta_2 \\
	-\iota \tau_2	& -\eta_2 & 0
	\end{matrix}\right]
	\left[\begin{matrix}
	e_1^a \\
	e_2^a \\
	e_3^a
	\end{matrix}\right]\:, \\
	e_3^b \nabla_b
	\left[\begin{matrix}
	e_1^a \\
	e_2^a \\
	e_3^a
	\end{matrix}\right]
	&~=~
	\left[\begin{matrix}
	0& \tau_3 & \iota \kappa_3 \\
	-\iota \tau_3 & 0 &\iota \eta_3 \\
	\iota \kappa_3& \eta_3 & 0
	\end{matrix}\right]
	\left[\begin{matrix}
	e_1^a \\
	e_2^a \\
	e_3^a
	\end{matrix}\right]\:,
	\end{align}
\label{eq:ricci_coeffs}
\end{subequations}
where $\kappa_i$, $\eta_i$, and $\tau_i$ are respectively the geodesic curvature,
normal curvature and relative torsion of an integral curve of $e_i^a$.
Their derivatives entail relations amongst each other,
which are collected in \ref{app:rels_ortho}.

By using \eqs \eqref{eq:frame_caseII}--\eqref{eq:ricci_coeffs},
the $11$-part of \eq \eqref{eq:Killing} can be  formally boiled down to 
\begin{align}
\kappa_1\:\omega_2+\eta_1\:\omega_3~=~0\:,
\label{eq:11_II}
\end{align}
yielding tests for $e_1^a$.
This implies that the analysis branches off,
depending on whether $e_1^a$
satisfies the geodesic equation $e_1^b \nabla_b e_1^a ~=0$.

\subsubsection{Branch where $e_1^a$ is not a geodesic tangent $\kappa_1\eta_1\ne 0$}
\label{subsec:caseII_nonnull_branch1}

From \eq \eqref{eq:11_II}, $\omega_2$ and $\omega_3$ are related to each other,
$\omega_3 = -(\kappa_1/\eta_1) \omega_2$, or $\omega_2 = -(\eta_1/\kappa_1) \omega_3$.
This allows us to rewrite \eq \eqref{eq:ansantz_II} as
\begin{align}
&K^a~=~ \omega_2 \left( e_2^a - \frac{\kappa_1}{\eta_1} e_3^a \right)\:,
&&\text{or}
&&K^a~=~ \omega_3 \left( e_3^a - \frac{\eta_1}{\kappa_1} e_2^a \right)\:,
\end{align}
which matches the assumption \eqref{eq:caseI} of the class 1.
It turns out that the annihilator of ${\boldsymbol R}_a$ is specified as $e_2^a - (\kappa_1/\eta_1) e_3^a$
or $e_3^a - (\eta_1/\kappa_1) e_2^a$.
As explained in \ref{sec:summary_I}, the sub-algorithm described in Figure \ref{fig:caseI} can be immediately testable.

\subsubsection{Branch where $e_1^a$ is a geodesic tangent $\kappa_1=\eta_1=0$}
\label{subsec:caseII_nonnull_branch2}

In this branch, the $11$-part of the Killing equation \eqref{eq:11_II}
is satisfied automatically.
The remaining parts read
\begin{subequations}
	\begin{align}
	\pounds_1 \omega_2 &~=~ -\kappa_2\:\omega_2 + (\tau_1+\tau_2)\omega_3\:,\\
	\pounds_1 \omega_3 &~=~\iota (\tau_1 - \tau_3) \omega_2 +\iota\:\kappa_3\:\omega_3\:,\\
	\pounds_2 \omega_2 &~=~ \eta_2\:\omega_3\:,\\
	\pounds_2 \omega_3 &~=~ \varpi \:, \label{eq:1-jet_II}\\
	\pounds_3 \omega_2 &~=~ -\eta_2\:\omega_2 - \eta_3\:\omega_3 + \iota\:\varpi\:,\\
	\pounds_3 \omega_3 &~=~ -\iota\:\eta_3\:\omega_2\:.
	\end{align}
	\label{eq:Killing_II}
\end{subequations}
where $\pounds_i$ denotes the Lie derivative along $e_i^a$ and
\eq \eqref{eq:1-jet_II} is the defining equation of the 1-jet variable $\varpi$.
Since the PDE system \eqref{eq:Killing_II} is not closed with respect to unknown scalars $\{\omega_2, \omega_3, \varpi\}$,
we need additional relations between them.
Such relations come from several parts of the identities $\nabla_{[a} \nabla_{b]} \omega_2 = \nabla_{[a} \nabla_{b]} \omega_3 = 0$ [c.f.  eq. (\ref{eq:app_comm_ortho})], 
leading to 
\begin{subequations}
	\begin{align}
	2(\tau_3 - \tau_2)\: \varpi~=~&
	-\Bigl(
	\pounds_2 (\kappa_2 -\iota \kappa_3)
	+\iota \pounds_3(\tau_3 - \tau_2)
	-2\iota \eta_2 (\tau_3-\tau_2)
	\Bigr)\omega_2 \notag\\
	&-\Bigl(\pounds_2 (\tau_3-\tau_2)+\pounds_3(\kappa_2-\iota \kappa_3) \Bigr) \omega_3\:,
	\label{eq:compatibility_II_1}\\
	2(\tau_3 + \tau_2)\: \varpi~=~&
	\Bigl(\pounds_2 (\kappa_2+\iota\kappa_3) -\iota\pounds_3(\tau_3 - \tau_2) \Bigr)\omega_2 \notag\\
	&+\Bigl(
	\pounds_2(\tau_3-\tau_2)
	+\pounds_3(\kappa_2 +\iota \kappa_3)
	+2 \iota \eta_3 (\tau_3+\tau_2)
	\Bigr) \omega_3\:,
	\label{eq:compatibility_II_2}
	\\
	(\kappa_2+\iota \kappa_3)\:\varpi~=~&
	\iota (\pounds_2 \tau_3) \omega_2
	+\iota \Bigl(\eta_3 (\kappa_2+ \iota \kappa_3) +\pounds_3 \tau_2\Bigr) \omega_3\:.
	\label{eq:compatibility_II_3}
	\end{align}
	\label{eq:compatibility_II}
\end{subequations}
This implies that $\varpi$ can be expressed in terms of $\omega_2$ and $\omega_3$
except for $\kappa_2+\iota \kappa_3 =\tau_3=\tau_2=0$.
Depending on the nonzeroness of the coefficients $\{\kappa_2+\iota \kappa_3, \tau_2+\tau_3, \tau_2-\tau_3\}$, the analysis further falls into four sub-branches.

\subsubsection*{\underline{Sub-branch where $\tau_2 = \tau_3 = \kappa_2+\iota \kappa_3 = 0$}}

In this sub-branch, the 1-jet variable $\varpi$ cannot be expressed in terms of $\omega_2$ and $\omega_3$.
The differential equations for $\varpi$ come from the remaining parts of the identities $\nabla_{[a} \nabla_{b]} \omega_2 = \nabla_{[a} \nabla_{b]} \omega_3 = 0$.
By combining this and \eqs \eqref{eq:Killing_II}, we obtain a PDE system
\begin{align}
&\nabla_a {\boldsymbol \omega} ~=~{\boldsymbol \Omega}_a^\iota\:{\boldsymbol \omega}\:,
&&{\boldsymbol \omega} ~\equiv~
[\omega_2~~ \omega_3~~ \varpi]^T\:,
\label{eq:Killing_II_1}
\end{align}
where
\begin{align}
{\boldsymbol \Omega}_a^\iota ~\equiv~&
\iota e_{1a}
\left[\begin{smallmatrix}
-\kappa_2 & \tau_1 & 0\\
\iota \tau_1 & -\kappa_2 & 0 \\
\iota \pounds_2 \tau_1 + \eta_3 \tau_1~
&~\iota \eta_2 \tau_1 - \pounds_2 \kappa_2~& 0
\end{smallmatrix}\right]
+e_{2a}
\left[\begin{smallmatrix}
0 & \eta_2 & 0\\
0 & 0 & 1 \\
\iota\pounds_2 \eta_2~
&~\eta_3^2 +\iota \pounds_3 \eta_2 +\iota \pounds_2 \eta_3~
& 0
\end{smallmatrix}\right]\notag\\
&
-\iota e_{3a}
\left[
\begin{smallmatrix}
-\eta_2 & -\eta_3 & \iota\\
-\iota \eta_3 & 0 & 0 \\
-\iota \pounds_2 \eta_3 -\eta_3^2~&
~-\iota \eta_2 \eta_3~&
\eta_2
\end{smallmatrix}
\right]
\:.
\end{align}
The compatibility condition, $( \nabla_{[a} {\boldsymbol \Omega}_{b]}^\iota
- {\boldsymbol \Omega}_{[a}^\iota {\boldsymbol \Omega}_{b]}^\iota
){\boldsymbol \omega} = 0$, for \eq \eqref{eq:Killing_II_1} reads
\begin{align}
&{\boldsymbol R}_{\mathrm{cls. 2}}^{\iota}\:{\boldsymbol \omega} ~=~ 0\:,
&&
{\boldsymbol R}_{\mathrm{cls. 2}}^{\iota}~\equiv~
\begin{bmatrix}
\pounds_2 \kappa_2 & \pounds_3 \kappa_2 & 0\\
\pounds_2 \lambda_2 & \pounds_3 \lambda_2 & 0
\end{bmatrix}\:,
\label{eq:compatibility_II_1_Sub}
\end{align}
where $\lambda_2\equiv R_{ab} e_2^a e_2^b$. Remark that 
the first line corresponds to \eq \eqref{eq:compatibility_II_1}, and 
some remaining components are derivable from its derivative.

In this sub-branch, the rank of ${\boldsymbol R}_{\mathrm{cls. 2}}^{\iota}$ controls the number of KVs:
If $\mathrm{rank} {\boldsymbol R}^{\iota}_{\mathrm{cls. 2}} = 0$, three KVs exist;
If $\mathrm{rank} {\boldsymbol R}^{\iota}_{\mathrm{cls. 2}} = 2$, there is no KV;
Otherwise $\mathrm{rank} {\boldsymbol R}^{\iota}_{\mathrm{cls. 2}} = 1$,
$\omega_2$ and $\omega_3$ are related to each other.
This implies that the KVs in this branch are proportional to an annihilator of ${\boldsymbol R}_a$.
As explained in \ref{sec:summary_I}, the sub-algorithm described in Figure \ref{fig:caseI} can be immediately testable.

\subsubsection*{\underline{Other sub-branches}}

Except for the case of $\tau_2 = \tau_3 = \kappa_2+\iota \kappa_3 = 0$,
\eqs \eqref{eq:compatibility_II} allow us to write the 1-jet variable $\varpi$ in terms of $\{\omega_2, \omega_3\}$.
In these sub-branches, \eq \eqref{eq:Killing_II} is closed with respect to $\{\omega_2, \omega_3\}$ and then
one needs to consider separately the compatibility of PDE systems of the form
\begin{align*}
\nabla_a
{\boldsymbol \omega}
~=~
{\boldsymbol \Omega}_a^\iota\:{\boldsymbol \omega}\:,
&&
{\boldsymbol \omega}~\equiv~
[\omega_2 ~~\omega_3]^T\:.
\end{align*}
Remark that explicit forms of the connection ${\boldsymbol \Omega}_a^\iota$ and
the obstruction matrix ${\boldsymbol R}_{\mathrm{cls. 2}}^\iota$
depend on the nonzeroness of the coefficients $\{\kappa_2+\iota \kappa_3, \tau_2, \tau_3\}$.
We thus number ${\boldsymbol R}^\iota_{\mathrm{cls. 2}}$ serially (\#1)--(\# 3) and 
each result is displayed as follows:

\begin{subequations}
	\noindent{\it (\#1) For the case of $\tau_2 = \tau_3 =0, \kappa_2+\iota \kappa_3 \neq 0$},
\begin{align}
\varpi &~=~ \iota \eta_3 \omega_3\:,\\
{\boldsymbol \Omega}_a^\iota &~=~
\iota e_{1a}
\begin{bmatrix}
-\kappa_2 & \tau_1 \\
\iota \tau_1 & \iota \kappa_3
\end{bmatrix}
+e_{2a}
\begin{bmatrix}
0 & \eta_2 \\
0 & \iota \eta_3
\end{bmatrix}
- \iota e_{3a}
\begin{bmatrix}
-\eta_2 & 0 \\
-\iota \eta_3 & 0
\end{bmatrix}\:,\\
{\boldsymbol R}_{\mathrm{cls. 2}}^{\iota~\# 1} &~=~
\begin{bmatrix}
\pounds_2 \kappa_2 & \pounds_3 \kappa_2 \\
\pounds_2 \kappa_3 & \pounds_3 \kappa_3 \\
\pounds_2 \eta_2 & \pounds_3 \eta_2 \\
\pounds_2 \eta_3 & \pounds_3 \eta_3 \\
\pounds_2 \tau_1 & \pounds_3 \tau_1 \\
\end{bmatrix}\:.
\end{align}
\label{eq:compatibility_II_1_1}
\end{subequations}
\begin{subequations}
	{\it (\#2) For the case of $\tau_2 = \tau_3 \neq 0$},
\begin{align}
\varpi ~=~& \left(\frac{\pounds_2 \kappa_\delta}{4 \tau_2}\right)\omega_2
+\left(\frac{\pounds_3  \kappa_\delta}{4\tau_2}
+ \iota \eta_3 \right)\omega_3\:,\\
{\boldsymbol \Omega}_a^\iota ~=~&
\iota e_{1a}
\left[
\begin{matrix}
- \frac{\kappa_\delta + \kappa_\sigma}{2} & \tau_1 + \tau_2 \\
\iota (\tau_1-\tau_2) & \frac{\kappa_\delta - \kappa_\sigma}{2}
\end{matrix}
\right]
+e_{2a}
\left[
\begin{matrix}
0 & \eta_2 \\
\frac{\pounds_2 \kappa_\delta}{4\tau_2} & \frac{\pounds_3 \kappa_\delta}{4\tau_2} + \iota \eta_3 
\end{matrix}
\right]
- \iota e_{3a}
\left[
\begin{matrix}
\frac{\iota \pounds_2 \kappa_\delta}{4\tau_2} -\eta_2 & \frac{\iota \pounds_3 \kappa_\delta}{4\tau_2} \\
-\iota \eta_3 & 0
\end{matrix}
\right]\:,\\
{\boldsymbol R}_{\mathrm{cls. 2}}^{\iota~\# 2} ~=~&
\left[
\begin{matrix}
\pounds_2 \kappa_\sigma & \pounds_3 \kappa_\sigma \\
\pounds_2 (\kappa_\delta^2 -4 \iota \tau_2^2) & \pounds_3 (\kappa_\delta^2 -4 \iota \tau_2^2) \\
\pounds_2 \left( \frac{\lambda_2 + \iota \lambda_3 + \iota \kappa_\delta \kappa_\sigma }{\tau_2} \right)
-\frac{2\iota (\pounds_{[2}\kappa_\delta)(\pounds_{1]} \tau_2)}{\tau_2^2}
&
\pounds_3 \left( \frac{\lambda_2 + \iota \lambda_3 + \iota \kappa_\delta \kappa_\sigma }{\tau_2} \right)
-\frac{2\iota (\pounds_{[3}\kappa_\delta)(\pounds_{1]} \tau_2)}{\tau_2^2}
\\
	\pounds_2 \psi_2
	+ \frac{\psi_3 \pounds_2 \kappa_\delta}{4 \tau_2}
&\pounds_3 \psi_2 + \frac{\psi_3 \pounds_3 \kappa_\delta}{4 \tau_2}
	+\frac{2\iota(\pounds_{[2} \kappa_\delta)(\pounds_{3]} \tau_2)}{\tau^2_2}
\\
\pounds_2 \psi_3 + \frac{ \iota \psi_2 \pounds_2 \kappa_\delta}{4 \tau_2}
	-\frac{2\iota(\pounds_{[2} \kappa_\delta)(\pounds_{3]} \tau_2)}{\tau^2_2}
&\pounds_3 \psi_3 + \frac{\iota \psi_2 \pounds_3 \kappa_\delta}{4 \tau_2}
\end{matrix}\right]\:,
\end{align}
where
\begin{align}
\lambda_2 &~\equiv~ R_{ab} e_2^a e_2^b\:,
&\lambda_3 &~\equiv~ R_{ab} e_3^a e_3^b\:, \\
\kappa_\delta &~\equiv~ \kappa_2 + \iota \kappa_3\:,
&\kappa_\sigma &~\equiv~ \kappa_2 -\iota \kappa_3\:,\\
\psi_\alpha &~\equiv~ \frac{\iota \pounds_\alpha \kappa_\delta}{\tau_2} + (-1)^{\alpha-1} 4 \eta_\alpha\:.
&(\alpha &~=~2,3)
\end{align}
\label{eq:compatibility_II_1_2}
\end{subequations}
\begin{subequations}
	{\it (\#3) For the case of $\tau_2 \neq \tau_3$},
	\begin{align}
	\varpi ~=~& - \left(\frac{\pounds_2 \kappa_\sigma + \iota \pounds_3 \tau_\delta - 2 \iota \tau_\delta \eta_2}{2 \tau_\delta} \right) \omega_2
	- \left(\frac{\pounds_3  \kappa_\sigma + \pounds_2 \tau_\delta}{2\tau_\delta}\right)\omega_3\:,\\
	{\boldsymbol \Omega}_a^\iota ~=~&
	\iota e_{1a}
	\left[\begin{matrix}
	-\frac{\kappa_\delta + \kappa_\sigma}{2} &
	\tau_1 - \frac{\tau_\delta - \tau_\sigma}{2} \\
	\iota \tau_1 - \iota \frac{\tau_\delta + \tau_\sigma}{2}  &
	\frac{\kappa_\delta - \kappa_\sigma}{2}
	\end{matrix}\right]
	+e_{2a}
	\left[\begin{matrix}
	0 & \eta_2 \\
	\iota \eta_2 - \frac{\pounds_2 \kappa_\sigma + \iota \pounds_3 \tau_\delta}{2\tau_\delta} &
	- \frac{\pounds_2 \tau_\delta + \pounds_3 \kappa_\sigma}{2\tau_\delta}
	\end{matrix}\right] \notag\\
	&- \iota e_{3a}
	\left[
	\begin{matrix}
	- \frac{ \iota  \pounds_2 \kappa_\sigma + \pounds_3 \tau_\delta}{2\tau_\delta} &
	-\eta_3 - \iota \frac{\pounds_3 \kappa_\sigma + \pounds_2 \tau_\delta}{2\tau_\delta}\\
	-\iota \eta_3 & 0
	\end{matrix}
	\right]\:, \\
	{\boldsymbol R}_{\mathrm{cls. 2}}^{\iota~\# 3} ~=~&
	\left[
	\begin{matrix}
	\pounds_2 \kappa_\delta - \iota \pounds_3 \tau_\delta + (\sigma_2 -2\iota \eta_2) \tau_\sigma &
	\pounds_3 \kappa_\delta + \pounds_2 \tau_\delta+ (\sigma_3+2\iota \eta_3) \tau_\sigma\\
	\pounds_2 (\tau_\sigma+\tau_\delta) + \iota (\sigma_2 -2\iota \eta_2) \kappa_\delta&
	\pounds_3 (\tau_\sigma-\tau_\delta) + \iota (\sigma_3+2\iota \eta_3) \kappa_\delta\\
	\pounds_2 R_{11} -2 \iota \tau_\delta R_{13} & \pounds_3 R_{11} -2 \tau_\delta R_{12}\\
	\pounds_2 \sigma_2
	-\frac{\sigma_3 (\sigma_2 -2 \iota \eta_2)}{2}
	-(\kappa_\delta+\kappa_\sigma) \tau_\delta&
	\pounds_2 \sigma_3
	-\frac{\sigma_3^2}{2}
	+ \sigma_2 \eta_2
	- \frac{(\tau_\sigma-\tau_\delta)^2}{2}
	- \Upsilon\\
	\pounds_3 \sigma_2
	-\frac{\iota \sigma_2^2}{2}
	- \iota \sigma_3 \eta_3
	+ \frac{(\tau_\sigma+\tau_\delta)^2}{2}
	+\Upsilon
	& \pounds_3 \sigma_3
	-\frac{\iota \sigma_2 (\sigma_3+2\iota \eta_3) }{2}
	-\iota (\kappa_\delta-\kappa_\sigma) \tau_\delta
	\end{matrix}\right]\:,
	\end{align}
where $R_{ij} \equiv R_{ab} e_i^a e_j^b$ and
\begin{align}
\tau_\sigma &~\equiv~ \tau_3 + \tau_2\:,
&\tau_\delta &~\equiv~ \tau_3 - \tau_2\:,\\
\sigma_2 &~\equiv~ \frac{\pounds_2 \kappa_\sigma+ \iota \pounds_3 \tau_\delta}{\tau_\delta}\:,
&\sigma_3 &~\equiv~ \frac{\pounds_3 \kappa_\sigma + \pounds_2 \tau_\delta}{\tau_\delta}\:,\\
\Upsilon & ~\equiv~R_{22} - \iota (R_{11} + R_{33} ) - \frac{\iota}{2}(\kappa_\delta^2 - \kappa_\sigma^2)\:.
\end{align}
\label{eq:compatibility_II_1_3}
\end{subequations}

In a way parallel to that of ${\boldsymbol R}^{\iota}_{\mathrm{cls. 2}}$, 
the rank of ${\boldsymbol R}^{\iota~\# 1}_{\mathrm{cls. 2}}, {\boldsymbol R}^{\iota~\# 2}_{\mathrm{cls. 2}}, {\boldsymbol R}^{\iota~\# 3}_{\mathrm{cls. 2}}$ 
controls the number of KVs.
For instance, in the case of $\tau_2 = \tau_3 =0, \kappa_2+\iota \kappa_3 \neq 0$,
$\mathrm{rank} {\boldsymbol R}^{\iota~\# 1}_{\mathrm{cls. 2}} = 0$ implies that two KVs exist.
If $\mathrm{rank} {\boldsymbol R}^{\iota~\# 1}_{\mathrm{cls. 2}}= 2$, we have no KVs.
Otherwise, $\mathrm{rank} {\boldsymbol R}^{\iota~\# 1}_{\mathrm{cls. 2}} = 1$ and then
the KVs in this branch are proportional to an annihilator of ${\boldsymbol R}_a$.
As explained in \ref{sec:summary_I}, the sub-algorithm described in Figure \ref{fig:caseI} can be immediately testable.

\subsection{Null case}
\label{subsec:caseII_null}

In this case, we suppose that $\{u^a,v^a,e^a\}$ spans a double-null basis of $T(M)$,
\begin{align}
g^{ab}~=~u^a v^b + v^a u^b + e^a e^b\:,
\label{eq:d_null_basis_II}
\end{align}
where $\{u^a, v^a\}$ are null vectors such that $g_{ab} u^a v^b=1$;
$e^a$ is a spacelike unit vector orthogonal to $u^a$ and $v^a$;
$\{u^a, e^a\}$ are two annihilators of ${\boldsymbol R}_a$.

When $\dim \ker {\boldsymbol R}_a =2$ and the 1-form in \eq \eqref{eq:ann_caseII} is null,
a null vector $u^a$ can be taken as its contravariant counterpart.
By constructing another null vector $v^a$ satisfying $g_{ab}u^a v^b = 1$,
a spacelike unit vector $e^a$ can also taken as $e^a \equiv \epsilon^{abc}u_b v_c$.

Given the above assumptions, any KV can be written in the form
\begin{align}
K^a~=~ \omega_u \:u^a + \omega_e \:e^a\:,
\label{eq:ansantz_II_null}
\end{align}
where $\{\omega_u, \omega_e\}$ are unknown scalars to be determined.

In order to write the components of \eq \eqref{eq:Killing} with \eq \eqref{eq:ansantz_II_null} explicitly,
we introduce the Ricci rotation coefficients as
\begin{subequations}
	\begin{align}
	u^b \nabla_b
	\left[\begin{matrix}
	u^a \\
	v^a \\
	e^a
	\end{matrix}\right]
	&~=~
	\left[\begin{matrix}
		\kappa_u & 0 & \eta_u \\
	0 &-\kappa_u & \tau_u \\
	-\tau_u & -\eta_u	& 0
	\end{matrix}\right]
	\left[\begin{matrix}
	u^a \\
	v^a \\
	e^a
	\end{matrix}\right]\:,\\
	v^b \nabla_b
	\left[\begin{matrix}
	u^a \\
	v^a \\
	e^a
	\end{matrix}\right]
	&~=~
	\left[\begin{matrix}
	-\kappa_v & 0 & \tau_v \\
	0 & \kappa_v & \eta_v \\
	-\eta_v & -\tau_v	& 0
	\end{matrix}\right]
	\left[\begin{matrix}
	u^a \\
	v^a \\
	e^a
	\end{matrix}\right]\:, \\
	e^b \nabla_b
	\left[\begin{matrix}
	u^a \\
	v^a \\
	e^a
	\end{matrix}\right]
	&~=~
	\left[\begin{matrix}
	\tau_e & 0 &-\kappa_e \\
	0 & -\tau_e & -\eta_e \\
	\eta_e &\kappa_e& 0
	\end{matrix}\right]
	\left[\begin{matrix}
	u^a \\
	v^a \\
	e^a
	\end{matrix}\right]\:.
	\end{align}
	\label{eq:ricci_coeffs_null}
\end{subequations}
where $\kappa_u$, $\eta_u$, and $\tau_u$ are respectively the geodesic,
normal curvature and relative torsion of an integral curve of $u^a$.
The same geometric interpretation is bestowed with quantities for $v^a$ and $e^a$.
Their derivatives entail relations amongst each other,
which are collected in \ref{app:rels_null}.

By using \eqs \eqref{eq:ansantz_II_null}--\eqref{eq:ricci_coeffs_null},
the $uu$-component of \eq \eqref{eq:Killing} reads
\begin{align}
\eta_u\:\omega_e ~=~0\:.
\label{eq:uu_II}
\end{align}
Depending on whether $u^a$ satisfies the geodesic equation $u^b \nabla_b u^a = \kappa_u u^a$,
the analysis branches off.

\subsubsection{Branch where $u^a$ is not a geodesic tangent}\label{subsec:caseII_null_branch1}
In this branch, $\eta_u \neq 0$. Hence \eq \eqref{eq:uu_II} implies $\omega_e = 0$ and then
\eq \eqref{eq:ansantz_II_null} takes the form
\begin{align}
K^a~=~ \omega_u \:u^a \:,
\end{align}
which conforms with the assumption \eqref{eq:caseI} of the class 1.
As explained in \ref{sec:summary_I}, the sub-algorithm described in Figure \ref{fig:caseI} can be immediately testable.

\subsubsection{Branch where $u^a$ is a geodesic tangent $\eta_u=0$}
In this branch, the remaining parts of \eq \eqref{eq:Killing} lead to
\begin{subequations}
	\begin{align}
	\pounds_u \omega_u &~=~ -\kappa_u\:\omega_u + (\tau_u + \tau_v)\:\omega_e\:,\\
	\pounds_u \omega_e &~=~ -\kappa_e\:\omega_e\:,\\
	\pounds_v \omega_u &~=~ \kappa_v\:\omega_u + \eta_v\:\omega_e\:,\\
	\pounds_v \omega_e &~=~ \varpi\:, \label{eq:1-jet_II_null}\\
	\pounds_e\omega_u &~=~ -(\tau_v+\tau_e)\:\omega_u - \eta_e\:\omega_e-\varpi\:,\\
	\pounds_e\omega_e &~=~\kappa_e\:\omega_u\:,
	\end{align}
	\label{eq:Killing_II_null}
\end{subequations}
where \eq \eqref{eq:1-jet_II_null} defines the 1-jet variable $\varpi$.
As is the case in Subsection \ref{subsec:caseII_nonnull},
the PDE system \eqref{eq:Killing_II_null} is not closed with respect to unknown scalars $\{\omega_u, \omega_e, \varpi\}$.
The supplementary equations follow from several parts of the identities $\nabla_{[a} \nabla_{b]} \omega_u=\nabla_{[a} \nabla_{b]} \omega_e=0$, yielding  
\begin{subequations}
\begin{align}
2\tau_v\:\varpi~=~&
\Bigl(
\pounds_u \kappa_v + \pounds_v \kappa_u +2 \kappa_u \kappa_v
+(\tau_u-\tau_v)(\tau_v+\tau_e)
\Bigr)\omega_u\notag \\
&-\Bigl(
\pounds_v (\tau_u+\tau_v)-\pounds_u \eta_v
-\eta_v(2\kappa_u-\kappa_e)
-\eta_e(\tau_u-\tau_v)
\Bigr)\omega_e\:, \label{eq:compatibility_II_null_1}\\
\kappa_e\:\varpi~=~&
\Bigl( \pounds_u \tau_v + \kappa_e (\tau_u-\tau_v)\Bigr)\omega_u
+\Bigl( \pounds_e \tau_v - \kappa_e \eta_e\Bigr)\omega_e\:.
\end{align}
\label{eq:compatibility_II_null}
\end{subequations}
This implies that  the 1-jet variable $\varpi$ is written in terms of $\omega_u$ and $\omega_e$ except for $\tau_v = \kappa_e = 0$.
Depending on the vanishing of the coefficients $\{\tau_v, \kappa_e\}$, the analysis falls into three sub-branches.

\subsubsection*{\underline{Sub-branch where $\tau_v = \kappa_e = 0$}}
In this sub-branch, the 1-jet variable $\varpi$ cannot be expressed in terms of $\omega_u$ and $\omega_e$.
The differential equations for $\varpi$ come from the remaining parts of the identities $\nabla_{[a} \nabla_{b]} \omega_u = \nabla_{[a} \nabla_{b]} \omega_e = 0$.
By combining these with \eqs \eqref{eq:Killing_II_null}, we obtain a PDE system of the form
\begin{align}
&\nabla_a {\boldsymbol \omega} ~=~{\boldsymbol \Omega}_a\:{\boldsymbol \omega}\:,
&&{\boldsymbol \omega} ~\equiv~
[\omega_u~~ \omega_e~~ \varpi]^T\:,
\label{eq:Killing_II_null_1}
\end{align}
where
\begin{align}
{\boldsymbol \Omega}_a ~\equiv~&
u_a
\left[\begin{matrix}
\kappa_v & \eta_v & 0 \\
0 & 0 & 1 \\
\pounds_v \tau_u - \pounds_u \eta_v -2\eta_v \kappa_u - \eta_e \tau_u ~&
-\pounds_v \eta_e - \pounds_e \eta_v +\eta_e (\eta_e+\kappa_v)+\eta_v(\tau_u-2\tau_e) ~&
\kappa_v \\
\end{matrix}\right]\notag\\
&+v_a
\left[\begin{matrix}
-\kappa_u & \tau_u & 0\\
0 & 0 & 0 \\
0 & 0 & -\kappa_u
\end{matrix}\right]
+e_a
\left[\begin{matrix}
-\tau_e & -\eta_e & -1\\
0 & 0 & 0 \\
0 & 0 & -\tau_e
\end{matrix}\right]\:.
\end{align}
The compatibility condition, $( \nabla_{[a} {\boldsymbol \Omega}_{b]}
- {\boldsymbol \Omega}_{[a} {\boldsymbol \Omega}_{b]}
){\boldsymbol \omega} = 0$,  for \eq \eqref{eq:Killing_II_null_1} reads
\begin{align}
&{\boldsymbol R}_{\mathrm{cls. 2}}\:{\boldsymbol \omega} ~=~ 0\:,
&&
{\boldsymbol R}_{\mathrm{cls. 2}}~\equiv~
\begin{bmatrix}
R_{uv}& R_{ve} & 0\\
\pounds_e R_{ve} + \tau_e R_{ve} & -\pounds_e R_{vv} -2 \tau_e R_{vv} -\eta_e R_{ve} & -2R_{ve}
\end{bmatrix}\:,
\label{eq:compatibility_II_null_1_Sub}
\end{align}
where $R_{uv}\equiv R_{ab} u^a v^b, R_{ve} \equiv R_{ab}v^a e^b$ and $R_{vv} \equiv R_{ab} v^a v^b$.
The first line corresponds to \eq \eqref{eq:compatibility_II_null_1}.

In this sub-branch, the rank of ${\boldsymbol R}_{\mathrm{cls. 2}}$ influences the number of KVs
in the same way as that presented in Subsection \ref{subsec:caseII_nonnull_branch2}.

\subsubsection*{\underline{Other sub-branches}}
Except for $\tau_v=\kappa_e=0$,
\eqs \eqref{eq:compatibility_II_null} allow us to express the 1-jet variable $\varpi$ in terms of $\{\omega_u, \omega_e\}$.
In these sub-branches, \eq \eqref{eq:Killing_II_null} is closed with respect to $\{\omega_u, \omega_e\}$ and then
the compatibility of PDE systems is of the form
\begin{align*}
\nabla_a
{\boldsymbol \omega}
~=~
{\boldsymbol \Omega}_a\:{\boldsymbol \omega}\:,
&&
{\boldsymbol \omega}~\equiv~
[\omega_u ~~\omega_e]^T\:. 
\end{align*}
The upshot is as follows:\\
\bigskip
\begin{subequations}
	\noindent{\it (\#1) For the case of $\tau_v =0, \kappa_e \neq 0$},
	\begin{align}
	\varpi &~=~ \tau_u\:\omega_u - \eta_e\:\omega_e\:,\\
	{\boldsymbol \Omega}_a &~=~
	u_a
	\begin{bmatrix}
	\kappa_v & \eta_v \\
	\tau_u & -\eta_e
	\end{bmatrix}
	+v_a
	\begin{bmatrix}
	-\kappa_u & \tau_u \\
	0 & -\kappa_e
	\end{bmatrix}
	+e_a
	\begin{bmatrix}
	-\tau_u-\tau_e& 0 \\
	\kappa_e & 0
	\end{bmatrix}\:,\\
	{\boldsymbol R}_{\mathrm{cls. 2}}^{~\# 1} &~=~
	\begin{bmatrix}
	R_{uv} - \tfrac{1}{2}R_{ee} & R_{ve} + \eta_v\:\kappa_e \\
	\pounds_u \tau_u & \pounds_e \tau_u \\
	\pounds_u \kappa_e -\kappa_u \:\kappa_e & \pounds_e \kappa_e - \tau_e\:\kappa_e\\
	\pounds_u\eta_v + 2\kappa_u\:\eta_v & \pounds_e \eta_v + 2 \tau_e\:\eta_v\\
	\pounds_u\eta_e + \kappa_u\:\eta_e & \pounds_e \eta_e + \tau_e\:\eta_e
	\end{bmatrix}\:,
	\end{align}
	where $R_{ee} \equiv R_{ab}e^a e^b$.
	\label{eq:compatibility_II_null_1_1}
\end{subequations}

\quad

\begin{subequations}
	\noindent{\it (\#2) For the case of $\tau_v \neq 0$},
	\begin{align}
	\varpi ~=~&
	\left(\frac{\pounds_u \kappa_v + \pounds_v \kappa_u + 2\kappa_u \kappa_v + (\tau_u-\tau_v)(\tau_v+\tau_e)}{2\tau_v}\right)\omega_u\notag \\
	&-\left(\frac{\pounds_v (\tau_u + \tau_v) -\pounds_u\eta_v -\eta_v (2\kappa_u - \kappa_e)-\eta_e (\tau_u - \tau_v)}{2\tau_v} \right)\omega_e\:,\\
	{\boldsymbol \Omega}_a ~=~&
	u_a
	\left[\begin{smallmatrix}
	\kappa_v & \eta_v \\
	\frac{\pounds_u \kappa_v + \pounds_v \kappa_u + 2\kappa_u \kappa_v + (\tau_u-\tau_v)(\tau_v+\tau_e)}{2\tau_v} &
	-\frac{\pounds_v (\tau_u + \tau_v) -\pounds_u\eta_v -\eta_v (2\kappa_u - \kappa_e)-\eta_e (\tau_u - \tau_v)}{2\tau_v}
	\end{smallmatrix}\right]
	+v_a
	\biggl[
	\begin{smallmatrix}
	-\kappa_u & \tau_u + \tau_v \\
	0 & -\kappa_e
	\end{smallmatrix}\biggr] \notag \\
	&+e_a
	\left[\begin{smallmatrix}
	-\frac{\pounds_u\kappa_v + \pounds_v \kappa_u + 2\kappa_u\kappa_v + (\tau_u+\tau_v)(\tau_v+\tau_e)}{2\tau_v}&
	\frac{\pounds_v (\tau_u+\tau_v)-\pounds_u \eta_v - \eta_v (2\kappa_u -\kappa_e) -\eta_e (\tau_u+\tau_v)}{2\tau_v}\\
	\kappa_e & 0
	\end{smallmatrix}\right]\:,
	\end{align}
	and ${\boldsymbol R}_{\mathrm{cls. 2}}^{~\# 2} {\boldsymbol \omega}=0$ can be written as
	\begin{align}
	0~=~&
	\Bigl( \pounds_u \kappa_e - \kappa_u \kappa_e \Bigr)\omega_u
	+ \Bigl( \pounds_e \kappa_e + (\tau_v-\tau_e)\kappa_e \Bigr)\omega_e\:,\\
	0~=~&
	\Bigl( 2\pounds_u \tau_v +(\psi_1 + 2 \tau_u) \kappa_e \Bigr)\omega_u
	+ \Bigl( 2 \pounds_e \tau_v + (\psi_2 - 2\eta_e) \kappa_e \Bigr)\omega_e\:,
	\label{eq:compatibility_II_null_1_22}\\
	0~=~&
	\Bigl(\pounds_u \psi_1 \Bigr)\omega_u
	+ \Bigl( \pounds_u \psi_2
	+(\kappa_u-\kappa_e)\psi_2
	+(\tau_u+\tau_v) \psi_1
	+2\tau_u\tau_v - R_{ee}
	\Bigr)\omega_e\:,\\
	0~=~&
	\Bigl(\pounds_v (\psi_1-2\tau_v) -\tfrac{\psi_2}{2}(\psi_1-2\tau_v)-2\eta_v \kappa_e\Bigr)\omega_u
	\notag \\
	&
	+ \Bigl(
	\pounds_v \psi_2
	-\tfrac{\psi_2}{2}(\psi_2+2\kappa_v)
	+(\psi_1+2\tau_v)\eta_v
	-2R_{vv}
	\Bigr)\omega_e\:,\\
	0~=~&
	\Bigl(\pounds_e \psi_1+\tfrac{\psi_1^2}{2} +\kappa_e \psi_2+R_{ee} \Bigr)\omega_u
	+ \Bigl(
	\pounds_e \psi_2
	+\tfrac{\psi_1}{2}(\psi_2-2\eta_e)
	+\tau_e \psi_2
	-2\eta_e \tau_v
	\Bigr)\omega_e\:,
	\end{align}
	where
	\begin{align}
	&\psi_1 ~\equiv~ \tfrac{1}{\tau_v} \Bigl(R_{uv} - \tfrac{1}{2} R_{ee} \Bigr) - \tau_v\:,
	&&\psi_2 ~\equiv~ \tfrac{1}{\tau_v} \Bigl( R_{ve} +\eta_v \kappa_e + \pounds_v \tau_v\Bigr)\:.
	\end{align}
	\label{eq:compatibility_II_null_1_2}
\end{subequations}
In these sub-branches, the rank of ${\boldsymbol R}^{~\# 1}_{\mathrm{cls. 2}}$ and ${\boldsymbol R}^{~\# 2}_{\mathrm{cls. 2}}$ governs the number of KVs
in the same way as that presented in Subsection \ref{subsec:caseII_nonnull_branch2}.

\subsection{Short summary of class 2}\label{sec:summary_II}

\begin{figure}[h]
	\begin{center}
		\begin{tikzpicture}
		[
		every node/.style={outer sep=0.15cm, inner sep=0cm},
		arrow/.style={-{Stealth[length=0.15cm]},thick},
		rblock/.style={rectangle, rounded corners,draw, minimum height = 0.5cm,
			minimum width=1.6cm, thick, outer sep = 0},
		rgblock/.style={rectangle, rounded corners,draw, minimum height = 0.5cm,
			minimum width=3.2cm, thick, outer sep = 0},
		point/.style={radius=2pt}
		]
		\node [rgblock] (caseII){class 2 non-null};
		\node [below=0.5 of caseII, xshift=3cm] (GeorNot){$\blacktriangleright \:$ The third vector is a geodesic tangent};
		\node [rblock, right=1.55 of GeorNot] (caseI1){class 1};
		\node [below=0.5 of GeorNot](No3){$\tau_2=\tau_3$};
				\node [right=1 of No3,xshift=-0.05cm](No3Obst){$\mathrm{rank} {\boldsymbol R}_{\mathrm{cls. 2}}^{\iota~ \#3}$};
						\node [right=1.5 of No3Obst](2KVs){2 KVs};
		\node [below=0.5 of No3](No2){$\tau_2=0$};
				\node [right=1 of No2,xshift=0.03cm](No2Obst){$\mathrm{rank} {\boldsymbol R}_{\mathrm{cls. 2}}^{\iota~ \#2}$};
						\node [right=1.5 of No2Obst](0KV){no KV};
		\node [below=0.5 of No2](No1){$\kappa_2+\iota\kappa_3=0$};
				\node [right=0.5 of No1](No1Obst){$\mathrm{rank} {\boldsymbol R}_{\mathrm{cls. 2}}^{\iota~ \#1}$};
						\node [right=1.5 of No1Obst](3KVs){3 KVs};
				\node [below=0.5 of No1Obst](Submaxi){$\mathrm{rank} {\boldsymbol R}_{\mathrm{cls. 2}}^{\iota}$};
						\node [rblock, right=1.47 of Submaxi] (caseI2){class 1};
		\draw[arrow] (GeorNot) -- (No3) node[right,pos=0.5] {{\scriptsize ~~yes}};
		\draw[arrow] (GeorNot) -- (caseI1.west) node[above,midway] {{\scriptsize no}};
		\draw[arrow] (No3) -- (No2) node[right,midway] {{\scriptsize ~~yes}};
		\draw[arrow] (No2) -- (No1) node[right,midway] {{\scriptsize ~~yes}};
		\draw[arrow] (No3) -- (No3Obst) node[above,midway] {{\scriptsize no}};
		\draw[arrow] (No2) -- (No2Obst) node[above,midway] {{\scriptsize no}};
		\draw[arrow] (No1) -- (No1Obst) node[above,midway] {{\scriptsize no}};
		\draw[arrow] (No1) |- (Submaxi) node[above,pos=0.4] {\qquad{\scriptsize ~yes}};
		\draw[arrow, name path=cross1] (No3Obst.east) -- (2KVs.west) node[above,pos=0.85,yshift=-0.5cm] {{\scriptsize $0$}};
		\draw[arrow, name path=cross2] (Submaxi.east) -- (0KV.west) node[above,pos=0.85,yshift=-0.5cm] {{\scriptsize $2$}};
		\draw[arrow, name path=cross4] (No1Obst.east) -- (caseI1.west) node[above,pos=0.9,yshift=-0.5cm] {{\scriptsize $1$}};
		\draw[arrow, name path=cross3, white] (No1Obst.east) -- (3KVs.west) ;
		\path[name intersections={of=cross1 and cross4,by={Point1}}];
		\path[name intersections={of=cross2 and cross3,by={Point2}}];
		\fill [point] (Point1) circle;
		\fill [point] (Point2) circle;
		\draw[thick] (No2Obst.east) -- (Point1) ;
		\draw[arrow] (Point1) -- (0KV.west) node[above,pos=0.6,yshift=-0.5cm] {{\scriptsize $2$}};
		\draw[arrow] (Point2) -- (3KVs.west) node[above,pos=0.7,yshift=-0.5cm] {{\scriptsize $0$}};
		\draw[arrow] (Point2) -- (caseI2.west) node[above,pos=0.7,yshift=-0.5cm] {{\scriptsize $1$}};
		\end{tikzpicture}
		\caption{The sub-algorithm for the class 2 non-null.
			See \eqs \eqref{eq:ricci_coeffs},
			\eqref{eq:compatibility_II_1_Sub}--\eqref{eq:compatibility_II_1_3} for notations.}
	\label{fig:caseII_nonnull}
	\end{center}
\end{figure}
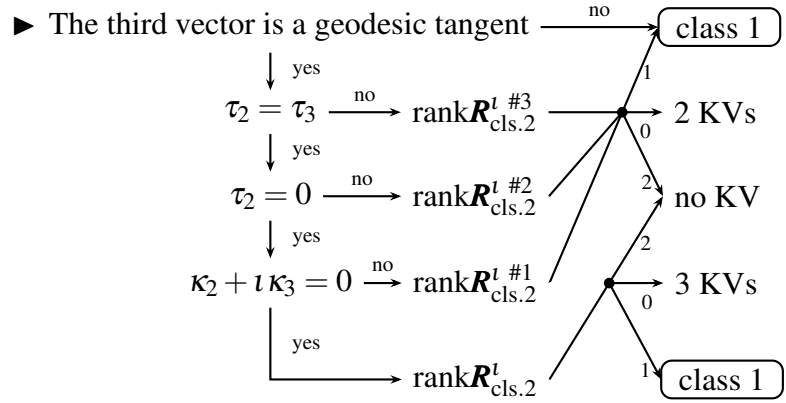

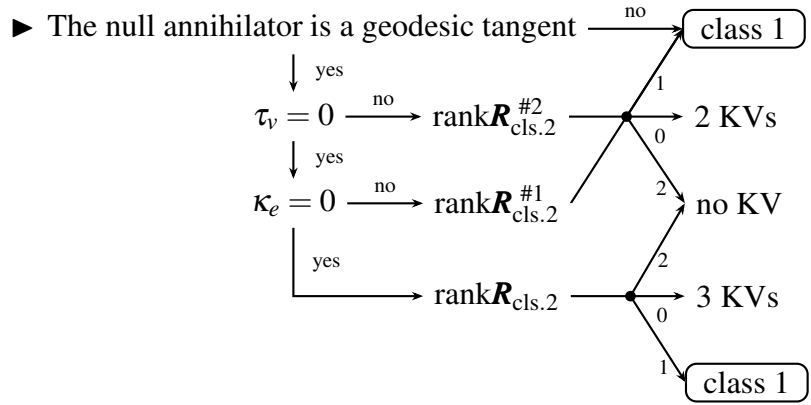
\begin{figure}[h]
	\begin{center}
		\begin{tikzpicture}
		[
		every node/.style={outer sep=0.15cm, inner sep=0cm},
		arrow/.style={-{Stealth[length=0.15cm]},thick},
		rblock/.style={rectangle, rounded corners,draw, minimum height = 0.5cm,
			minimum width=1.6cm, thick, outer sep = 0},
		rgblock/.style={rectangle, rounded corners,draw, minimum height = 0.5cm,
			minimum width=3.2cm, thick, outer sep = 0},
		point/.style={radius=2pt}
		]
		\node [rgblock] (caseII){class 2 null};
		\node [right=5.5 of caseII] (Empty1){};
		\node [below=0.5 of caseII, xshift=3cm] (GeorNot){$\blacktriangleright \:$ The null annihilator is a geodesic tangent};
		\node [rblock, right=1.25 of GeorNot] (caseI1){class 1};
		\node [below=0.5 of GeorNot](NullNo2){$\tau_v=0$};
		\node [right=1 of NullNo2](NullNo2Obst){$\mathrm{rank} {\boldsymbol R}_{\mathrm{cls. 2}}^{~ \#2}$};
		\node [right=1.5 of NullNo2Obst](2KVs){2 KVs};
		\node [below=0.5 of NullNo2](NullNo1){$\kappa_e=0$};
		\node [right=1 of NullNo1](NullNo1Obst){$\mathrm{rank} {\boldsymbol R}_{\mathrm{cls. 2}}^{~ \#1}$};
		\node [right=1.5 of NullNo1Obst](0KV){no KV};
		\node [below=0.5 of NullNo1Obst](NullSubmaxi){$\mathrm{rank} {\boldsymbol R}_{\mathrm{cls. 2}}$};
		\node [right=1.5 of NullSubmaxi](3KVs){3 KVs};
		\node [rblock,below=0.6 of 3KVs,xshift=0.12cm](caseI2){class 1};
		\node [below=0.5 of NullSubmaxi](Empty2){};
		\draw[arrow] (GeorNot) -- (NullNo2) node[right,pos=0.5] {{\scriptsize ~~yes}};
		\draw[arrow] (GeorNot) -- (caseI1) node[above,midway] {{\scriptsize no}};
		\draw[arrow] (NullNo2) -- (NullNo1) node[right,midway] {{\scriptsize ~~yes}};
		\draw[arrow] (NullNo2) -- (NullNo2Obst) node[above,midway] {{\scriptsize no}};
		\draw[arrow] (NullNo1) -- (NullNo1Obst) node[above,midway] {{\scriptsize no}};
		\draw[arrow] (NullNo1) |- (NullSubmaxi) node[above,pos=0.4] {\qquad{\scriptsize yes}};
		\draw[thick, name path=invisible1] (NullNo1Obst.east) -- (caseI1.west) ;
		\draw[arrow, name path=invisible2, white] (NullNo1Obst.east) -- (caseI2.west) ;
		
		\draw[arrow, name path=Nullcross] (NullNo2Obst.east) -- (2KVs.west) node[above,pos=0.8,yshift=-0.5cm] {{\scriptsize $0$}};
		\draw[arrow, name path=Nullsubcross] (NullSubmaxi.east) -- (3KVs.west) node[above,pos=0.8,yshift=-0.5cm] {{\scriptsize $0$}};
		\path[name intersections={of=Nullcross and invisible1,by={Point1}}];
		\path[name intersections={of=Nullsubcross and invisible2,by={Point2}}];
		\fill [point] (Point1) circle;
		\fill [point] (Point2) circle;
		\draw[arrow] (Point1) -- (0KV.west) node[above,pos=0.6,yshift=-0.5cm] {{\scriptsize $2$}};
		\draw[arrow] (Point1) -- (caseI1.west) node[above,pos=0.6,yshift=-0.5cm] {{\scriptsize $1$}};
		\draw[arrow] (Point2) -- (0KV.west) node[above,pos=0.6,yshift=-0.5cm] {{\scriptsize $2$}};
		\draw[arrow] (Point2) -- (caseI2.west) node[above,pos=0.6,yshift=-0.5cm] {{\scriptsize $1$}};
		\end{tikzpicture}
		\caption{The sub-algorithm for the class 2 null.
			See \eqs \eqref{eq:ricci_coeffs_null}, \eqref{eq:compatibility_II_null_1_Sub}--\eqref{eq:compatibility_II_null_1_2}
			for notations.}
	\label{fig:caseII_null}
	\end{center}
\end{figure}

We summarise the results obtained in this section in Figures \ref{fig:caseII_main}--\ref{fig:caseII_null}.

\begin{figure}[h]
	\begin{center}
						\begin{tikzpicture}
						[
						every node/.style={outer sep=0.15cm, inner sep=0cm},
						arrow/.style={-{Stealth[length=0.15cm]},thick},
						rblock/.style={rectangle, rounded corners,draw, minimum height = 0.5cm,
							minimum width=1.6cm, thick, outer sep = 0},
						rgblock/.style={rectangle, rounded corners,draw, minimum height = 0.5cm,
							minimum width=3.2cm, thick, outer sep = 0},
						point/.style={radius=2pt}
						]
						\node [] (NullorNot){$\blacktriangleright \:$ Either of two annihilators of ${\boldsymbol R}_a$ is null};
						\node [rblock,above=0.5 of NullorNot, xshift=-3.25cm] (title){class 2};
						\node [rgblock,right=1 of NullorNot] (Null){class 2 null};
						\node [rgblock,below=0.5 of Null] (Non){class 2 non-null};
						\draw[arrow] (NullorNot) -- (Null) node[above,midway] {{\scriptsize yes}};
						\draw[arrow] (NullorNot) |- (Non) node[right,pos=0.3] {{\scriptsize ~~no}};
						\end{tikzpicture}
		\caption{The sub-algorithm for the class 2.
			For details, see the beginning of this section.}
	\label{fig:caseII_main}
	\end{center}
\end{figure}
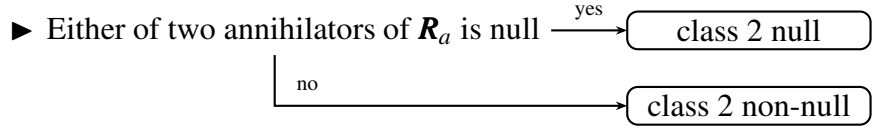

It should be emphasised that the sub-algorithm in Figures \ref{fig:caseII_main}--\ref{fig:caseII_null} can be again applicatory for some cases in class 3 as explained in Section \ref{sec:introduction}:
At the outset, KVs in class 3 is expressed as a linear combination of the three annihilators of ${\boldsymbol R}_a$. In some branches, the KVs eventuates into the form of a linear combination of two (or less) annihilators of ${\boldsymbol R}_a$, while keeping the property $\dim \ker {\boldsymbol R}_a = 3$.
Since $K^a=\sum_\alpha^2 \omega_\alpha u_\alpha^a$ accords with the prerequisite of class 2 \eqref{eq:caseII},
the results in this section is adapted to such branches as well.

\section{Analysis of class 3}\label{sec:caseIII}

\begin{figure}[h]
	\begin{center}
		\begin{tikzpicture}
		[
		every node/.style={outer sep=0.15cm, inner sep=0cm},
		arrow/.style={-{Stealth[length=0.15cm]},thick},
		rblock/.style={rectangle, rounded corners, draw, minimum height = 0.5cm,
			minimum width=1.6cm, thick, outer sep = 0},
		rgblock/.style={rectangle, rounded corners, draw, minimum height = 0.5cm,
			minimum width=2.5cm, thick, outer sep = 0},
		point/.style={radius=2pt}
		]
		\node [] (MaxorN){$\blacktriangleright \:S = 0$};
						\node [right=1.5 of MaxorN] (6KVs){6 KVs};
\node [rblock,above=0.25 of MaxorN, xshift=-2cm] (title){class 3};
		\node [below=0.5 of MaxorN] (typeNor){$S^2 = 0$};
						\node [rgblock,right=1.5 of typeNor] (typeN){type $[(21)]$};
		\node [below=0.5 of typeNor] (type3or){$S^3  = 0$};
				\node [rgblock,right=1.5 of type3or] (type3){type $[3]$};
		\node [below=0.5 of type3or] (type1or){$(S^{(2)})^3 = 6 (S^{(3)})^2$};
				\node [rgblock,right=1.5 of type1or] (type1d){type $[1,11]$};
				\node [right=0.2 of type1d] (OR1){or};
				\node [rgblock,right=0.2 of OR1] (typezz){type $[z\bar{z}1]$};
		\node [below=0.5 of type1or] (typeDor2){$S^2 = (S^{(3)}/S^{(2)}) S + (S^{(2)}/3)$};
				\node [rgblock,right=1.5 of typeDor2] (type11){type $[(1,1)1]$};
				\node [right=0.2 of type11] (OR2){or};
				\node [rgblock,right=0.2 of OR2] (type12){type $[1,(11)]$};
				\node [rgblock,below=0.5 of type11] (type21d){type $[21]$};
		\draw[arrow] (MaxorN) -- (6KVs) node[above,midway] {{\scriptsize yes}};
		\draw[arrow] (MaxorN) -- (typeNor) node[right,midway] {{\scriptsize ~~no}};
		\draw[arrow] (typeNor) -- (typeN) node[above,midway] {{\scriptsize yes}};
		\draw[arrow] (typeNor) -- (type3or) node[right,midway] {{\scriptsize ~~no}};
		\draw[arrow] (type3or) -- (type3) node[above,midway] {{\scriptsize yes}};
		\draw[arrow] (type3or) -- (type1or) node[right,midway] {{\scriptsize ~~no}};
		\draw[arrow] (type1or) -- (type1d) node[above,midway] {{\scriptsize no}};
		\draw[arrow] (type1or) -- (typeDor2) node[right,midway] {{\scriptsize ~~yes}};
		\draw[arrow] (typeDor2) -- (type11) node[above,midway] {{\scriptsize yes}};
		\draw[arrow] (typeDor2.east) -- (type21d.west) node[right,midway] {{\scriptsize ~~no}};
		\end{tikzpicture}
		\caption{The sub-algorithm for the class 3.
			The notation is as follows:
			The two indices on the traceless Ricci operator $S^a{}_b$ are dropped for short,
			e.g. $S^2$ denotes $S^a{}_b S^b{}_c$;
			$S^{(i)}$ denotes the $g$-trace of $S^i (i=2,3)$.
			}
		\label{fig:Segre}
	\end{center}
\end{figure}
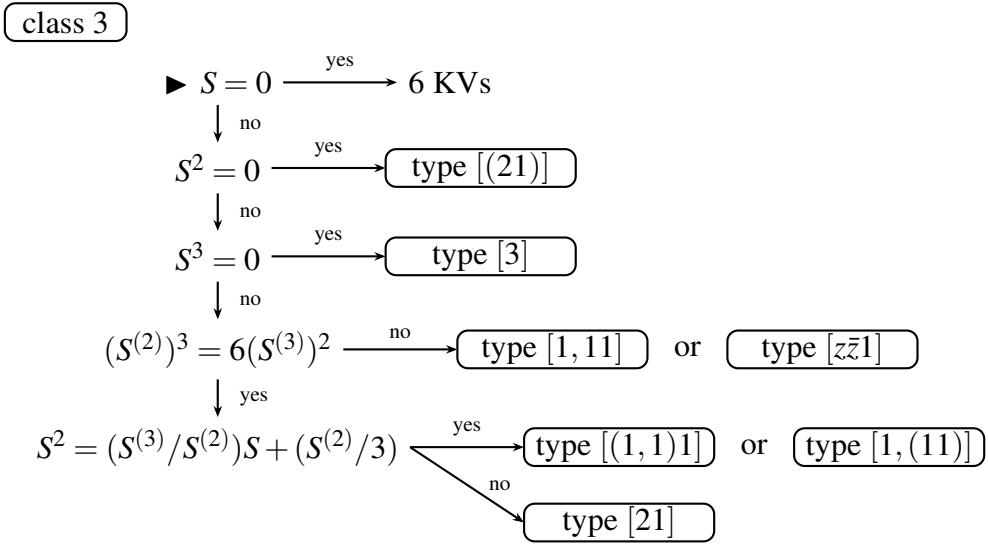

In this section, we address the case of class 3, for which all criteria in \eqs \eqref{eq:ann_cases} are vanishing.
This implies ${\boldsymbol R}_a$ is a zero matrix, whence any vector can be an annihilator of ${\boldsymbol R}_a$.
Since the first obstruction matrix ${\boldsymbol R}_a$ has been intentionally designed to ensure that
all the eigenvalues of the traceless Ricci operator $S^a{}_b$ are constants if $\mathrm{rank} {\boldsymbol R}_a $ is zero, 
it is thereby reasonable to resort to the Jordan basis of $S^a{}_b$. Perhaps the other choices for the basis of $T(M)$ fail to lessen the burden of computations, despite the fact that Jordan basis  inevitably demands us to solve the eigenvalue problem. Thus, 
our proposed formulation is based on the Jordan decomposition of the matrix $S^a{}_b$,
which is nothing but the classification of the Segre type of $S^a{}_b$.
It can be found in \cite{ONeill:1995} that the Segre classification is carried out by an examination of the minimal polynomial of $S^a{}_b$ as shown in Figure \ref{fig:Segre}.
See also \cite{SantosEtAl:1995} for the Segre classification of
symmetric tensors in Lorentzian geometry.

Let us pause here to declare the Segre notation \cite{Stephani:2003tm}. 
The eigenvalue equation $S^a{}_b V^b =\lambda V^a$
determines the orders of elementary divisors which belong to the several eigenvalues. 
A characteristic feature in Lorentzian geometry is that the elementary divisors
can be non-simple and the eigenvalues can be complex. 
The Segre notation stands for the orders of elementary divisors with 
the round brackets specifying which eigenvalues coincide. If two eigenvalues are 
complex conjugates, they are denoted by $z$ and $\bar z$.

With these notations in mind, 
we are proceeding along four cases:
We discuss the Segre type $[1,11]$ in 
Subsection \ref{subsec:caseIII_111}, 
the Segre type $[21]$ in 
Subsection \ref{subsec:caseIII_21}, 
the Segre type $[3]$ in  
Subsection \ref{subsec:caseIII_3} and 
the Segre type $[z \bar{z}1]$ in 
Subsection \ref{subsec:caseIII_zz1}.

\subsection{Type $[1,11]$ and its degeneracies}\label{subsec:caseIII_111}

In this case, we have the following Jordan chains:
\begin{align}
&S^a{}_b\:e_\alpha^b ~=~ \lambda_\alpha\:e_\alpha^a\:,
&&(\alpha=1,2,3)
\end{align}
with $\sum_\alpha \lambda_\alpha = 0$.
Here $e_\alpha^a$ is an eigenvector of $S^a{}_b$ belong to the eigenvalue $\lambda_\alpha$.
In this subsection, it is assumed that $\{ e_\alpha^a \}$ are normalised and $e_1^a$ is timelike,
so that $\{e_\alpha^a\}$ form an orthonormal basis of $T(M)$,
\begin{align}
g^{ab} ~=~ -e_1^a e_1^b + e_2^a e_2^b + e_3^a e_3^b\:.
\label{eq:frame_caseIII_111}
\end{align}
Then, any KV can be written as
\begin{align}
K^a ~=~ \sum_{\alpha=1}^3 \omega_\alpha \: e^a_\alpha\:,
\label{eq:caseIII_111}
\end{align}
where scalars $\{ \omega_\alpha\}$ are yet indeterminate.

It is an elementary computation to write down the 
the first compatibility condition, $\pounds_K S_{ab}=0$,
of \eq \eqref{eq:Killing}, giving
\begin{subequations}
	\begin{align}
	0&~=~
	(\lambda_1-\lambda_2) \Bigl(\varpi_2 +\kappa_2\:\omega_2 -(\tau_1+\tau_3)\omega_3 \Bigr)\:,\\
	0&~=~
	(\lambda_2-\lambda_3) \Bigl(\varpi_3 -(\tau_1-\tau_2) \omega_1 +\eta_3\:\omega_3 \Bigr)\:,\\
	0&~=~
	(\lambda_3-\lambda_1) \Bigl(\varpi_1 -\eta_1\:\omega_1 -(\tau_2-\tau_3)\omega_3 \Bigr)\:,
	\end{align}
	\label{eq:caseIII_111_compa1}
\end{subequations}
where the Ricci rotation coefficients are defined by \eqs \eqref{eq:ricci_coeffs},
and the 1-jet variables $\{\varpi_\alpha\}$ are respectively defined as
\begin{align}
&\varpi_1~\equiv~\pounds_3 \omega_1\:,
&&\varpi_2~\equiv~\pounds_1 \omega_2\:,
&&\varpi_3~\equiv~\pounds_2 \omega_3\:.
\end{align}
The eigenvalues $\lambda _ \alpha$ are constrained by 
the second Bianchi identity $\nabla_a R^a{}_b -(1/2) \nabla_bR=0$ as 
\begin{subequations}
\begin{align}
0&~=~(\lambda_1 -\lambda_2)\kappa_2 - (\lambda_3-\lambda_1)\kappa_3\:,\\
0&~=~(\lambda_1 -\lambda_2)\kappa_1 + (\lambda_2-\lambda_3)\eta_3\:,\\
0&~=~(\lambda_2 -\lambda_3)\eta_2 + (\lambda_3-\lambda_1)\eta_1\:.
\end{align}
\label{eq:caseIII_111_Bianchi}
\end{subequations}

The compatibility conditions \eqref{eq:caseIII_111_compa1} are fulfilled trivially
if the Segre type is $[(1,11)]$, $\lambda_1=\lambda_2=\lambda_3=0$,
thereby yielding the result shown in Figure \ref{fig:Segre}.
In the remaining parts of this subsection,
we investigate the Segre types $[1,(11)]$, $[(1,1)1]$ and $[1,11]$ separately.

\subsubsection{Branch where the Segre type is $[1,(11)]$}\label{subsec:caseIII_111_21_t}

In this branch, two eigenvalues in the spacelike direction coincide, i.e.
$\lambda_2=\lambda_3=-(1/2)\lambda_1$.
Then, it immediately follows from \eqs \eqref{eq:caseIII_111_compa1} and \eqref{eq:caseIII_111_Bianchi} that
\begin{subequations}
	\begin{align}
	\kappa_1 &~=~0\:,
	&\eta_1 &~=~0\:,
	&\kappa_3&~=~-\kappa_2\:,\\
	\varpi_1 &~=~ 
	(\tau_2-\tau_3)\omega_2\:,
	&\varpi_2 &~=~ - \kappa_2\:\omega_2 + (\tau_1+\tau_3)\omega_2\:.
	\end{align}
	\label{eq:caseIII_111_21_t_cond}
\end{subequations}
Given these conditions \eqref{eq:caseIII_111_21_t_cond}, the Killing equation \eqref{eq:Killing} and the identities $\nabla_{[a}\nabla_{b]}\omega_1=\nabla_{[a}\nabla_{b]}\omega_2=\nabla_{[a}\nabla_{b]}\omega_3=0$ produce a PDE system of the form
\begin{align}
&\nabla_a {\boldsymbol \omega} ~=~{\boldsymbol \Omega}_a\:{\boldsymbol \omega}\:,
&&{\boldsymbol \omega} ~\equiv~
[\omega_1~~ \omega_2~~ \omega_3~~\varpi_3]^T\:,
\label{eq:Killing_III_111_21_t}
\end{align}
where
\begin{align}
{\boldsymbol \Omega}_a ~\equiv~&
-e_{1a}
\left[\begin{smallmatrix}
0 & 0 & 0 & 0\\
0 & -\kappa_2 & \tau_1 + \tau_3 & 0 \\
0 & -\tau_1+\tau_2 & \kappa_2 & 0 \\
-2\kappa_2 \tau_1 & ~\kappa_2 \eta_2 - \pounds_1 \eta_2 &~ -(\tau_1+\tau_3)\eta_2 & 0
\end{smallmatrix}\right] \notag \\
&+e_{2a}
\left[\begin{smallmatrix}
0 & 0 & -\tau_2+\tau_3 & 0 \\
\kappa_2 & 0 & \eta_2 & 0\\
0 & 0 & 0 & 1\\
-\kappa_2 \eta_2 - \pounds_2 \tau_2 &
~-\pounds_2 \eta_2 &
~\eta_3^2 - (\tau_1 -\tau_2)(\tau_2-\tau_3) - \pounds_2 \eta_3 -\pounds_3 \eta_2 & 0
\end{smallmatrix}\right]\notag\\
&
+e_{3a}
\left[
\begin{smallmatrix}
0 & \tau_2-\tau_3 & 0 & 0 \\
-\tau_2 - \tau_3 & -\eta_2 & -\eta_3 & -1\\
-\kappa_2 & \eta_3 & 0 & 0 \\
2\kappa_2\eta_3 - \pounds_2\kappa_2 &
~-\eta_3^2 +(\tau_1-\tau_2)(\tau_2 -\tau_3) + \pounds_2 \eta_3 ~&
\eta_2\eta_3 &
\eta_2
\end{smallmatrix}
\right]
\:.
\end{align}
The compatibility condition for \eq \eqref{eq:Killing_III_111_21_t} leads to
\begin{subequations}
	\begin{align}
	(\tau_2 + \tau_3)\:\varpi_3~=~&
	-\tau_2 (\tau_2 + \tau_3 )\omega_1
	+\Bigl( \pounds_3 \tau_2 - \eta_2 (\tau_2 + \tau_3) +2\eta_3 \kappa_2\Bigr) \omega_2
	-\Bigl( \pounds_2 \tau_3 -2\eta_2 \kappa_2 \Bigr) \omega_3\:, \\
	4\kappa_2\:\varpi_3~=~&
	-4\kappa_2 \tau_2\:\omega_1
	- \Bigl(\pounds_2 (\tau_2+\tau_3)\Bigr)\omega_2
	- \Bigl( \pounds_3(\tau_2+\tau_3) + 4\eta_3 \kappa_2\Bigr) \omega_3
	\:.
	\end{align}
	\label{eq:compatibility_III_111_t}
\end{subequations}
This implies that $\varpi_3$ can be expressed in terms of $\{\omega_1, \omega_2, \omega_3\}$
except for $\kappa_2=\tau_2+\tau_3=0$.
Depending on the nonzeroness of the coefficients $\{\kappa_2, \tau_2+\tau_3\}$, the analysis falls into three sub-branches.

\subsubsection*{\underline{Sub-branch where $\kappa_2=\tau_2+\tau_3=0$}}
In this sub-branch, the conditions \eqref{eq:compatibility_III_111_t} gives
\begin{align}
(\pounds_2 \tau_2)\omega_2 + (\pounds_3 \tau_2) \omega_3 ~=~0\:,
\end{align}
whilst the relations in \eqs \eqref{eq:app_ricci_ortho} imply that $\pounds_2 \tau_2 = \pounds_3 \tau_2=0$.
We conclude that four KVs exist in this sub-branch, since the compatibility is trivially 
fulfilled.

\subsubsection*{\underline{Other sub-branches}}
Except for $\kappa_2=\tau_2+\tau_3=0$, the 1-jet variable $\varpi_3$
is not an independent variable, but is expressed in 
in terms of $\{\omega_1, \omega_2, \omega_3\}$ by virtue of \eqs \eqref{eq:compatibility_III_111_t}.
In these sub-branches, \eq \eqref{eq:Killing_III_111_21_t} is closed with respect to $\{ \omega_1, \omega_2, \omega_3\}$ and then
the compatibility of PDE systems is of the form
\begin{align*}
\nabla_a
{\boldsymbol \omega}
~=~
{\boldsymbol \Omega}_a\:{\boldsymbol \omega}\:,
&&
{\boldsymbol \omega}~\equiv~
[\omega_1 ~~\omega_2~~\omega_3]^T\:.
\end{align*}
The results for two cases (\#1) and (\#2) are presented as follows:

\begin{subequations}
	\noindent{\it (\#1) For the case of $\tau_2+\tau_3 =0, \kappa_2 \neq 0$},
	\begin{align}
	\varpi_3 &~=~ -\tau_2 \omega_1 - \eta_3 \omega_3\:,\\
	{\boldsymbol \Omega}_a &~=~
	-e_{1a}
	\left[
	\begin{matrix}
	0 & 0 & 0 \\
	0 & -\kappa_2 & -\tau_2 \\
	0 & \tau_2 & \kappa_2
	\end{matrix}\right]
	+e_{2a}
	\left[
	\begin{matrix}
	0 & 0 & -2\tau_2 \\
	\kappa_2 & 0 & \eta_2 \\
	-\tau_2 & 0 & -\eta_3
	\end{matrix}\right]
	+e_{3a}
	\left[
	\begin{matrix}
	0 & 2\tau_2 & 0 \\
	\tau_2 & -\eta_2 & 0 \\
	-\kappa_2 & \eta_3 & 0
	\end{matrix}\right]\:,\\
	{\boldsymbol R}_{[1,(11)]}^{~\# 1} &~=~
	\begin{bmatrix}
	0 & \pounds_2 \tau_2 & \pounds_3 \tau_2 \\
	\pounds_1 \eta_2 & \pounds_2 \eta_2 & \pounds_3 \eta_2 \\
	\pounds_1 \eta_3 & \pounds_2 \eta_3 & \pounds_3 \eta_3
	\end{bmatrix}\:,
	\end{align}
	\label{eq:compatibility_III_111_t_1}
\end{subequations}

\begin{subequations}
	\noindent{\it (\#2) For the case of $\tau_2+\tau_3 \neq 0$},
	\begin{align}
	\varpi_3 ~=~& \left(\frac{\tau_\delta - \tau_\sigma}{2}\right) \omega_1
	+\left( \frac{\pounds_2 \kappa_2}{\tau_\sigma}\right) \omega_2
	+\left( \frac{\pounds_3 \kappa_2}{\tau_\sigma} - \eta_3\right) \omega_3
	\:,\\
	{\boldsymbol \Omega}_a ~=~&
	-e_{1a}
	\left[
	\begin{matrix}
	0 & 0 & 0 \\
	0 & -\kappa_2 & \tau_1 + \frac{\tau_\delta+\tau_\sigma}{2} \\
	0 & -\tau_1 - \frac{\tau_\delta - \tau_\sigma}{2} & \kappa_2
	\end{matrix}\right]
	+e_{2a}
	\left[
	\begin{matrix}
	0 & 0 & \tau_\delta \\
	\kappa_2 & 0 & \eta_2 \\
	\frac{\tau_\delta-\tau_\sigma}{2} &
	\frac{\pounds_2 \kappa_2}{\tau_\sigma} &
	\frac{\pounds_3 \kappa_2}{\tau_\sigma} - \eta_3
	\end{matrix}\right] \notag \\
	&+e_{3a}
	\left[
	\begin{matrix}
	0 & - \tau_\delta & 0 \\
	-\frac{\tau_\delta+\tau_\sigma}{2} &
	-\frac{\pounds_2 \kappa_2}{\tau_\sigma} - \eta_2 &
	- \frac{\pounds_3 \kappa_2}{\tau_\sigma} \\
	-\kappa_2 & \eta_3 & 0
	\end{matrix}\right]\:,\\
	{\boldsymbol R}_{[1,(11)]}^{~\# 2} ~=~&
	\begin{bmatrix}
	0 & \pounds_2 \tau_\delta & \pounds_3 \tau_\delta \\
	\Pi_1 & \Pi_2 & \Pi_3 \\
	\Xi_1 & \Xi_2 & \Xi_3 \\
	\end{bmatrix}\:,
	\end{align}
	where
	\begin{align}
	\tau_\delta &~\equiv~ \tau_3 - \tau_2\:, \\
	\tau_\sigma &~\equiv~ \tau_3+\tau_2\:, \\
	\Pi_\alpha & ~\equiv~ \pounds_\alpha \eta_2 +\frac{\pounds_\alpha \pounds_2 \kappa_2}{\tau_\sigma}
	-\Bigl( \pounds_2 (\tau_\delta + 3\tau_\sigma) -4\eta_2\kappa_2\Bigr)
	\frac{\pounds_\alpha \kappa_2}{2\tau_\sigma^2}  \:, \\
	\Xi_\alpha &~\equiv~\pounds_\alpha \eta_3 -\frac{\pounds_\alpha \pounds_3 \kappa_2}{\tau_\sigma}
	-\Bigl( \pounds_3 (\tau_\delta -3 \tau_\sigma)-4\eta_3\kappa_2 \Bigr)
	\frac{\pounds_\alpha \kappa_2}{2\tau_\sigma^2} \:.
	\end{align}
	\label{eq:compatibility_III_111_t_2}
\end{subequations}

In these sub-branches, the rank of ${\boldsymbol R}_{[1,(11)]}^{~\# 1}$ and ${\boldsymbol R}_{[1,(11)]}^{~\# 2}$ is linked to the number of KVs as follows:
If $\mathrm{rank} {\boldsymbol R}_{[1,(11)]}^{~\# 1} = 0$, three KVs exist;
If $\mathrm{rank} {\boldsymbol R}_{[1,(11)]}^{~\# 1} = 3$, there is no KV;
If $\mathrm{rank} {\boldsymbol R}_{[1,(11)]}^{~\# 1} = 2$,
the sub-algorithm described in Figure \ref{fig:caseI} can be testable
since the KVs in this case are proportional to an annihilator of ${\boldsymbol R}_a$;
Otherwise $\mathrm{rank} {\boldsymbol R}_{[1,(11)]}^{~\# 1} = 1$,
the sub-algorithm described in Figures \ref{fig:caseII_main}--\ref{fig:caseII_null} can be testable
since the KVs in this case can be written as a linear combination of two annihilators of ${\boldsymbol R}_a$.
The argument for ${\boldsymbol R}_{[1,(11)]}^{~\# 2}$ parallels with above.

\subsubsection{Branch where the Segre type is $[(1,1)1]$}\label{subsec:caseIII_111_21_s}

In this branch, we have $\lambda_1=\lambda_3=-(1/2)\lambda_2$.
Then, it immediately follows from \eqs \eqref{eq:caseIII_111_compa1} and \eqref{eq:caseIII_111_Bianchi} that
	\begin{subequations}
		\begin{align}
		\kappa_2 &~=~0\:,
		&\eta_2 &~=~0\:,
		&\kappa_1&~=~\eta_3\:,\\
		\varpi_2 &~=~ 
		(\tau_1+\tau_3)\omega_3\:,
		&\varpi_3 &~=~ (\tau_1-\tau_2)\omega_1 - \eta_3\:\omega_3\:.
		\end{align}
		\label{eq:caseIII_111_21_s_cond}
	\end{subequations}
	Given these conditions \eqref{eq:caseIII_111_21_s_cond}, the Killing equation \eqref{eq:Killing} and the identities $\nabla_{[a}\nabla_{b]}\omega_1=\nabla_{[a}\nabla_{b]}\omega_2=\nabla_{[a}\nabla_{b]}\omega_3=0$ produce a PDE system of the form
	\begin{align}
	&\nabla_a {\boldsymbol \omega} ~=~{\boldsymbol \Omega}_a\:{\boldsymbol \omega}\:,
	&&{\boldsymbol \omega} ~\equiv~
	[\omega_1~~ \omega_2~~ \omega_3~~\varpi_1]^T\:,
	\label{eq:Killing_III_111_21_s}
	\end{align}
	where
	\begin{align}
	{\boldsymbol \Omega}_a ~\equiv~&
	-e_{1a}
	\left[\begin{smallmatrix}
	0 & -\kappa_1 & -\eta_1 & 0\\
	0 & 0 & \tau_1 + \tau_3 & 0\\
	-\eta_1 & -\tau_1+\tau_3 & -\kappa_3 & 1 \\
	-\eta_1 \kappa_3 &
	-\pounds_3 \kappa_1 -2 \eta_1 \kappa_1 &
	-\eta_1^2 + (\tau_1 + \tau_3) (\tau_2 - \tau_3) - \pounds_3 \eta_1 & \kappa_3
	\end{smallmatrix}\right] \notag \\
	&+e_{2a}
	\left[\begin{smallmatrix}
	\kappa_1 & 0 & -\tau_2+\tau_3 & 0\\
	0 & 0 & 0 & 0 \\
	\tau_1-\tau_2 & 0 & -\kappa_1 & 0\\
	(\tau_1-\tau_2)\kappa_3 &
	-2\kappa_1 \tau_2&
	\pounds_2 \kappa_3-\kappa_1 \kappa_3&0
	\end{smallmatrix}\right]\notag\\
	&
	+e_{3a}
	\left[
	\begin{smallmatrix}
	0 & 0 & 0 & 1\\
	-\tau_1 - \tau_3 & 0 & 0 & 0 \\
	\kappa_3 & \kappa_1 & 0 & 0 \\
	\eta_1^2 - (\tau_1+\tau_3) (\tau_2 - \tau_3) + \pounds_1 \kappa_3 + \pounds_3 \eta_1 &
	\kappa_1\kappa_3 - \pounds_3 \tau_3 &
	\pounds_3 \kappa_3 & 0
	\end{smallmatrix}
	\right]
	\:.
	\end{align}
The compatibility condition for \eq \eqref{eq:Killing_III_111_21_s} leads to
	\begin{subequations}
		\begin{align}
		(\tau_1 - \tau_3)\:\varpi_1~=~&
		\Bigl( 2\kappa_1 \kappa_3 - \pounds_3 \tau_1\Bigr)\omega_1
		-\tau_3 (\tau_1-\tau_3) \omega_2
		-\Bigl( 2\eta_1 \kappa_1- \kappa_3(\tau_1 - \tau_3) - \pounds_1 \tau_3\Bigr)\omega_3
		\:, \\
		4\kappa_1\:\varpi_1~=~& \Bigl( 4\eta_1 \kappa_1 + \pounds_1 (\tau_1-\tau_3) \Bigr) \omega_1
		-4\kappa_1 \tau_3 \omega_2
		+\Big( \pounds_3 (\tau_1 - \tau_3)\Bigr) \omega_3
		\:.
		\end{align}
		\label{eq:compatibility_III_111_s}
	\end{subequations}
	This implies that $\varpi_1$ can be expressed in terms of $\{\omega_1, \omega_2, \omega_3\}$
	except for $\kappa_1=\tau_1-\tau_3=0$.
	Depending on whether the coefficients $\{\kappa_1, \tau_1-\tau_3\}$ of $\varpi$ are vanishing, the analysis falls into three sub-branches.
	
\subsubsection*{\underline{Sub-branch where $\kappa_1=\tau_1-\tau_3=0$}}
In this sub-branch, the conditions \eqref{eq:compatibility_III_111_s} gives
\begin{align}
(\pounds_1 \tau_1)\omega_1 + (\pounds_3 \tau_1) \omega_3 ~=~0\:,
\end{align}
whilst the relations in \eqs \eqref{eq:app_ricci_ortho} imply that $\pounds_1 \tau_1 = \pounds_3 \tau_1=0$.
We conclude that four KVs exist in this sub-branch, as the compatibility conditions are trivially met.

\subsubsection*{\underline{Other sub-branches}}
Except when $\kappa_1=\tau_1-\tau_3=0$,
\eqs \eqref{eq:compatibility_III_111_s} allow us to write the 1-jet variable $\varpi_1$ in terms of $\{\omega_1, \omega_2, \omega_3\}$.
In these sub-branches, \eq \eqref{eq:Killing_III_111_21_s} is closed with respect to $\{ \omega_1, \omega_2, \omega_3\}$ and then
the compatibility of PDE systems takes the form
\begin{align*}
\nabla_a
{\boldsymbol \omega}
~=~
{\boldsymbol \Omega}_a\:{\boldsymbol \omega}\:,
&&
{\boldsymbol \omega}~\equiv~
[\omega_1 ~~\omega_2~~\omega_3]^T\:.
\end{align*}
The individual results are described as follows:

	\begin{subequations}
		\noindent{\it (\#1) For the case of $\tau_1=\tau_3, \kappa_1 \neq 0$},
		\begin{align}
		\varpi_1 &~=~ \eta_1\:\omega_1 - \tau_1\:\omega_2\:,\\
		{\boldsymbol \Omega}_a &~=~
		-e_{1a}
		\begin{bmatrix}
		0 & -\kappa_1 & -\eta_1 \\
		0 & 0 & 2\tau_1 \\
		0 & -\tau_1 & -\kappa_3
		\end{bmatrix}
		+e_{2a}
		\begin{bmatrix}
		\kappa_1 & 0 & \tau_1 \\
		0 & 0 & 0 \\
		\tau_1 & 0 & -\kappa_1
		\end{bmatrix}
		+e_{3a}
		\begin{bmatrix}
		\eta_1 & -\tau_1 & 0 \\
		-2\tau_1 & 0 & 0 \\
		\kappa_3 & \kappa_1 & 0
		\end{bmatrix}\:,\\
		{\boldsymbol R}_{[(1,1)1]}^{~\# 1} &~=~
		\begin{bmatrix}
		\pounds_1 \tau_1 & 0 & \pounds_3 \tau_1 \\
		\pounds_1 \eta_1 & \pounds_2 \eta_1 & \pounds_3 \eta_1\\
		\pounds_1 \kappa_3 & \pounds_2 \kappa_3 & \pounds_3 \kappa_3
		\end{bmatrix}\:,
		\end{align}
		\label{eq:compatibility_III_111_s_1}
	\end{subequations}
	
		\begin{subequations}
			\noindent{\it (\#2) For the case of $\tau_1 \neq \tau_3$},
			\begin{align}
			\varpi_1 ~=~& \left( \eta_1 - \frac{\pounds_1 \kappa_1}{\tau_\delta}\right) \omega_1
			- \left(\frac{\tau_\delta+\tau_\sigma}{2}\right)\omega_2
			-\left(\frac{\pounds_3 \kappa_1}{\tau_\delta} \right) \omega_3
			\:,\\
			{\boldsymbol \Omega}_a ~=~&
			-e_{1a}
			\begin{bmatrix}
			0 & -\kappa_1 & -\eta_1 \\
			0 & 0 & \tau_\sigma \\
			-\tfrac{\pounds_1 \kappa_1}{\tau_\delta} &
			\tfrac{\tau_\delta-\tau_\sigma}{2} &
			-\tfrac{\pounds_3 \kappa_1}{\tau_\delta} -\kappa_3
			\end{bmatrix}
			+e_{2a}
			\begin{bmatrix}
			\kappa_1 & 0 & \tfrac{\tau_\delta+\tau_\sigma}{2} - \tau_2 \\
			0 & 0 & 0 \\
			-\tfrac{\tau_\delta-\tau_\sigma}{2}-\tau_2 & 0 & -\kappa_1
			\end{bmatrix} \notag\\
			&
			+e_{3a}
			\begin{bmatrix}
			\eta_1 - \tfrac{\pounds_1 \kappa_1}{\tau_\delta} & -\tfrac{\tau_\delta+\tau_\sigma}{2} & -\tfrac{\pounds_3\kappa_1}{\tau_\delta}\\
			-\tau_\sigma & 0 & 0 \\
			\kappa_3 & \kappa_1 & 0
			\end{bmatrix}
			\:,\\
			{\boldsymbol R}_{[(1,1)1]}^{~\# 2} ~=~&
			\begin{bmatrix}
			\pounds_1 \tau_\sigma & 0 & \pounds_3 \tau_\sigma\\
			\Pi_1 & \Pi_2 & \Pi_3 \\
			 \Xi_1 & \Xi_2 & \Xi_3
			\end{bmatrix}\:,
			\end{align}
			where
			\begin{align}
			\tau_\sigma ~\equiv~& \tau_3+ \tau_1 \:,\\
			\tau_\delta ~\equiv~& \tau_3 - \tau_1\:,\\
			\Pi_\alpha ~\equiv~&
			\pounds_\alpha \eta_1-\frac{\pounds_\alpha \pounds_1 \kappa_1}{\tau_\delta}
			+\Bigl( \pounds_1 (\tau_\sigma + 3 \tau_\delta)-4 \eta_1 \kappa_1 \Bigr)
			\frac{\pounds_\alpha \kappa_1}{2\tau_\delta^2}\:,\\
			\Xi_\alpha ~\equiv~&\pounds_\alpha \kappa_3
			+\frac{\pounds_\alpha \pounds_3 \kappa_1}{\tau_\delta}
			+\Bigl( \pounds_3 (\tau_\sigma-3 \tau_\delta)-4 \kappa_1 \kappa_3 \Bigr)
			\frac{\pounds_\alpha \kappa_1}{2\tau_\delta^2}\:.
			\end{align}
			\label{eq:compatibility_III_111_s_2}
		\end{subequations}

In these sub-branches, the rank of ${\boldsymbol R}^{~\# 1}_{[1,(11)]}$ and ${\boldsymbol R}^{~\# 2}_{[1,(11)]}$ controls the number of KVs
in the same way as that presented in Sub-subsection \ref{subsec:caseIII_111_21_t}.

\subsubsection{Branch where the Segre type is $[1,11]$}\label{subsec:caseIII_111_dis}

In this branch,  the eigenvalues $\{\lambda_\alpha \}$ differ from each other.
Then, it immediately follows from \eqs \eqref{eq:caseIII_111_compa1} and \eqref{eq:caseIII_111_Bianchi} that
	\begin{subequations}
		\begin{align}
		\kappa_3 &~=~ \delta \lambda_1\:\kappa_2\:,
		&\kappa_1 &~=~- \delta \lambda_2\:\eta_3\:,
		& \eta_2 &~=~-\delta \lambda_3\:\eta_1\:,\\
		\varpi_1 &~=~ \eta_1\:\omega_1 +(\tau_2-\tau_3)\omega_2\:,
		&\varpi_2 &~=~ -\kappa_2\:\omega_2 + (\tau_1+\tau_3)\omega_3\:,
		&\varpi_3 &~=~ (\tau_1-\tau_2)\omega_1 - \eta_3\:\omega_3\:,
		\end{align}
		\label{eq:caseIII_111_dis_cond}
	\end{subequations}
	where $\delta \lambda_1\equiv (\lambda_1-\lambda_2) / (\lambda_3-\lambda_1 )$,
	$\delta \lambda_2 \equiv (\lambda_2-\lambda_3) / (\lambda_1-\lambda_2)$ and 
	$\delta \lambda_3 \equiv (\lambda_3 - \lambda_1) / (\lambda_2-\lambda_3)$.
	
	Given these conditions \eqref{eq:caseIII_111_dis_cond}, the Killing equation \eqref{eq:Killing}
	produces a PDE system of the form
	\begin{align}
	&\nabla_a {\boldsymbol \omega} ~=~{\boldsymbol \Omega}_a\:{\boldsymbol \omega}\:,
	&&{\boldsymbol \omega} ~\equiv~
	[\omega_1~~ \omega_2~~ \omega_3]^T\:,
	\label{eq:Killing_III_111_dis}
	\end{align}
	where
		\begin{align}
		{\boldsymbol \Omega}_a ~\equiv~&
		-e_{1a}
		\left[\begin{smallmatrix}
		0 &  \delta \lambda_2\:\eta_3 & -\eta_1 \\
		0 & -\kappa_2 & \tau_1 + \tau_3 \\
		0 & -\tau_1+\tau_2 & -\delta \lambda_1 \:\kappa_2 \\
		\end{smallmatrix}\right]
		+e_{2a}
		\left[\begin{smallmatrix}
		-\delta \lambda_2 \:\eta_3 & 0 & -\tau_2+\tau_3 \\
		\kappa_2 & 0 & -\delta \lambda_3\:\eta_1 \\
		\tau_1 - \tau_2 & 0 & -\eta_3 \\
		\end{smallmatrix}\right]
		+e_{3a}
		\left[
		\begin{smallmatrix}
		\eta_1 & \tau_2-\tau_3 & 0 \\
		-\tau_1 - \tau_3 & \delta \lambda_3\:\eta_1 & 0\\
		\delta \lambda_1\:\kappa_2 & \eta_3 & 0 \\
		\end{smallmatrix}
		\right]
		\:.
		\end{align}
	The compatibility condition for \eq \eqref{eq:Killing_III_111_dis} leads to
		\begin{align}
		{\boldsymbol R}_{[1,11]}&~=~
		\begin{bmatrix}
		\pounds_1 \kappa_1 & \pounds_2 \kappa_1 & \pounds_3 \kappa_1 \\
		\pounds_1 \kappa_2 & \pounds_2 \kappa_2 & \pounds_3 \kappa_2 \\
		\pounds_1 \eta_1 & \pounds_2 \eta_1 & \pounds_3 \eta_1 \\
		\pounds_1 \tau_1 & \pounds_2 \tau_1 & \pounds_3 \tau_1 \\
		\pounds_1 \tau_2 & \pounds_2 \tau_2 & \pounds_3 \tau_2 \\
		\pounds_1 \tau_3 & \pounds_2 \tau_3 & \pounds_3 \tau_3 \\
		\end{bmatrix}\:.
		\label{eq:compatibility_III_111_dis}
		\end{align}

In this sub-branch, the rank of ${\boldsymbol R}_{[1,11]}$ governs the number of KVs
in the same way as that presented in Sub-subsection \ref{subsec:caseIII_111_21_t}.

\subsubsection{Short summary of class 3 type $[1,11]$}\label{sec:summary_type_111}

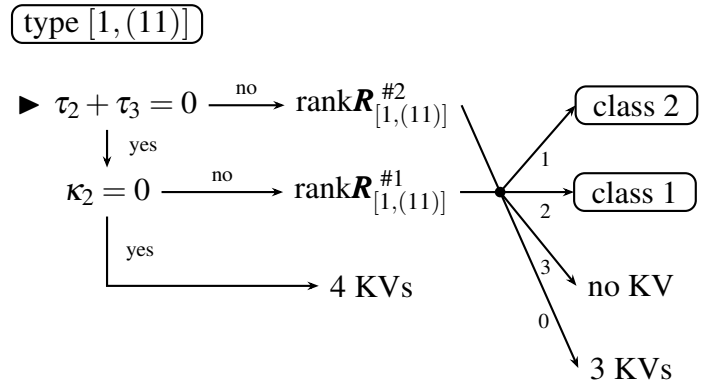
\begin{figure}[p]
	\begin{center}
		\begin{tikzpicture}
		[
		every node/.style={outer sep=0.15cm, inner sep=0cm},
		arrow/.style={-{Stealth[length=0.15cm]},thick},
		rblock/.style={rectangle, rounded corners,draw, minimum height = 0.5cm,
			minimum width=1.6cm, thick, outer sep = 0},
		rgblock/.style={rectangle, rounded corners,draw, minimum height = 0.5cm,
			minimum width=2.5cm, thick, outer sep = 0},
		point/.style={radius=2pt}
		]
		\node [rgblock] (title){type $[1,(11)]$};
		\node [below=0.5 of title] (test1){$\blacktriangleright \:\tau_2 + \tau_3=0$};
		\node [below=0.5 of test1] (test2){$\kappa_2=0$};
		\node [right=1 of test1] (Obst2){$\mathrm{rank}{\boldsymbol R}_{[1,(11)]}^{~\# 2}$};
		\node [right=1.6 of test2] (Obst1){$\mathrm{rank}{\boldsymbol R}_{[1,(11)]}^{~\# 1}$};
		\node [rblock,right=1.5 of Obst2] (caseII){class 2};
		\node [rblock,right=1.5 of Obst1] (caseI){class 1};
		\node [below=0.5 of Obst1] (4KVs){4 KVs};
		\node [right=2 of 4KVs] (noKV){no KV};
		\node [below=0.5 of noKV] (3KVs){3 KVs};
		\draw[arrow] (test1) -- (test2) node[right,pos=0.5] {{\scriptsize ~~yes}};
		\draw[arrow] (test2) |- (4KVs) node[above,pos=0.4] {\qquad{\scriptsize yes}};
		\draw[arrow] (test1) -- (Obst2) node[above,midway] {{\scriptsize no}};
		\draw[arrow] (test2) -- (Obst1) node[above,midway] {{\scriptsize no}};
		\draw[arrow, name path=cross1] (Obst1.east) -- (caseI.west) node[above,pos=0.75,yshift=-0.5cm] {{\scriptsize $2$}};
		\draw[arrow, name path=cross2] (Obst2.east) -- (3KVs.west) node[above,pos=0.75,yshift=-0.5cm] {{\scriptsize $0~~$}};
		\path[name intersections={of=cross1 and cross2,by={Point1}}];
		\fill [point] (Point1) circle;
		\draw[arrow] (Point1) -- (caseII.west) node[above,pos=0.6,yshift=-0.5cm] {{\scriptsize $1$}};
		\draw[arrow] (Point1) -- (noKV.west) node[above,pos=0.6,yshift=-0.5cm] {{\scriptsize $3$}};
		\end{tikzpicture}
		\caption{The sub-sub-algorithm for the class 3 type $[1,(11)]$.
		See \eqs \eqref{eq:compatibility_III_111_t_1} and \eqref{eq:compatibility_III_111_t_2} for notations.
			}
		\label{fig:type111_t}
	\end{center}
\end{figure}

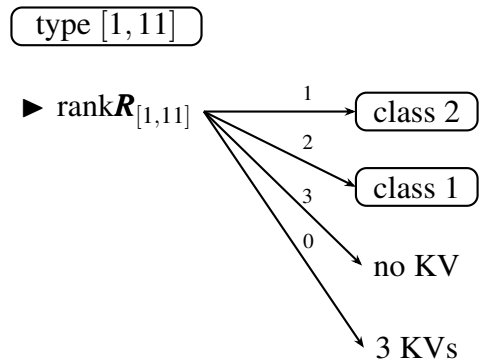
\begin{figure}[p]
	\begin{center}
		\begin{tikzpicture}
		[
		every node/.style={outer sep=0.15cm, inner sep=0cm},
		arrow/.style={-{Stealth[length=0.15cm]},thick},
		rblock/.style={rectangle, rounded corners,draw, minimum height = 0.5cm,
			minimum width=1.6cm, thick, outer sep = 0},
		rgblock/.style={rectangle, rounded corners,draw, minimum height = 0.5cm,
			minimum width=2.5cm, thick, outer sep = 0},
		point/.style={radius=2pt}
		]
		\node [rgblock] (title){type $[1,11]$};
		\node [below=0.5 of title] (Obst){$\blacktriangleright \:\mathrm{rank}{\boldsymbol R}_{[1,11]}$};
		\node [rblock,right=2 of Obst] (caseII){class 2};
		\node [rblock,below=0.5 of caseII] (caseI){class 1};
		\node [below=0.5 of caseI] (0KV){no KV};
		\node [below=0.5 of 0KV] (3KVs){3 KVs};
		\draw[arrow] (Obst.east) -- (caseII.west) node[above,pos=0.65] {{\scriptsize ~~$1$}};
		\draw[arrow] (Obst.east) -- (caseI.west) node[above,pos=0.65] {{\scriptsize ~~$2$}};
		\draw[arrow] (Obst.east) -- (3KVs.west) node[right,pos=0.55] {{\scriptsize $0$}};
		\draw[arrow] (Obst.east) -- (0KV.west) node[right,pos=0.55] {{\scriptsize $3$}};
		\end{tikzpicture}
		\caption{The sub-sub-algorithm for the class 3 type $[1,11]$.
		See \eq \eqref{eq:compatibility_III_111_dis} for notations.
			}
		\label{fig:type111_dis}
	\end{center}
\end{figure}

Let us visually abridge the results obtained in this subsection in Figures \ref{fig:type111_t}--\ref{fig:type111_dis}.

\begin{figure}[p]
	\begin{center}
		\begin{tikzpicture}
		[
		every node/.style={outer sep=0.15cm, inner sep=0cm},
		arrow/.style={-{Stealth[length=0.15cm]},thick},
		rblock/.style={rectangle, rounded corners,draw, minimum height = 0.5cm,
			minimum width=1.6cm, thick, outer sep = 0},
		rgblock/.style={rectangle, rounded corners,draw, minimum height = 0.5cm,
			minimum width=2.5cm, thick, outer sep = 0},
		point/.style={radius=2pt}
		]
		\node [rgblock] (title){type $[(1,1)1]$};
		\node [below=0.5 of title] (test1){$\blacktriangleright \:\tau_1 - \tau_3=0$};
		\node [below=0.5 of test1] (test2){$\kappa_1=0$};
		\node [right=1 of test1] (Obst2){$\mathrm{rank}{\boldsymbol R}_{[(1,1)1]}^{~\# 2}$};
		\node [right=1.6 of test2] (Obst1){$\mathrm{rank}{\boldsymbol R}_{[(1,1)1]}^{~\# 1}$};
		\node [rblock,right=1.5 of Obst2] (caseII){class 2};
		\node [rblock,right=1.5 of Obst1] (caseI){class 1};
		\node [below=0.5 of Obst1] (4KVs){4 KVs};
		\node [right=2 of 4KVs] (noKV){no KV};
		\node [below=0.5 of noKV] (3KVs){3 KVs};
		\draw[arrow] (test1) -- (test2) node[right,pos=0.5] {{\scriptsize ~~yes}};
		\draw[arrow] (test2) |- (4KVs) node[above,pos=0.4] {\qquad{\scriptsize yes}};
		\draw[arrow] (test1) -- (Obst2) node[above,midway] {{\scriptsize no}};
		\draw[arrow] (test2) -- (Obst1) node[above,midway] {{\scriptsize no}};
		\draw[arrow, name path=cross1] (Obst1.east) -- (caseI.west) node[above,pos=0.75,yshift=-0.5cm] {{\scriptsize $2$}};
		\draw[arrow, name path=cross2] (Obst2.east) -- (3KVs.west) node[above,pos=0.75,yshift=-0.5cm] {{\scriptsize $0~~$}};
		\path[name intersections={of=cross1 and cross2,by={Point1}}];
		\fill [point] (Point1) circle;
		\draw[arrow] (Point1) -- (caseII.west) node[above,pos=0.6,yshift=-0.5cm] {{\scriptsize $1$}};
		\draw[arrow] (Point1) -- (noKV.west) node[above,pos=0.6,yshift=-0.5cm] {{\scriptsize $3$}};
		\end{tikzpicture}
		\caption{The sub-sub-algorithm for the class 3 type $[(1,1)1]$.
		See \eqs \eqref{eq:compatibility_III_111_s_1} and \eqref{eq:compatibility_III_111_s_2} for notations.
			}
		\label{fig:type111_s}
	\end{center}
\end{figure}

\subsection{Type $[21]$ and its degeneracy}\label{subsec:caseIII_21}

In this case, we have the following Jordan chains:
\begin{subequations}
\begin{align}
S^a{}_b\:j_1^b &~=~ \lambda_1\:j_1^a\:,\\
S^a{}_b\:j_2^b &~=~ \lambda_1\:j_2^a + j_1^a\:,\\
S^a{}_b\:j_3^b &~=~ \lambda_3\:j_3^a\:,
\end{align}
\label{eq:chain_21}
\end{subequations}
with $2\lambda_1+\lambda_3 = 0$.
Here $j_\alpha^a$ is an generalised eigenvector of $S^a{}_b$.
It can be shown that the eigenvectors $j_1^a, j_3^a$ are respectively null and spacelike.
The causal nature of $j_2$ is indeterminate.
Taking the double-null basis $\{u^a,v^a,e^a\}$  of $T(M)$ as 
\begin{align}
g^{ab}~=~u^a v^b + v^a u^b + e^a e^b\:,
\label{eq:d_null_basis_III_21}
\end{align}
it turns out useful to choose 
$u^a \equiv j_1^a$ and $e^a \equiv j_3^a$, with 
$v^a$ being a null vector such that $g_{ab} u^a v^b=1$ and $g_{ab} v^a e^b = 0$.
Then, any KV can be written as
\begin{align}
K^a ~=~ \omega_u\:u^a + \omega_v\:v^a + \omega_e\:e^a\:,
\label{eq:caseIII_21}
\end{align}
where scalars $\{ \omega_u, \omega_v, \omega_e\}$ are yet indeterminate.

The first compatibility condition, $\pounds_K S_{ab}=0$, of \eq \eqref{eq:Killing}
yields 
	\begin{subequations}
		\begin{align}
		\lambda\:\varpi_e~=~& -\lambda (\tau_u-\tau_v) \omega_v - \lambda \kappa_e\:\omega_e\:,\\
		3\lambda\: \varpi_u-S_{vv}\:\varpi_e ~=~&-3\lambda(\tau_u+\tau_e)\omega_u
		-\Bigl( 3 \eta_v \lambda - (\tau_u-\tau_v) S_{vv} \Bigr) \omega_v
		+\kappa_e S_{vv}\:\omega_e\:,\\
		2S_{vv}\:\varpi_v~=~& -\Bigl( \pounds_u S_{vv} + 2 \kappa_u S_{vv}\Bigr) \omega_u
		-(\pounds_v S_{vv}) \omega_v
		-\Bigl( \pounds_e S_{vv} - 2 (\tau_v-\tau_e) S_{vv}\Bigr) \omega_e
		\:,
		\end{align}
		\label{eq:caseIII_21_compa1}
	\end{subequations}
	where $\lambda \equiv \lambda_1 = -(1/2)\lambda_3$ and $S_{vv}\equiv S_{ab} v^a v^b$.
	The Ricci rotation coefficients are defined by \eqs \eqref{eq:ricci_coeffs_null},
	and the 1-jet variables $\{\varpi_u, \varpi_v, \varpi_e\}$ are respectively defined as
	\begin{align}
	&\varpi_u~\equiv~\pounds_e \omega_u\:,
	&&\varpi_v~\equiv~\pounds_v \omega_v\:,
	&&\varpi_e~\equiv~\pounds_u\omega_e\:.
	\end{align}
	The second Bianchi identity $\nabla_a R^a{}_b -(1/2) \nabla_bR=0$ puts the following constraints
	\begin{subequations}
		\begin{align}
		0&~=~\lambda\kappa_e\:,\\
		0&~=~\pounds_u S_{vv} - 3\eta_e\lambda + (2\kappa_u-\kappa_e) S_{vv}\:,\\
		0&~=~\eta_uS_{vv} + 3\lambda(\tau_u+\tau_v) \:.
		\end{align}
		\label{eq:caseIII_21_Bianchi}
	\end{subequations}
	
	Suppose $S_{vv} = 0$. Then \eqs \eqref{eq:chain_21} and \eqref{eq:d_null_basis_III_21} imply that 
	the basis $\{u^a,v^a,e^a\}$ satisfies
	\begin{align*}
	&S^a{}_b\:u^b ~=~ \lambda \:u^a\:,
	&&S^a{}_b\:v^b ~=~ \lambda \:v^a \:,
	&&S^a{}_b\:e^b ~=~ -2 \lambda \:e^a\:,
	\end{align*}
	which contradicts the assumption that the Segre type of $S^a{}_b$ is $[21]$.
	It therefore follows that $S_{vv} \neq 0$.
	In the remaining parts of this subsection,
	we investigate the Segre types $[(21)]$ and $[21]$ separately.

\subsubsection{Branch where the Segre type is $[(21)]$}\label{subsec:caseIII_21_deg}

In this branch, three eigenvalues of $S^a{}_b$ are coincident and then  
$\lambda = \lambda_1=\lambda_3 =0$ follows from traceless property.
Then,  \eqs \eqref{eq:caseIII_21_compa1} and \eqref{eq:caseIII_21_Bianchi} are combined to give
	\begin{subequations}
		\begin{align}
		\varpi_v~=~&-\left(\frac{\kappa_e}{2}\right)\omega_u
		- \Bigl( \pounds_v \varphi\Bigr)\omega_v
		-\Bigl( \pounds_e \varphi -(\tau_v-\tau_e) \Bigr) \omega_e\:,\\
		\varpi_e~=~&-(\tau_u-\tau_v)\omega_v - \kappa_e\:\omega_e\:,
		\end{align}
		\label{eq:caseIII_21_deg}
	\end{subequations}
	where $\varphi \equiv (1/2) \log S_{vv}$.
	Given these conditions \eqref{eq:caseIII_21_deg}, the Killing equation \eqref{eq:Killing} and the identities $\nabla_{[a}\nabla_{b]}\omega_u=\nabla_{[a}\nabla_{b]}\omega_v=\nabla_{[a}\nabla_{b]}\omega_e=0$ produce a PDE system of the form
	\begin{align}
	&\nabla_a {\boldsymbol \omega} ~=~{\boldsymbol \Omega}_a\:{\boldsymbol \omega}\:,
	&&{\boldsymbol \omega} ~\equiv~
	[\omega_u~~ \omega_v~~ \omega_e~~\varpi_u]^T\:,
	\label{eq:Killing_III_21_deg}
	\end{align}
	where
	\begin{align}
	&{\boldsymbol \Omega}_a ~\equiv~
	u_a
	\left[\begin{smallmatrix}
	\kappa_v & 0 & \eta_v & 0 \\
	-\frac{\kappa_e}{2} &
	- \pounds_v \varphi&
	\tau_v-\tau_e-\pounds_e \varphi & 0\\
	-\tau_v-\tau_e & -\eta_v & -\eta_e & -1 \\
	\frac{\eta_v (2\kappa_u +\kappa_e) }{2} - \kappa_v (\tau_v - \tau_e) + \pounds_e \kappa_v &
	~\eta_v (\eta_e + \kappa_v-\pounds_v \varphi) ~&
	-\eta_v (\tau_u + \tau_v+ \pounds_e \varphi )+ \pounds_e \eta_v ~&
	\kappa_v + \eta_e
	\end{smallmatrix}\right]\notag\\
	&+v_a
	\left[\begin{smallmatrix}
	-\frac{2\kappa_u-\kappa_e}{2}&
	\pounds_v \varphi-\kappa_v &
	\pounds_e \varphi + \tau_u + \tau_e
	& 0\\
	0 & \kappa_u & 0 & 0 \\
	0 & -\tau_u+\tau_v & -\kappa_e & 0 \\
	\frac{\kappa_e (\tau_u-3\tau_e)}{2} + \kappa_u (\tau_u+ \tau_e)-\pounds_e \kappa_u &
	\eta_e(\tau_u-\tau_v) - (\tau_u+\tau_e) \bigl( \pounds_v \varphi - \kappa_v \bigr) &
	\pounds_e\tau_u  - (\tau_u+\tau_e) \bigl( \pounds_e \varphi +\tau_u + \tau_e \bigr)&
	-\kappa_u
	\end{smallmatrix}\right] \notag \\
	&
	+e_a
	\left[
	\begin{smallmatrix}
	0 & 0 & 0 & 1 \\
	0 & -\tau_v + \tau_e & 0 & 0 \\
	\kappa_e & \eta_e & 0 & 0 \\
	\frac{\eta_e (\kappa_e - 2\kappa_u)}{2} + \tau_u^2 - \tau_e^2 - \pounds_u \eta_e - \pounds_e (\tau_u+\tau_e) &
	\eta_v (\tau_u + \tau_v - 2\tau_e) - \pounds_v \eta_e - \pounds_e \eta_v + \eta_e \pounds_v \varphi &
	\eta_e \pounds_e \varphi - \pounds_e \eta_e & - 2\tau_e
	\end{smallmatrix}
	\right]\:.
	\end{align}
Several parts of the compatibility condition for \eq \eqref{eq:Killing_III_21_deg} lead to
	\begin{subequations}
		\begin{align}
		\sigma\:\varpi_u~=~&
		-\Bigl( \e^{\varphi} \pounds_u \Sigma + (\tau_u + \tau_e) \sigma \Bigr) \omega_u
		-\Bigl( \e^{\varphi} \pounds_v \Sigma + \eta_v\sigma \Bigr) \omega_v
		-\Bigl( \e^{\varphi}\pounds_e \Sigma \Bigr) \omega_e
		\:,\\
		\kappa_e\:\varpi_u~=~&
		-\kappa_e (\tau_v+\tau_e) \omega_u
		-(\pounds_v \tau_v + \eta_v \kappa_e) \omega_v
		-(\pounds_e \tau_v) \omega_e
		\:,
		\end{align}
		where
		\begin{align}
		&\sigma ~\equiv~ \pounds_e \varphi + \tau_v + \tau_e\:,
		&&\Sigma ~\equiv~ \pounds_v ( \e^{-\varphi} )+ \kappa_v e^{-\varphi}\:.
		\end{align}
		\label{eq:compatibility_III_21_deg}
	\end{subequations}
	This implies that $\varpi_u$ can be expressed in terms of $\{\omega_u, \omega_v, \omega_e\}$
	except when $\sigma=\kappa_e=0$.
	Depending on the nonzeroness of the coefficients $\{\sigma,\kappa_e\}$, the analysis falls into three sub-branches.

\subsubsection*{\underline{Sub-branch where $\sigma=\kappa_e=0$}}

In this sub-branch, the 1-jet variable $\varpi_u$ cannot be expressed in terms of $\{\omega_u, \omega_v, \omega_e\}$.
The remaining parts of the compatibility condition for \eq \eqref{eq:Killing_III_21_deg} read
\begin{align}
&{\boldsymbol R}_{[(21)]}\:{\boldsymbol \omega} ~=~ 0\:,
&&
{\boldsymbol R}_{[(21)]}~\equiv~
\begin{bmatrix}
0 & \pounds_v \tau_v & \pounds_e \tau_v & 0\\
\e^{\varphi} \pounds_u \Sigma & \e^{\varphi} \pounds_v \Sigma & \e^{\varphi} \pounds_e \Sigma & 0
\end{bmatrix}\:.
\label{eq:compatibility_III_21_def_submaxi}
\end{align}
Remark that some remaining components are derivable from its derivative.
In this sub-branch, the rank of ${\boldsymbol R}_{[(21)]}$ governs the number of KVs
in the same way as that presented in Sub-subsection \ref{subsec:caseIII_111_21_t}.

\subsubsection*{\underline{Other sub-branches}}

Except when $\sigma=\kappa_e=0$,
\eqs \eqref{eq:compatibility_III_21_deg} allow us to write the 1-jet variable $\varpi_u$ in terms of $\{\omega_u, \omega_v, \omega_e\}$.
In these sub-branches, \eq \eqref{eq:Killing_III_21_deg} reduces to a PDE system of the form
\begin{align*}
\nabla_a
{\boldsymbol \omega}
~=~
{\boldsymbol \Omega}_a\:{\boldsymbol \omega}\:,
&&
{\boldsymbol \omega}~\equiv~
[\omega_u ~~\omega_v~~\omega_e]^T\:.
\label{eq:Killing_III_21_deg_others}
\end{align*}
Here, the compatibility condition of \eq \eqref{eq:Killing_III_21_deg_others} is considered collectively.
The results are displayed as follows:

\begin{subequations}
	\noindent{\it (\#1) For the case of $\sigma =0, \kappa_e \neq 0$},
	\begin{align}
	\varpi_u ~=~& -(\tau_v+\tau_e)\omega_u -\left( \frac{\pounds_v \tau_v}{\kappa_e} + \eta_v \right) \omega_v
	-\left( \frac{\pounds_e \tau_v}{\kappa_e} \right) \omega_e
	\:,\\
	{\boldsymbol \Omega}_a ~=~&
	u_a
	\left[
	\begin{matrix}
	\kappa_v & 0 & \eta_v \\
	-\frac{\kappa_e}{2} & - \pounds_v \varphi & \tau_v - \tau_e - \pounds_e \varphi \\
	0 & \frac{\pounds_v \tau_v}{\kappa_e} & \frac{\pounds_e \tau_v}{\kappa_e} -\eta_e
	\end{matrix}\right]
	+v_a
	\left[
	\begin{matrix}
	-\kappa_u + \frac{\kappa_e}{2} & \pounds_v \varphi - \kappa_v & \pounds_e \varphi + \tau_u + \tau_e\\
	0 & \kappa_u & 0 \\
	0 & -\tau_u+\tau_v & -\kappa_e
	\end{matrix}\right] \notag \\
	&+e_a
	\left[
	\begin{matrix}
	-\tau_v-\tau_e &
	-\eta_v - \frac{\pounds_v \tau_v}{\kappa_e} &
	- \frac{\pounds_e \tau_v}{\kappa_e} \\
	0 & \tau_e-\tau_v & 0\\
	\kappa_e & \eta_e & 0
	\end{matrix}\right]\:,\\
	{\boldsymbol R}_{[(21)]}^{~\# 1} ~=~&
	\begin{bmatrix}
	\pounds_u \tau_v & \pounds_v \tau_v & \pounds_e \tau_v \\
	\e^{\varphi} \pounds_u \Sigma & \e^{\varphi} \pounds_v \Sigma & \e^{\varphi} \pounds_e \Sigma \\
	\pounds_u\kappa_e+\kappa_e \pounds_u \varphi& 
	\pounds_v\kappa_e+\kappa_e \pounds_v\varphi & \pounds_e\kappa_e+\kappa_e \pounds_e \varphi\\
	\pounds_u \eta_e-\eta_e \pounds _u \varphi &
	\pounds_v \eta_e-\eta_e \pounds _v \varphi &
	\pounds_e \eta_e-\eta_e \pounds _e \varphi \\
	\pounds _u \eta_v-2\eta_v\pounds _u \varphi &
	\pounds _v \eta_v-2\eta_v\pounds _v \varphi &
	\pounds _e \eta_v-2\eta_v\pounds _e \varphi
	\end{bmatrix}\:.
	\end{align}
	\label{eq:compatibility_III_21_deg_1}
\end{subequations}

\begin{subequations}
	\noindent{\it (\#2) For the case of $\sigma \neq 0$},
	\begin{align}
	\varpi_u ~=~& -\left( \frac{\e^\varphi \pounds_u \Sigma}{\sigma} + \tau_u + \tau_e\right) \omega_u
	-\left( \frac{\e^\varphi \pounds_v \Sigma}{\sigma} + \eta_v\right) \omega_v
	-\left( \frac{\e^\varphi \pounds_e \Sigma}{\sigma} \right) \omega_e
	\:,\\
	{\boldsymbol \Omega}_a ~=~&
	u_a
	\left[
	\begin{smallmatrix}
	\kappa_v & 0 & \eta_v \\
	-\frac{\kappa_e}{2} & - \pounds_v \varphi & \tau_v - \tau_e - \pounds_e \varphi \\
	\tau_u-\tau_v + \frac{\e^\varphi \pounds_u \Sigma}{\sigma} &
	\frac{\e^\varphi \pounds_v \Sigma}{\sigma} &
	\frac{\e^\varphi \pounds_e \Sigma}{\sigma} - \eta_e
	\end{smallmatrix}\right]
	+v_a
	\left[
	\begin{smallmatrix}
	-\kappa_u + \frac{\kappa_e}{2} & \pounds_v \varphi - \kappa_v & \pounds_e \varphi + \tau_u + \tau_e\\
	0 & \kappa_u & 0 \\
	0 & -\tau_u+\tau_v & -\kappa_e
	\end{smallmatrix}\right] \notag \\
	&+e_a
	\left[
	\begin{smallmatrix}
	-\tau_u - \tau_e - \frac{\e^\varphi \pounds_u \Sigma}{\sigma} &
	-\eta_v - \frac{\e^\varphi \pounds_v \Sigma}{\sigma} &
	- \frac{\e^\varphi \pounds_e \Sigma}{\sigma} \\
	0 & -\tau_v + \tau_e & 0 \\
	\kappa_e & \eta_e & 0 
	\end{smallmatrix}\right]\:,\\
	{\boldsymbol R}_{[(21)]}^{~\# 2} ~=~&
	\begin{bmatrix}
	\pounds_u \left(\sigma - \frac{5}{2}\tau_v \right) &
	\pounds_v \left(\sigma - \frac{5}{2}\tau_v \right) &
	\pounds_e \left(\sigma - \frac{5}{2}\tau_v \right) \\
	\pounds_u \tau_v - \frac{\e^\varphi \kappa_e \pounds_u \Sigma}{\sigma} &
	\pounds_v \tau_v - \frac{\e^\varphi \kappa_e \pounds_v \Sigma}{\sigma} &
	\pounds_e \tau_v - \frac{\e^\varphi \kappa_e \pounds_e \Sigma}{\sigma} \\
	\pounds_u \kappa_e+\kappa_e \pounds _u \varphi &
	\pounds_v \kappa_e+\kappa_e \pounds _v \varphi &
	\pounds_e \kappa_e+\kappa_e \pounds _e \varphi  \\
	\Phi_u & \Phi_v & \Phi_e \\
	\Theta_u & \Theta_v & \Theta_e
	\end{bmatrix}\:,
	\end{align}
	where 
	\begin{align}
	\Phi_\alpha ~\equiv~&
	\pounds_\alpha \eta_v
	- 2 \eta_v \pounds_\alpha \varphi
	+\frac{\e^\varphi \pounds_\alpha \pounds_v \Sigma}{\sigma} \notag \\
	&
	-\frac{\e^\varphi (\pounds_\alpha \varphi) (\pounds_v \Sigma)}{\sigma}
	+\Bigl(
	\e^\varphi \pounds_e \Sigma
	-\pounds_v \sigma 
	+(\pounds_v \varphi -\kappa_v-\eta_e) \sigma
	\Bigr)
	\frac{\e^\varphi \pounds_\alpha \Sigma}{\sigma^2}\:, \\
	\Theta_\alpha ~\equiv~& \pounds_\alpha \eta_e
	-\eta_e \pounds_\alpha \varphi
	-\frac{\e^\varphi \pounds_\alpha \pounds_e \Sigma}{\sigma}
	+
	\Bigl(
	\e^{\varphi} \pounds_u \Sigma
	+ \pounds_e \sigma 
	+(\tau_u + \tau_v - \sigma)\sigma
	\Bigr) \frac{ \e^\varphi \pounds_\alpha \Sigma}{\sigma^2}\:,
	\end{align}
	\label{eq:compatibility_III_21_deg_2}
\end{subequations}

In these sub-branches, the rank of ${\boldsymbol R}_{[(21)]}^{~\# 1}$ and ${\boldsymbol R}_{[(21)]}^{~\# 2}$ governs the number of KVs
in the same way as that presented in Sub-subsection \ref{subsec:caseIII_111_21_t}.

\subsubsection{Branch where the Segre type is $[21]$}\label{subsec:caseIII_21_dis}

In this branch, it is assumed that $\lambda = \lambda_1 = -(1/2)\lambda_3 \neq 0$.
Then, it immediately follows from \eqs \eqref{eq:caseIII_21_compa1} and \eqref{eq:caseIII_21_Bianchi} that
\begin{subequations}
	\begin{align}
	\varpi_u &~=~
	-(\tau_u + \tau_e) \omega_u
	-\eta_v \omega_v\:, \\
	\varpi_v &~=~
	-\Bigl( \pounds_u \varphi + \kappa_u\Bigr) \omega_u
	-\Bigl( \pounds_v \varphi \Bigr) \omega_v
	-\Bigl( \pounds_e \varphi - \tau_v + \tau_e\Bigr) \omega_e
	\:,\\
	\varpi_e &~=~ -(\tau_u-\tau_v) \omega_v\:,
	\end{align}
	and
	\begin{align}
	&\kappa_e ~=~0\:,
	&&\eta_u ~=~ -\left(\frac{3\lambda}{S_{vv}}\right) (\tau_u+\tau_v)\:.
	\end{align}
	\label{eq:caseIII_21_dis_cond}
\end{subequations}

Given these conditions \eqref{eq:caseIII_21_dis_cond}, the Killing equation \eqref{eq:Killing}
produces a PDE system of the form
\begin{align}
&\nabla_a {\boldsymbol \omega} ~=~{\boldsymbol \Omega}_a\:{\boldsymbol \omega}\:,
&&{\boldsymbol \omega} ~\equiv~
[\omega_u~~ \omega_v~~ \omega_e]^T\:,
\label{eq:Killing_III_21_dis}
\end{align}
where
\begin{align}
{\boldsymbol \Omega}_a ~\equiv~&
u_a
\left[\begin{smallmatrix}
\kappa_v & 0 & \eta_v \\
-\kappa_u - \pounds_u \varphi &
-\pounds_v \varphi &
-\pounds_e \varphi + \tau_v -\tau_e\\
\tau_u-\tau_v & 0 & -\eta_e
\end{smallmatrix}\right]
+v_a
\left[\begin{smallmatrix}
\pounds_u \varphi & \pounds_v \varphi -\kappa_v & \pounds_e \varphi + \tau_u+ \tau_e\\
0 & \kappa_u & - \frac{3\lambda}{S_{vv}} (\tau_u+\tau_v) \\
0 & -\tau_u+\tau_v & 0
\end{smallmatrix}\right]
+e_a
\left[
\begin{smallmatrix}
-\tau_u-\tau_e & -\eta_v & 0 \\
\frac{3\lambda}{S_{vv}} (\tau_u+\tau_v) & -\tau_v+\tau_e & 0 \\
0 & \eta_e & 0
\end{smallmatrix}
\right]
\:.
\end{align}
with $\varphi \equiv (1/2) \log S_{vv}$.
The compatibility condition for \eq \eqref{eq:Killing_III_21_dis} leads to
\begin{subequations}
\begin{align}
{\boldsymbol R}_{[21]}&~=~
\begin{bmatrix}
\pounds_u \tau_u & \pounds_v \tau_u & \pounds_e \tau_u \\
\pounds_u \tau_v & \pounds_v \tau_v & \pounds_e \tau_v \\
\pounds_u \sigma&
\pounds_v \sigma&
\pounds_e \sigma\\
\e^\varphi \pounds_u \Sigma &
\e^\varphi \pounds_v \Sigma &
\e^\varphi \pounds_e \Sigma \\
\pounds_u \eta_v - 2\eta_v \pounds_u \varphi &
\pounds_v \eta_v - 2\eta_v \pounds_v \varphi  &
\pounds_e \eta_v - 2\eta_v \pounds_e \varphi \\
\pounds_u \eta_e-\eta_e \pounds_u \varphi &
\pounds_v \eta_e-\eta_e \pounds_v \varphi  &
\pounds_e \eta_e-\eta_e \pounds_e \varphi  \\
\end{bmatrix}\:,
\label{eq:compatibility_III_21_dis}
\end{align}
where
\begin{align}
\Sigma ~\equiv~ \pounds_v (\e^{-\varphi}) + \kappa_v e^{-\varphi}\:.
\end{align}
\end{subequations}

In this sub-branch, the rank of ${\boldsymbol R}_{[21]}$ governs the number of KVs
in the same way as that presented in Sub-subsection \ref{subsec:caseIII_111_21_t}.

\subsubsection{Short summary of class 3 type $[21]$}\label{sec:summary_type_21}

\begin{figure}[h]
	\begin{center}
		\begin{tikzpicture}
		[
		every node/.style={outer sep=0.15cm, inner sep=0cm},
		arrow/.style={-{Stealth[length=0.15cm]},thick},
		rblock/.style={rectangle, rounded corners,draw, minimum height = 0.5cm,
			minimum width=1.6cm, thick, outer sep = 0},
		rgblock/.style={rectangle, rounded corners,draw, minimum height = 0.5cm,
			minimum width=2.5cm, thick, outer sep = 0},
		point/.style={radius=2pt}
		]
		\node [rgblock] (title){type $[21]$};
		\node [below=0.5 of title] (Obst){$\blacktriangleright \:\mathrm{rank}{\boldsymbol R}_{[21]}$};
		\node [rblock,right=2 of Obst] (caseII){class 2};
		\node [rblock,below=0.5 of caseII] (caseI){class 1};
		\node [below=0.5 of caseI] (0KV){no KV};
		\node [below=0.5 of 0KV] (3KVs){3 KVs};
		\draw[arrow] (Obst.east) -- (caseII.west) node[above,pos=0.65] {{\scriptsize ~~$1$}};
		\draw[arrow] (Obst.east) -- (caseI.west) node[above,pos=0.65] {{\scriptsize ~~$2$}};
		\draw[arrow] (Obst.east) -- (3KVs.west) node[right,pos=0.55] {{\scriptsize $0$}};
		\draw[arrow] (Obst.east) -- (0KV.west) node[right,pos=0.55] {{\scriptsize $3$}};
		\end{tikzpicture}
		\caption{The sub-sub-algorithm for the class 3 type $[21]$.
			See \eq \eqref{eq:compatibility_III_21_dis} for notations.
			}
		\label{fig:type21_dis}
	\end{center}
\end{figure}
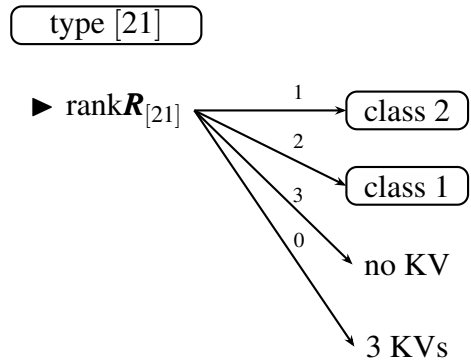

We synopsise the results obtained in this subsection in Figures \ref{fig:type21_deg}--\ref{fig:type21_dis}.

\begin{figure}[h]
	\begin{center}
		\begin{tikzpicture}
		[
		every node/.style={outer sep=0.15cm, inner sep=0cm},
		arrow/.style={-{Stealth[length=0.15cm]},thick},
		rblock/.style={rectangle, rounded corners,draw, minimum height = 0.5cm,
			minimum width=1.6cm, thick, outer sep = 0},
		rgblock/.style={rectangle, rounded corners,draw, minimum height = 0.5cm,
			minimum width=2.5cm, thick, outer sep = 0},
		point/.style={radius=2pt}
		]
		\node [rgblock] (title){type $[(21)]$};
		\node [below=0.5 of title] (test1){$\blacktriangleright \:\sigma=0$};
		\node [below=0.5 of test1] (test2){$\quad\kappa_e=0$};
		\node [right=1 of test1] (Obst2){$\mathrm{rank}{\boldsymbol R}_{[(21)]}^{~\# 2}$};
		\node [right=1 of test2] (Obst1){$\mathrm{rank}{\boldsymbol R}_{[(21)]}^{~\# 1}$};
		\node [rblock,right=1.5 of Obst2] (caseII1){class 2};
		\node [rblock,right=1.5 of Obst1] (caseI1){class 1};
		\node [below=0.5 of Obst1] (empty){\qquad \quad};
		\node [right=2 of empty] (noKV){no KV};
		\node [below=0.5 of noKV] (3KVs){3 KVs};
		\node [below=1.5 of empty] (Obst){$\mathrm{rank}{\boldsymbol R}_{[(21)]}$};
		\node [rblock,right=1.5 of Obst] (caseII2){class 2};
		\node [rblock,below=0.5 of caseII2] (caseI2){class 1};
		\node [below=0.5 of caseI2] (0KV2){no KV};
		\node [below=0.5 of 0KV2] (4KVs2){4 KVs};
		\draw[arrow] (test1) -- (test2) node[right,pos=0.5] {{\scriptsize ~~yes}};
		\draw[arrow] (test2) |- (Obst) node[above,pos=0.1] {\qquad ~{\scriptsize yes}};
		\draw[arrow] (test1) -- (Obst2) node[above,midway] {{\scriptsize no}};
		\draw[arrow] (test2) -- (Obst1) node[above,midway] {{\scriptsize no}};
		\draw[arrow, name path=cross1] (Obst1.east) -- (caseI1.west) node[above,pos=0.75,yshift=-0.5cm] {{\scriptsize $2$}};
		\draw[arrow, name path=cross2] (Obst2.east) -- (3KVs.west) node[above,pos=0.75,yshift=-0.5cm] {{\scriptsize $0~$}};
		\path[name intersections={of=cross1 and cross2,by={Point1}}];
		\fill [point] (Point1) circle;
		\draw[arrow] (Point1) -- (caseII1.west) node[above,pos=0.6,yshift=-0.5cm] {{\scriptsize $1$}};
		\draw[arrow] (Point1) -- (noKV.west) node[above,pos=0.6,yshift=-0.5cm] {{\scriptsize $3$}};
		\draw[arrow] (Obst.east) -- (caseII2.west) node[above,pos=0.65] {{\scriptsize ~~$1$}};
		\draw[arrow] (Obst.east) -- (caseI2.west) node[above,pos=0.65] {{\scriptsize ~~$2$}};
		\draw[arrow] (Obst.east) -- (4KVs2.west) node[right,pos=0.525] {{\scriptsize $0$}};
		\draw[arrow] (Obst.east) -- (0KV2.west) node[right,pos=0.525] {{\scriptsize $3$}};
		\end{tikzpicture}
		\caption{The sub-sub-algorithm for the class 3 type $[(21)]$.
		See \eqs \eqref{eq:compatibility_III_21_def_submaxi}--\eqref{eq:compatibility_III_21_deg_2} for notations.
			}
		\label{fig:type21_deg}
	\end{center}
\end{figure}

\subsection{Type $[3]$}\label{subsec:caseIII_3}

In this case, we have the following Jordan chain:
\begin{subequations}
	\begin{align}
	S^a{}_b\:j_1^b &~=~ \lambda\:j_1^a\:,\\
	S^a{}_b\:j_2^b &~=~ \lambda\:j_2^a + j_1^a\:,\\
	S^a{}_b\:j_3^b &~=~ \lambda\:j_3^a + j_2^a\:,
	\end{align}
	\label{eq:chain_3}
\end{subequations}
with $\lambda = 0$.
Here $j_\alpha^a$ is an generalised eigenvector of $S^a{}_b$.
It can be shown that the vectors $j_1^a, j_2^a$ are respectively null and spacelike,
whereas the causal nature of $j_3^a$ is free from restriction.
In this subsection, it is assumed that $\{u^a,v^a,e^a\}$ forms a double-null basis of $T(M)$,
\begin{align}
g^{ab}~=~u^a v^b + v^a u^b + e^a e^b\:,
\label{eq:d_null_basis_III_3}
\end{align}
where $u^a \equiv j_1^a$; $e^a \equiv j_2^a$;
$v^a$ is defined as a null vector such that $g_{ab} u^a v^b=1$ and $g_{ab} v^a e^b = 0$.
Then, any KV can be written as
\begin{align}
K^a ~=~ \omega_u\:u^a + \omega_v\:v^a + \omega_e\:e^a\:,
\label{eq:caseIII_3}
\end{align}
where scalars $\{ \omega_u, \omega_v, \omega_e\}$ are yet indeterminate.

Calculating the second Bianchi identity, $\nabla_a R^a{}_b - (1/2) \nabla_b R = 0$, leads to
\begin{align}
& \eta_u ~=~ 0\:,
&& \kappa_u ~=~ 2 \kappa_e\:,
&& \pounds_u S_{vv} + 3 \kappa_e\:S_{vv} ~=~ 2 \tau_u + \tau_v - \tau_e\:,
\label{eq:caseIII_3_Bianchi}
\end{align}
where $S_{vv}\equiv S_{ab} v^a v^b$ and
the Ricci rotation coefficients are defined by \eqs \eqref{eq:ricci_coeffs_null}.
Note that $S_{ab}v^a e^b = 1$ by \eqs \eqref{eq:chain_3} and \eqref{eq:d_null_basis_III_3}.
Using this identity, the first compatibility of \eq \eqref{eq:Killing}, $\pounds_K S_{ab}=0$, can be written in components
\begin{subequations}
	\begin{align}
	\varpi_u ~=~&
	\tfrac{1}{2} (\tau_v-3\tau_e-3 \kappa_e S_{vv}) \omega_u
	+\tfrac{1}{2} (\pounds_v S_{vv} - 2 \eta_v) \omega_v
	+\tfrac{1}{2}(\pounds_e S_{vv})\omega_e\:,\\
	\varpi_v ~=~& - 2\kappa_e\:\omega_u + (\tau_v-\tau_e)\omega_e\:,\\
	\varpi_e ~=~& - (\tau_u-\tau_v)\omega_v - \kappa_e\:\omega_e\:,
	\end{align}
	\label{eq:caseIII_3_compa1}
\end{subequations}
where the 1-jet variables $\{\varpi_u, \varpi_v, \varpi_e\}$ are respectively defined as
\begin{align}
&\varpi_u~\equiv~\pounds_e \omega_u\:,
&&\varpi_v~\equiv~\pounds_v \omega_v\:,
&&\varpi_e~\equiv~\pounds_u\omega_e\:.
\end{align}
Then, the Killing equation \eqref{eq:Killing} produces a PDE system of the form
\begin{align}
&\nabla_a {\boldsymbol \omega} ~=~{\boldsymbol \Omega}_a\:{\boldsymbol \omega}\:,
&&{\boldsymbol \omega} ~\equiv~
[\omega_u~~ \omega_v~~ \omega_e]^T\:,
\label{eq:Killing_III_3}
\end{align}
where
\begin{align}
{\boldsymbol \Omega}_a ~\equiv~&
u_a
\left[\begin{smallmatrix}
\kappa_v & 0 & \eta_v \\
-2\kappa_e & 0 & \tau_v-\tau_e \\
\frac{\tau_e - 3\tau_v + 3\kappa_e S_{vv}}{2} & -\frac{\pounds_v S_{vv}}{2} & -\frac{\pounds_e S_{vv}}{2} -\eta_e
\end{smallmatrix}\right]
+v_a
\left[\begin{smallmatrix}
0 & -\kappa_v & \tau_u + \tau_e \\
0 & 2\kappa_e & 0\\
0 & -\tau_u + \tau_v & -\kappa_e
\end{smallmatrix}\right] \notag \\
&+e_a
\left[
\begin{smallmatrix}
\frac{\tau_v - 3\tau_e - 3 \kappa_e S_{vv}}{2} & \frac{\pounds_v S_{vv}}{2} - \eta_v & \frac{\pounds_e S_{vv}}{2} \\
0 & -\tau_v + \tau_e & 0\\
\kappa_e & \eta_e & 0
\end{smallmatrix}
\right]
\:,
\end{align}
The compatibility condition for \eq \eqref{eq:Killing_III_3} leads to
\begin{subequations}
\begin{align}
{\boldsymbol R}_{[3]}&~=~
\begin{bmatrix}
\pounds_u \kappa_e & \pounds_v \kappa_e & \pounds_e \kappa_e \\
\pounds_u (\tau_e - 3 \tau_v) & \pounds_v (\tau_e - 3 \tau_v) & \pounds_e (\tau_e - 3 \tau_v) \\
\pounds_u (\kappa_e S_{vv}+2 \tau_v) & \pounds_v (\kappa_e S_{vv}+2 \tau_v) & \pounds_e (\kappa_e S_{vv}+2 \tau_v) \\
\pounds_u \kappa_v + \tfrac{\tau_v + \tau_e}{2} \pounds_u S_{vv} & \pounds_v \kappa_v + \tfrac{\tau_v + \tau_e}{2} \pounds_v S_{vv} & \pounds_e \kappa_v + \tfrac{\tau_v + \tau_e}{2} \pounds_e S_{vv}\\
\Xi_u & \Xi_v & \Xi_e \\
\Theta_u & \Theta_v & \Theta_e 
\end{bmatrix}\:,
\label{eq:compatibility_III_3}
\end{align}
where
\begin{align}
\Xi_\alpha ~\equiv~&
\pounds_\alpha \eta_e
+ \frac{\pounds_\alpha \pounds_e S_{vv}}{2}
+\left( \tau_v + \tau_e - 3 \kappa_e S_{vv} \right) \frac{\pounds_\alpha S_{vv}}{4}\:, \\
\Theta_\alpha ~\equiv~&
\pounds_\alpha \eta_v
- \frac{\pounds_\alpha \pounds_v S_{vv}}{2}
+\left( \pounds_e S_{vv} +2 \kappa_v + 2 \eta_e \right) \frac{\pounds_\alpha S_{vv}}{4}\:.
\end{align}
\end{subequations}

In this sub-branch, the rank of ${\boldsymbol R}_{[3]}$ governs the number of KVs
in the way shown in Figure \ref{fig:type3}.

\begin{figure}[h]
	\begin{center}
		\begin{tikzpicture}
		[
		every node/.style={outer sep=0.15cm, inner sep=0cm},
		arrow/.style={-{Stealth[length=0.15cm]},thick},
		rblock/.style={rectangle, rounded corners,draw, minimum height = 0.5cm,
			minimum width=1.6cm, thick, outer sep = 0},
		rgblock/.style={rectangle, rounded corners,draw, minimum height = 0.5cm,
			minimum width=2.5cm, thick, outer sep = 0},
		point/.style={radius=2pt}
		]
		\node [rgblock] (title){type $[3]$};
		\node [below=0.5 of title] (Obst){$\blacktriangleright \:\mathrm{rank}{\boldsymbol R}_{[3]}$};
		\node [rblock,right=2 of Obst] (caseII){class 2};
		\node [rblock,below=0.5 of caseII] (caseI){class 1};
		\node [below=0.5 of caseI] (0KV){no KV};
		\node [below=0.5 of 0KV] (3KVs){3 KVs};
		\draw[arrow] (Obst.east) -- (caseII.west) node[above,pos=0.65] {{\scriptsize ~~$1$}};
		\draw[arrow] (Obst.east) -- (caseI.west) node[above,pos=0.65] {{\scriptsize ~~$2$}};
		\draw[arrow] (Obst.east) -- (3KVs.west) node[right,pos=0.55] {{\scriptsize $0$}};
		\draw[arrow] (Obst.east) -- (0KV.west) node[right,pos=0.55] {{\scriptsize $3$}};
		\end{tikzpicture}
		\caption{The sub-sub-algorithm for the class 3 type $[3]$.
		See \eq \eqref{eq:compatibility_III_3} for notations.
		}
		\label{fig:type3}
	\end{center}
\end{figure}

\subsection{Type $[z\bar{z}1]$}\label{subsec:caseIII_zz1}
In this case, we have the following Jordan chains:
\begin{subequations}
	\begin{align}
	S^a{}_b\:j_+^b &~=~ \lambda_+\:j_+^a\:,\\
	S^a{}_b\:j_-^b &~=~ \lambda_-\:j_-^a\:,\\
	S^a{}_b\:j^b &~=~ \lambda \:j^a\:,
	\end{align}
	\label{eq:chain_zz1}
where $\lambda_\pm = \alpha \pm i \beta (\beta\neq0)$ are complex eigenvalues
corresponding to the complex eigenvectors $j_\pm^a = x^a \pm i y^a$.
It follows from the symmetric traceless property of $S_{ab}$ that
$2\alpha + \lambda=0$ and $g_{ab} (x^a x^b + y^a y^b) = 0$.
On the other hand, the real/imaginary parts of \eqs \eqref{eq:chain_zz1} give
\begin{align}
S^a{}_b\:x^b ~=~& \alpha\:x^a - \beta\:y^a\:,\\
S^a{}_b\:y^b ~=~& \beta\:x^a + \alpha\:y^a\:.
\end{align}
\end{subequations}
This implies that the real vectors $\{x^a, y^a\}$ span a timelike surface.
One can then fix $x^a$ to be timelike and $y^a$ to be spacelike without loss of generality.
In this subsection, it is supposed that $\{e_\alpha^a\}$ forms an orthonormal basis of $T(M)$,
\begin{align}
g^{ab} ~=~ -e_1^a e_1^b + e_2^a e_2^b + e_3^a e_3^b\:,
\label{eq:frame_caseIII_zz1}
\end{align}
where $e_1^a \propto x^a$; $e_2^a \propto y^a$ and $e_3^a \propto j^a$.
Then, any KV can be written as
\begin{align}
K^a ~=~ \sum_{\alpha=1}^3 \omega_\alpha \: e^a_\alpha\:,
\label{eq:caseIII_zz1}
\end{align}
where scalars $\{ \omega_\alpha\}$ are yet indeterminate.

The first compatibility condition, $\pounds_K S_{ab}=0$,  of \eq \eqref{eq:Killing}, gives rise to
	\begin{subequations}
		\begin{align}
		\varpi_1~=~& \eta_1\:\omega_1 + (\tau_2-\tau_3)\omega_2\:,\\
		\varpi_2~=~&-\kappa_2\:\omega_2+(\tau_1+\tau_3) \omega_3\:,\\
		\varpi_3~=~&(\tau_1-\tau_2)\omega_1 - \eta_3\:\omega_3\:,
		\end{align}
		\label{eq:caseIII_zz1_compa1}
	\end{subequations}
	where the Ricci rotation coefficients are defined by \eqs \eqref{eq:ricci_coeffs}.
	The 1-jet variables $\{\varpi_u, \varpi_v, \varpi_e\}$ are respectively defined as
	\begin{align}
	&\varpi_1~\equiv~\pounds_3 \omega_1\:,
	&&\varpi_2~\equiv~\pounds_1 \omega_2\:,
	&&\varpi_3~\equiv~\pounds_2\omega_3\:.
	\end{align}
	Then, the Killing equation \eqref{eq:Killing} produces a PDE system of the form
\begin{align}
&\nabla_a {\boldsymbol \omega} ~=~{\boldsymbol \Omega}_a\:{\boldsymbol \omega}\:,
&&{\boldsymbol \omega} ~\equiv~
[\omega_1~~ \omega_2~~ \omega_3]^T\:,
\label{eq:Killing_III_zz1}
\end{align}
where
\begin{align}
{\boldsymbol \Omega}_a ~\equiv~&
-e_{1a}
\left[\begin{smallmatrix}
0 & - \kappa_1 & -\eta_1 \\
0 & -\kappa_2 & \tau_1 + \tau_3 \\
0 & \tau_2 -\tau_1 & - \kappa_3
\end{smallmatrix}\right]
+e_{2a}
\left[\begin{smallmatrix}
\kappa_1 & 0 & \tau_3 - \tau_2 \\
\kappa_2 & 0 & \eta_2 \\
\tau_1-\tau_2 & 0 & -\eta_3
\end{smallmatrix}\right]
+e_{3a}
\left[
\begin{smallmatrix}
\eta_1 & \tau_2 - \tau_3 & 0\\
-\tau_1 - \tau_3 & -\eta_2 & 0 \\
\kappa_3 & \eta_3 & 0
\end{smallmatrix}
\right]
\:,
\end{align}
where the Ricci rotation coefficients are defined by \eqs \eqref{eq:ricci_coeffs}.
Note that the second Bianchi identity imposes
\begin{align}
&\eta_3 -2 \kappa_1 ~=~ \frac{3 \alpha}{\beta}\:\kappa_3\:,
&&\kappa_3 +2 \kappa_2 ~=~ -\frac{3 \alpha}{\beta}\:\eta_3\:,
&&\tau_1+\tau_2 ~=~ \frac{3 \alpha}{\beta} (\eta_1 - \eta_2)\:,
\end{align}
The compatibility condition for \eq \eqref{eq:Killing_III_zz1} leads to
	\begin{align}
	{\boldsymbol R}_{[z\bar{z}1]}&~=~
	\begin{bmatrix}
	\pounds_1 \kappa_3 & \pounds_2 \kappa_3 & \pounds_3 \kappa_3\\
	\pounds_1 \eta_1 & \pounds_2 \eta_1 & \pounds_3 \eta_1\\
	\pounds_1 \eta_2 & \pounds_2 \eta_2 & \pounds_3 \eta_2\\
	\pounds_1 \eta_3 & \pounds_2 \eta_3 & \pounds_3 \eta_3\\
	\pounds_1 \tau_2 & \pounds_2 \tau_2 & \pounds_3 \tau_2\\
	\pounds_1 \tau_3 & \pounds_2 \tau_3 & \pounds_3 \tau_3\\
	\end{bmatrix}\:.
	\label{eq:compatibility_III_zz1}
	\end{align}

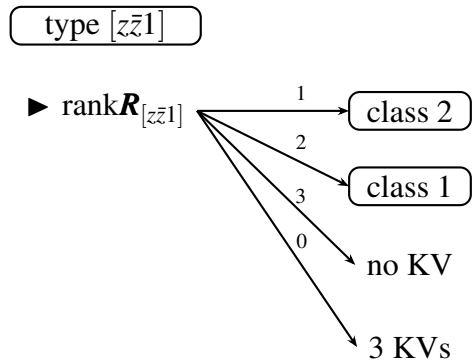
\begin{figure}[h]
	\begin{center}
		\begin{tikzpicture}
		[
		every node/.style={outer sep=0.15cm, inner sep=0cm},
		arrow/.style={-{Stealth[length=0.15cm]},thick},
		rblock/.style={rectangle, rounded corners,draw, minimum height = 0.5cm,
			minimum width=1.6cm, thick, outer sep = 0},
		rgblock/.style={rectangle, rounded corners,draw, minimum height = 0.5cm,
			minimum width=2.5cm, thick, outer sep = 0},
		point/.style={radius=2pt}
		]
		\node [rgblock] (title){type $[z\bar{z}1]$};
		\node [below=0.5 of title] (Obst){$\blacktriangleright \:\mathrm{rank}{\boldsymbol R}_{[z\bar{z}1]}$};
		\node [rblock,right=2 of Obst] (caseII){class 2};
		\node [rblock,below=0.5 of caseII] (caseI){class 1};
		\node [below=0.5 of caseI] (0KV){no KV};
		\node [below=0.5 of 0KV] (3KVs){3 KVs};
		\draw[arrow] (Obst.east) -- (caseII.west) node[above,pos=0.65] {{\scriptsize ~~$1$}};
		\draw[arrow] (Obst.east) -- (caseI.west) node[above,pos=0.65] {{\scriptsize ~~$2$}};
		\draw[arrow] (Obst.east) -- (3KVs.west) node[right,pos=0.55] {{\scriptsize $0$}};
		\draw[arrow] (Obst.east) -- (0KV.west) node[right,pos=0.55] {{\scriptsize $3$}};
		\end{tikzpicture}
		\caption{The sub-sub-algorithm for the class 3 type $[z\bar{z}1]$.
			See \eq \eqref{eq:compatibility_III_zz1} for notations.
		}
		\label{fig:typezz1}
	\end{center}
\end{figure}

In this sub-branch, the rank of ${\boldsymbol R}_{[z\bar{z}1]}$ governs the number of KVs
in the way shown in Figure \ref{fig:typezz1}.

\section{Application}\label{sec:application}

In this section, a couple of examples is provided to illustrate how useful our prescription is. 
First, we shall consider the Lifshitz spacetime (see e.g. \cite{Kachru:2008yh} and references therein) in Subsection \ref{subsec:Lifshitz},
whose metric has an arbitrary constant $z$.
Afterwards, we deal with a pp-wave spacetime in Subsection \ref{subsec:ppwave},
which serves as a typical example of vanishing scalar invariant spaces and
is characterised by a single function $h$.
As we will see below, a complete classification of their local isometry groups depends respectively on the values of $z$ and the profile of $h$.
This demonstrates the power of the present formulation.

\subsection{The Lifshitz spacetime}\label{subsec:Lifshitz}

\begin{figure}[h]
	\begin{center}
		\begin{tikzpicture}
		[
		every node/.style={outer sep=0.15cm, inner sep=0cm},
		arrow/.style={-{Stealth[length=0.15cm]},thick},
		rblock/.style={rectangle, rounded corners,draw, minimum height = 0.5cm,
			minimum width=1.6cm, thick, outer sep = 0},
		rgblock/.style={rectangle, rounded corners,draw, minimum height = 0.5cm,
			minimum width=2.5cm, thick, outer sep = 0},
		point/.style={radius=2pt}
		]
		\node [] (z1){$\blacktriangleright \:z=1$};
		\node [below=0.75 of z1] (z0){$z=0$};
		\node [right=1.75 of z1] (6KVs){6 KVs};
		\node [right=2.00 of z0] (4KVs){4 KVs};
		\node [below=0.75 of 4KVs] (3KVs){3 KVs};
		
		\draw[arrow] (z1) -- (6KVs.west) node[above,pos=0.7] {{\scriptsize yes}};
		\draw[arrow] (z1) -- (z0) node[right,pos=0.5] {{\scriptsize ~~no}};
		\draw[arrow] (z0.east) -- (4KVs) node[above,pos=0.725] {{\scriptsize yes}};
		\draw[arrow] (z0.east) -- (3KVs.west) node[right,pos=0.5] {{\scriptsize ~~no}};
		\end{tikzpicture}
		\caption{
			A flowchart to classify the number of KVs of the Lifshitz spacetime \eqref{eq:Lifshitz} in 3D.
		}
		\label{fig:Lifshitz}
	\end{center}
\end{figure}
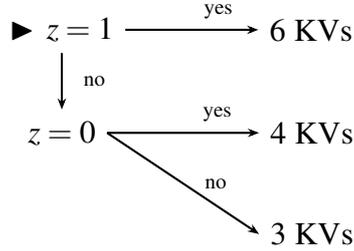

In condensed matter systems,
many phase transitions are governed by the fixed points admitting the anisotropic dynamical scaling.  In light of holography, a great deal of attention has been recently focused on the gravity dual with this dynamical scaling, which is modelled by the Lifshitz metric~\cite{Kachru:2008yh}.
The three-dimensional Lifshitz metric reads
\begin{align}
g_{\mathrm{Lifshitz}}~=~ -\frac{r^{2z}}{L^{2z}}\:\D t^2 + \frac{L^2}{r^2}\:\D r^2 + \frac{r^2}{L^2}\:\D x^2\:,
\label{eq:Lifshitz}
\end{align}
where $z$ is an arbitrary constant corresponding to the dynamical exponent and $L$ is related to the curvature scale. For this metric, the scalar curvature $R$ and
the principal traces of of powers of the traceless Ricci operator $S^a{}_b$ are all constants, 
which evaluate to 
\begin{align}
&R~=~-\frac{2(z^2+z+1)}{L^2}\:,
&&S^{(2)}~=~ \frac{2(z-1)^2(z^2+z+1)}{3L^4}\:,
&&S^{(3)}~=~ \frac{(z-1)^4 (2z^2 + 5z +2)}{9L^6}\:.
\end{align}
As the metric \eqref{eq:Lifshitz} belongs to the class 3, the criteria in Figure \ref{fig:Segre} must be checked.
After simple calculations, we find that $S^a{}_b = 0$ if $z=1$.
It then follows from the result in Figure \ref {fig:Segre} that 6 KVs exist and the AdS metric is recovered.
By solving the Killing equation \eqref{eq:Killing}, their explicit form can be read as
\begin{align}
&\partial_t\,, \qquad
\partial_x\,, \qquad
x \partial_t+ t \partial_x\,, \qquad
t \partial_t -r \partial_r +x \partial_x\,, \notag \\
&
\frac{L^4 + r^2 (t^2 + x^2)}{2r^2}\:\partial_t -r t \:\partial_r +t x \:\partial_x\,, \qquad 
t x\partial_t -r x \partial_r -\frac{L^4-r^2(t^2+x^2)}{2r^2} \partial_x \,.
\end{align}
If $z=0$, the Segre type of $S^a{}_b$ is $[1,(11)]$.
Consequently, the number of KVs can be computed by the algorithm described in Figure \ref{fig:type111_t} and it equals $4$. One sees that the metric culminates in $\mathbb R\times H^2$. 
Once again, solving \eq \eqref{eq:Killing} gives their explicit form
\begin{align}
&\partial_t\:,
&&\partial_x\:,
&&r\partial_r -x \partial_x \:,
&& rx \partial_r +\frac{L^4-r^2 x^2}{2r^2}\partial_x\,.
\end{align}
If $z \neq 0, 1$, the Segre type of $S^a{}_b$ is either $[(1,1)1]$ for $z=-1$ or $[1,11]$ for $z \neq -1$.
Then the number of KVs can be computed by the algorithm described in Figure \ref{fig:type111_s} or \ref{fig:type111_dis}.
In either case, there are 3 KVs in the form
\begin{align}
\partial_t \,, \qquad \partial_x \,, \qquad 
tz \partial_t -r \partial_r +x\partial_x
\end{align}
The last one captures the anisotropic scaling 
$t \to \lambda ^z t$, $r\to \lambda^{-1} r$, $x\to \lambda x$.
This completes a classification of the metric \eqref{eq:Lifshitz} based on their level of symmetry,
which is summarised in Figure \ref{fig:Lifshitz}.

\subsection{The pp-wave spacetime}\label{subsec:ppwave}
For Lorentzian manifolds, a natural question to ask is whether our theorem works for the metric with vanishing scalar invariant (VSI) property.
Here we show that it does.

A VSI spacetime is a Lorentzian manifold $M$ in which scalar Weyl invariants of any order vanish identically,
yet the Riemann--Christoffel tensor $R_{abc}{}^d$ is nonvanishing.
Note that scalar Weyl invariants (or polynomial curvature invariants) of order $p$
are scalars on $M$ obtained from the first $p$ covariant derivatives of the Riemann--Christoffel tensor
$\nabla_{a_1} \cdots \nabla_{a_p} R_{bcd}{}^e$ by tensor products and complete contractions \cite{Weyl:1946}.
There are nontrivial spacetimes with a VSI property
which have received some attention  in the context of general relativity, see e.g. \cite{KoutrasMcIntosh:1996,ColeyEtAl:2004,ColeyEtAl:2009}.

As a classical example of VSI spacetimes, we deal with a pp-wave spacetime 
which admits a covariantly constant null Killing vector $V^a$ satisfying $\nabla_a V^b =0$, $V_aV^a=0$. 
In dimension 3, the general form of the pp-wave metric takes the following form
\begin{align}
g_{\mathrm{pp}}~=~ h(u,x) \D u^2 + 2\D u \D v +\D x^2\:,
\label{eq:pp-wave}
\end{align}
where $h$ is a function of $u$ and $x$.
It is obvious that the covariantly constant null vector is given by $V=K_1 = \partial_v$.
Our aim here is to obtain a complete classification of KVs of $g_{\mathrm{pp}}$ based on the tournure of the function $h$.

By the definition of VSI spacetimes, the metric \eqref{eq:pp-wave} belongs to the class 3.
After simple calculations, we find that the Segre type of the traceless Ricci operator $S^a{}_b$ depends on
whether the function $h$ satisfies a PDE $h_{,xx} = \tfrac{\partial^2h}{\partial x^2} = 0$ or not.
If $h_{,xx} = 0$ holds, the Segre type is $[(1,11)]$ and $h$ takes the form
\begin{align}
h(u,x)~=~ h_0 (u) + 2x\:h_1 (u)\:,
\end{align}
where $h_0$ and $h_1$ are arbitrary functions of $u$.
Consequently it follows from the result in Figure \ref{fig:Segre} that 6 KVs exist and the spacetime is locally reduced 
to the Minkowski $\mathbb R^{1,2}$.
By solving the Killing equation \eqref{eq:Killing}, the explicit expressions of the set of KVs can be written as
\begin{subequations}
\begin{align}
K_1~=~& \partial_v\,,\\
K_2~=~& H_1 \partial_v - \partial_x\,, \\ 
K_3~=~& \left(x + u H_1 -\mathcal{H}_1\right) \partial_v - u \partial_x\:, \\
K_4~=~&\partial_u - \frac{1}{2} \left( h + H_1 ^2\right)\partial_v + H_1 \partial_x\:, \\ 
K_5~=~& u\partial_u - \frac{1}{2} \left(
2v +u h + H + u H_1 ^2 - \int \D u\:H_1^2\right) \partial_v
+ u H_1 \partial_x\:, \\
K_6~=~&\left( x-\mathcal{H}_1 \right) \partial_u
+\frac{1}{2} \left(
(x+\mathcal{H}_1)H_1^2
-(x- \mathcal{H}_1) h
+2v H_1
+H_0 H_1 - H_1 \int \D u\:H_1^2
\right) \partial_v \notag \\
&-\frac{1}{2}\left(H_0 + 2H_1 \mathcal{H}_1 -\int \D u\:H_1^2 + 2v\right) \partial_x\:,
\end{align}
where
\begin{align}
H(u,x) &~\equiv~ \int \D u\:h(u,x)\:,
&H_0(u) &~\equiv~ \int \D u\:h_0(u)\:,\notag \\
H_1(u) &~\equiv~ \int \D u\:h_1(u)\:,
&\mathcal{H}_1(u) &~\equiv~ \int \D u\:H_1(u)\:.
\end{align}
\end{subequations}
The nonvanishing commutation relations for these KVs are 
\begin{subequations}
\begin{align}
\label{}
[K_1,K_5]~=~&-K_1 \,, &
[K_1,K_6]~=~&K_2 \,, &
[K_2,K_3]~=~&-K_1 \,,  \\
[K_2, K_6]~=~&-K_4 \,, &
[K_3,K_4]~=~&-K_2 \,, &
[K_3,K_5]~=~&-K_3 \,, \\ 
[K_3, K_6]~=~& -K_5\,, &
[K_4,K_5]~=~&K_4 \,, &
[K_5,K_6]~=~&-K_6 \,. 
\end{align}
\end{subequations}
These KVs constitute the 3-dimensional Poincar\'e algebra.

If $h_{,xx}=0$ fails to be fulfilled, 
the Segre type of $S^a{}_b$ is $[(21)]$ with the invariants $\{ \sigma, \kappa_e \} = \{ \tfrac{h_{,xxx}}{2h_{,xx}}, 0\}$
(see Figure \ref{fig:type21_deg} and \eq \eqref{eq:compatibility_III_21_deg} for notations).
The number of KVs is controlled by either $\mathrm{rank} {\boldsymbol R}_{[(21)]}^{~\# 2}$ if $h_{,xxx} \neq 0$ or $\mathrm{rank} {\boldsymbol R}_{[(21)]}$ if $h_{,xxx} =0$.
In either case, any KV can be identified at the outset
$K^a=\omega_u u^a + \omega_v v^a + \omega_e e^a$,
where $\{u^a, v^a, e^a\}$ is the double-null basis defined as
\begin{align}
&u^a ~\equiv~ (\partial_v)^a\:,
&&v^a~\equiv~(\partial_u)^a - \frac{h}{2} (\partial_v)^a\:,
&&e^a~\equiv~(\partial_x)^a\:,
\label{eq:ppbasis}
\end{align}
whose nonvanishing rotation coefficient consists only of $\eta_v= -h_{,x}/2$.

\subsubsection{The case $\sigma \propto h_{,xxx} = 0$}

The solution to the equation $h_{,xxx}=0$ leads to the general form of $h$ as
\begin{align}
h(u,x)~=~ h_0 (u) + 2x\:h_1 (u) + x^2\:h_2(u)\:,
\label{eq:h_submaximal}
\end{align}
where $\{h_0, h_1, h_2\}$ are arbitrary functions of $u$.
By combining \eqs \eqref{eq:h_submaximal} and \eqref{eq:ppbasis},
the obstruction matrix ${\boldsymbol R}_{[(21)]}$ acting on ${\boldsymbol \omega} = [\omega_u, \omega_v, \omega_e, \varpi_u]^T$ reads
\begin{align}
{\boldsymbol R}_{[(21)]}~=~
\begin{bmatrix}
0 & \pounds_v \tau_v & \pounds_e \tau_v & 0\\
\e^{\varphi} \pounds_u \Sigma & \e^{\varphi} \pounds_v \Sigma & \e^{\varphi} \pounds_e \Sigma & 0
\end{bmatrix}
~\propto~
\begin{bmatrix}
0 & 0 & 0 & 0\\
0 & (h_{,xxu}/(h_{,xx})^{3/2})_{,u} & 0 & 0
\end{bmatrix}\:.
\label{eq:ppwave_matrix_eq_xxx}
\end{align}
Thereby $\mathrm{rank}{\boldsymbol R}_{[(21)]} = 0$ if $h_2$ solves the ODE
$(h_{,xxu}/(h_{,xx})^{3/2})_{,u} = (h_{2,u}/h^{3/2}_{2})_{,u} = 0$ whose
solution is given by $h_2(u) = 1/(c_1 u + c_2)^2$, where $\{c_1, c_2\}$ are constants.
Notice that a coordinate shift $u \rightarrow u + c_0$ $(c_0=\text{const.})$ allows us to classify the solution as either
$h_2(u) = \text{const.}$ for $c_1 = 0$ or $h_2(u) \propto u^{-2}$ for $c_1 \neq 0$.

For the former case $h_2 (u) = c = \text{const.} \ne 0$, there are 4 KVs in the form
\begin{subequations}
\begin{align}
K_1~=~& \partial_v\,,\\
K_2~=~& \left(\sqrt{c} x \e^{-\sqrt{c}u}- \langle h_1\rangle^-\right)\partial_v
+\e^{-\sqrt{c}u} \partial_x\,, \\
K_3~=~& \left(\sqrt{c} x \e^{\sqrt{c}u} +\langle h_1\rangle^+\right)\partial_v
-\e^{\sqrt{c}u} \partial_x\:, \\
K_4~=~&
2\partial_u
-\left(
h_0+
2xh_1+
\sqrt{c} x
\left(\e^{\sqrt{c} u} \langle h_1\rangle^-
-\e^{-\sqrt{c} u}\langle h_1\rangle^{+}\right)
+\langle h_1\rangle^+ \langle h_1\rangle^-
\right)\partial_v \notag \\
&
+\left( \e^{\sqrt{c} u} \langle h_1\rangle^-
+\e^{-\sqrt{c} u} \langle h_1\rangle^+\
\right)\partial_x
\:,
\end{align}
where we have assumed $c>0$ and defined
\begin{align}
\langle h_1\rangle^\pm ~\equiv~ \int \D u\:\e^{\pm\sqrt{c} u}h_1\:.
\end{align}
The nonzero commutators for these KVs are 
\begin{align}
\label{}
[K_2,K_3]~=~&2\sqrt c K_1 \,, \qquad 
[K_2,K_4]~=~2\sqrt c K_2 \,, \qquad 
[K_3,K_4]~=~-2\sqrt c  K_3 \,.
\end{align}
These correspond to the $\mathfrak{sl}(2,\mathbb R)$ algebra.
One can deduce the explicit expressions of KVs also for the $c<0$ case.
\end{subequations}

For the latter case $h_2 (u) = c\:u^{-2} (c = \text{const.} \neq 0)$, 4 KVs exist in the form
\begin{subequations}
\begin{align}
K_1~=~& \partial_v\:,\\
K_2~=~& \left(2c x u^{-\frac{1+k}{2}}
-\left(1+k \right) \langle h_1\rangle_{1-k}
\right)\partial_v
+\left(1+k\right)u^{\frac{1-k}{2}}\partial_x\:,\\
K_3~=~&
\left(2c x u^{-\frac{1-k}{2}}
-\left(1-k \right) \langle h_1\rangle_{1+k}
\right)\partial_v
+\left(1 - k \right)u^{\frac{1+k}{2}}\partial_x\:, \\
K_4 ~=~&2k u \partial_u -\Biggl(2kv +x \Bigl(
2ku h_1+\frac 12 (1+k)^2 u^{\frac{k-1}2}\langle h_1\rangle_{1-k} 
-\frac 12 (1-k)^2 u^{-\frac{k+1}2}\langle h_1\rangle_{1+k} 
\Bigr) 
\notag \\
&+k \Bigl(u h_0 +H_0\Bigr)+(1+k) \int \D u\:u^{\frac{1+k}2} h_1 \langle h_1\rangle _{1-k} 
-(1-k) \int \D u\:u^{\frac{1-k}2} h_1 \langle h_1\rangle _{1+k} 
\Biggr)\partial_v \notag \\
& +\Biggl(
(1+k )u^{\frac{1+k}2} \langle h_1 \rangle _{1-k} -(1-k )u^{\frac{1-k}2} \langle h_1 \rangle _{1+k} 
\Biggr) \partial_x \,,
\end{align}
where we have used abbreviations
\begin{align}
k &~\equiv~ \sqrt{1+4c}\:,
&H_0 &~\equiv~ \int \D u\: h_0\:,
& \langle h_1\rangle_p &~\equiv~ \int \D u \: u^{\frac{p}{2}} h_1 \,.
\end{align}
\end{subequations}
In the above expressions, we have tentatively assumed $c>-1/4$. The commutation relations are given by
\begin{subequations}
\begin{align}
\label{}
[ K_1, K_4]~=~&-2k K_1 \,, &
[K_2,K_4]~=~&-k(1-k)K_2 \,,  \\
[K_2, K_3]~=~& 4ck K_1 \,, & 
[K_3,K_4]~=~& -k(1+k)K_3 \,. 
\end{align}
\end{subequations}
The KVs for the $c=-1/4$ and $c<-1/4$  cases can be obtained in a similar fashion, but
we shall not attempt to do this here.

If $\sigma \propto h_{,xxx}=0$ but $\Sigma_{,u} \propto (h_{,xxu}/(h_{,xx})^{3/2})_{,u} \neq 0$,
it follows from \eq \eqref{eq:ppwave_matrix_eq_xxx} that $\mathrm{rank} {\boldsymbol R}_{[(21)]} = 1$ and $\omega_v$ has to be zero.
As any KV takes the form $K^a = \omega_u u^a + \omega_e e^a$, the results in Section \ref{subsec:caseII_null} are reusable.
Since all the spin coefficients are vanishing except for $\eta_v$,
$\mathrm{rank} {\boldsymbol R}_{\mathrm{cls. 2}}$ governs the number of KVs (see Figure \ref{fig:caseII_null}).
For the function $h$ in the form of \eq \eqref{eq:h_submaximal}, one can verify that 
$\mathrm{rank} {\boldsymbol R}_{\mathrm{cls. 2}}$ is always zero.
By solving \eq \eqref{eq:Killing} directly, it can be ascertained that 3 KVs exist in the form
\begin{align}
&K_1~=~\partial_v\:,
&& K_\pm ~=~ -\left( \int \D u \: h_1 \phi_\pm 
+ x \int \D u\:h_2 \phi_\pm \right) \partial_v
+\phi_\pm  \partial_x\:,
\end{align}
where $\phi_\pm (u)$ are the two linearly independent solutions to the following ODE
\begin{align}
\phi_{,uu} ~=~ h_2\:\phi \:.
\label{eq:pp-wave:3KV_diffeq}
\end{align}
By the conservation of Wronskian $\phi_{+,u}\phi_--\phi_+\phi_{-,u}={\rm const.} \equiv W$, 
the only nonvanishing commutator is 
\begin{align}
\label{}
[ K_+, K_-]~=~W K_ 1 \,. 
\end{align}

\subsubsection{The case $\sigma \propto h_{,xxx} \neq 0$}

For the case in question, the obstruction matrix ${\boldsymbol R}_{[(21)]}^{~ \#2}$ controls the number of KVs primarily (see Figure \ref{fig:type21_deg}).
As a strategy for the classification, we focus on the first row of ${\boldsymbol R}_{[(21)]}^{~ \#2}$
\begin{align}
\begin{bmatrix}
\pounds_u (\sigma- \tfrac{5}{2} \tau_v) &\pounds_v (\sigma- \tfrac{5}{2} \tau_v) &\pounds_e (\sigma- \tfrac{5}{2} \tau_v)
\end{bmatrix}
~\propto~
\begin{bmatrix}
0 &(h_{,xxx}/h_{,xx})_{,u} &(h_{,xxx}/h_{,xx})_{,x}
\end{bmatrix}\:.
\label{eq:ppwave_matrixeq1}
\end{align}
It can be shown that if $(h_{,xxx}/h_{,xx})_{,u}=(h_{,xxx}/h_{,xx})_{,x}=0$
all entries of ${\boldsymbol R}_{[(21)]}^{~ \#2}$ are zero except for the fifth row
\begin{align}
\begin{bmatrix}
\Phi_u &\Phi_v&\Phi_e
\end{bmatrix}
~\propto~
\begin{bmatrix}
0
& \varsigma h_{,xxu} - \varsigma_{,u} h_{,xx}
& \varsigma h_{,xxx}
\end{bmatrix}\:,
\label{eq:ppwave_matrixeq2}
\end{align}
where $\varsigma \equiv (h_{,xxu}/h_{,xx})_{,u} - (h_{,xx}/2) (h_{,x}/h_{,xx})_{,x}$.
The obstruction elements for $\mathrm{rank}{\boldsymbol R}_{[(21)]}^{~ \#2} = 0$ are therefore given by
\begin{subequations}
	\begin{align}
	&\left( \frac{h_{,xxx}}{h_{,xx}} \right)_{,x}\:,
	&&\left( \frac{h_{,xxx}}{h_{,xx}} \right)_{,u}\:,
	\end{align}
	and collaterally
	\begin{align}
	&\left( \frac{h_{,xxu}}{h_{,xx}} \right)_{,u} - \frac{h_{,xx}}{2} \left(\frac{h_{,x}}{h_{,xx}}\right)_{,x}\:.
	\end{align}
	\label{eq:pp-wave_3criteria}
\end{subequations}

If these criteria \eqref{eq:pp-wave_3criteria} are all vanishing,  $\mathrm{rank}{\boldsymbol R}_{[(21)]}^{~ \#2} = 0$ and we can parameterise $h$ by a nonzero function $h_1$ as
\begin{align}
h(u,x)~=~h_0(u) + \e^{c_1 (x + h_1 (u))} - 2x h_{1,uu} (u)\:,
\end{align}
where $h_0$ is an arbitrary functions of $u$ and $c_1$ is a nonzero constant, thereby allowing us to obtain 3 KVs
\begin{align}
K_1~=~&\partial_v\:,\\
K_2~=~&\partial_u - \frac{1}{2}\bigl(
h_0 + h_{1,u}^2
-2x h_{1,uu} \bigr) \partial_v
-h_{1,u} \partial_x\:,\\
K_3~=~&u\partial_u
-\biggl(v + \frac{u h_0 + \int \D u \:h_0}{2}
-ux h_{1,uu}
+\frac{u h_{1,u}^2 - \int \D u\: h_{1,u}^2}{2} \notag \\
&+\biggl(\frac{2}{c_1} -x\biggr) h_{1,u}
\biggr)\partial_v
-\biggl( \frac{2}{c_1} + u h_{1,u}\biggr) \partial_x
\:,
\end{align}
together with their commutators
\begin{align}
&[K_1, K_2]~=~0\:,
&&[K_2, K_3] ~=~ K_2\:,
&&[K_3, K_1] ~=~K_1\:.
\end{align}

\begin{figure}[h]
	\begin{center}
		\begin{tikzpicture}
		[
		every node/.style={outer sep=0.15cm, inner sep=0cm},
		arrow/.style={-{Stealth[length=0.15cm]},thick},
		rblock/.style={rectangle, rounded corners,draw, minimum height = 0.5cm,
			minimum width=1.6cm, thick, outer sep = 0},
		rgblock/.style={rectangle, rounded corners,draw, minimum height = 0.5cm,
			minimum width=2.5cm, thick, outer sep = 0},
		point/.style={radius=2pt}
		]
		
		\node [] (xx){$\blacktriangleright \:h_{,xx}=0$};
		\node [below=0.75 of xx] (xxx){$h_{,xxx}=0$};
		\node [below=0.75 of xxx] (xxxx){$\bigl(\tfrac{h_{,xxx}}{h_{,xx}}\bigr)_{,x}=0$};
		\node [right=1 of xxxx] (xxxu){$\bigl(\tfrac{h_{,xxx}}{h_{,xx}}\bigr)_{,u}=0$};
		\node [right=1 of xxxu] (varsigma1){\small{$\Bigl( \tfrac{h_{,xxu}}{h_{,xx}} \Bigr)_{,u} = \frac{h_{,xx}}{2} \Bigl(\tfrac{h_{,x}}{h_{,xx}}\Bigr)_{,x}$}};
		\node [below=0.75 of varsigma1] (varsigma2){\small{$\Bigl(\tfrac{h_{2,u}}{h_2^{3/2}}\Bigr)_{,u}=0$}};
		
		\node [right=1 of xxx] (h2){\small{$\Bigl(\tfrac{h_{2,u}}{h_2^{3/2}}\Bigr)_{,u}=0$}};
		\node [below=2.5 of xxxx] (xxxu2){$\bigl(\tfrac{h_{,xxx}}{h_{,xx}}\bigr)_{,u}=0$};
		\node [right=1.5 of xxxu2] (hi){\small{$\Bigl(\tfrac{h_{2,u}}{h_2^{3/2}}\Bigr)_{,u}=\tfrac{h_{1,u}}{h_1}-\tfrac{h_{2,u}}{h_2} =0$}};
		\node [below=0.75 of xxxu2] (special){\small{$\Bigl(\tfrac{h_{2,u}}{h_2^{3/2}}\Bigr)_{,u}=\tfrac{h_{1,u}}{h_1}-\tfrac{h_{2,u}}{h_2} =0$}};
		
		\node [right=1.5 of xx] (6KVs){6 KVs};
		\node [right=1.5 of h2] (4KVs){4 KVs};
		\node [below=0.25 of 4KVs] (3KVs1){3 KVs};
		\node [right=1.5 of varsigma1] (3KVs2){3 KVs};
		\node [below=0.75 of xxxu] (1KV1){1 KV};
		\node [right=1.5 of varsigma2] (2KVs1){2 KVs};
		\node [below=0.25 of 2KVs1] (1KV2){1 KV};
		\node [right=1.5 of hi] (2KVs2){2 KVs};
		\node [below=0.25 of 2KVs2] (1KV3){1 KV};
		\node [right=1.5 of special] (2KVs3){2 KVs};
		\node [below=0.25 of 2KVs3] (1KV4){1 KV};

		\draw[arrow] (xx.east) -- (6KVs.west) node[above,pos=0.5] {{\scriptsize yes}};
		\draw[arrow] (xx) -- (xxx) node[right,pos=0.5] {{\scriptsize ~~no}};
		\draw[arrow] (xxx) -- (h2) node[above,pos=0.5] {{\scriptsize yes}};
		\draw[arrow] (h2.east) -- (4KVs.west) node[above,pos=0.5] {~~{\scriptsize yes}};
		\draw[arrow] (h2.east) -- (3KVs1.west) node[above,pos=0.5] {~~{\scriptsize no}};
		\draw[arrow] (xxx) -- (xxxx) node[right,pos=0.5] {{\scriptsize ~~no}};
		\draw[arrow] (xxxx.east) -- (xxxu.west) node[above,pos=0.5] {{\scriptsize yes}};
		\draw[arrow] (xxxx) -- (xxxu2) node[right,pos=0.5] {{\scriptsize ~~no}};
		\draw[arrow] (xxxu.east) -- (varsigma1.west) node[above,pos=0.5] {{\scriptsize yes}};
		\draw[arrow] (xxxu) -- (1KV1) node[right,pos=0.5] {{\scriptsize ~~no [{type B}]}};
		\draw[arrow] (varsigma1) -- (varsigma2) node[right,pos=0.5] {{\scriptsize ~~no [{type A}]}};
		\draw[arrow] (varsigma1.east) -- (3KVs2.west) node[above,pos=0.5] {{\scriptsize yes}};
		\draw[arrow] (varsigma2.east) -- (2KVs1.west) node[above,pos=0.5] {~~{\scriptsize yes}};
		\draw[arrow] (varsigma2.east) -- (1KV2.west) node[above,pos=0.5] {~~{\scriptsize no}};
		\draw[arrow] (xxxu2.east) -- (hi.west) node[above,pos=0.5] {{\scriptsize yes [{type C}]}};
		\draw[arrow] (hi.east) -- (2KVs2.west) node[above,pos=0.5] {~~{\scriptsize yes}};
		\draw[arrow] (hi.east) -- (1KV3.west) node[above,pos=0.5] {~~{\scriptsize no}};
		\draw[arrow] (xxxu2) -- (special) node[right,pos=0.5] {{\scriptsize ~~no [{type D}]}};
		\draw[arrow] (special.east) -- (2KVs3.west) node[above,pos=0.5] {~~{\scriptsize yes}};
		\draw[arrow] (special.east) -- (1KV4.west) node[above,pos=0.5] {~~{\scriptsize no}};
		
		\end{tikzpicture}
		\caption{
			A flowchart to classify the number of KVs of the pp-wave spacetime \eqref{eq:pp-wave} in 3D.
		}
		\label{fig:ppwave}
	\end{center}
\end{figure}
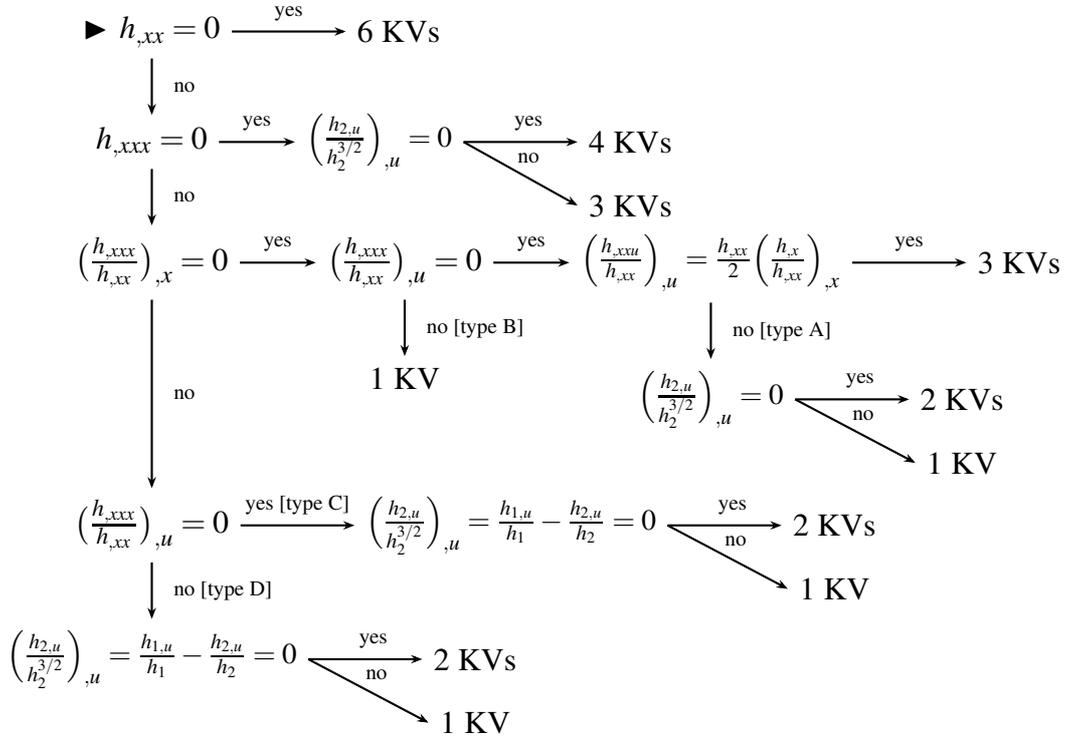

Let us next consider the case in which ${\rm rank}{\boldsymbol R}_{[(21)]}^{~ \#2} \ne 0$.
Since ${\rm rank}{\boldsymbol R}_{[(21)]}^{~ \#2}=2$ implies that there exists a single KV $\partial_v$,
we shall concentrate on the case ${\rm rank}{\boldsymbol R}_{[(21)]}^{~ \#2}=1$.
As a result, the conditions for which the metric \eqref{eq:pp-wave} admits 2 KVs are identified as shown in Figure \ref{fig:ppwave}.
This will be achieved by separating our analysis into the four types (A, B, C, D) based on the nonzeroness of \eq \eqref{eq:pp-wave_3criteria}.
We shall fix the explicit forms of $h(u,x)$, the 2nd KV and its commutator in the rest of this subsection.

\subsubsection*{\underline{Type A:}}

	In this type, the function $h$ solves the PDEs $(h_{,xxx}/h_{,xx})_{,x} = 0$ and $(h_{,xxx}/h_{,xx})_{,u} = 0$ simultaneously,
	whereas $\varsigma = (h_{,xxu}/h_{,xx})_{,u} - (h_{,xx}/2) (h_{,x}/h_{,xx})_{,x} \neq 0$.
	Thus we have
	\begin{align}
	h(u,x)~=~h_0(u) + \e^{c_1 (x + h_1(u))} - 2x\left( h_{1,uu} (u) - h_2(u)\right)\:,
	\label{eq:ppwave:h_typeA}
	\end{align}
	where $\{h_0, h_1, h_2\}$ are functions of $u$ and $c_1$ is a constant. $h_2$ and $c_1$ are nonvanishing.
	From \eq \eqref{eq:ppwave_matrixeq2}, ${\boldsymbol R}_{[(21)]}^{~ \#2} {\boldsymbol \omega} = 0$ imposes
	\begin{align}
	&\omega_e ~=~ \gamma\:\omega_v\:,
	&&
	\gamma~\equiv~ - \frac{\varsigma h_{,xxu} - \varsigma_{,u} h_{,xx}}{\varsigma h_{,xxx}}
	~=~ \frac{h_{2,u}- c_1 h_2 h_{1,u}}{c_1 h_2}\:.
	\label{eq:pp-wave_typeA1}
	\end{align}
	In a nod to \eq \eqref{eq:pp-wave_typeA1} and the basis \eqref{eq:ppbasis},
	we take an orthonormal basis $\{e_1^a, e_2^a, e_3^a\}$ as
	\begin{align}
	e_1^a &~\equiv~- \gamma\:u^a + e^a\:,
	& e_2^a &~\equiv~ \gamma^{-1}\:v^a + e^a\:,
	& e_3^a &~\equiv~ \gamma \: u^a -\gamma^{-1}\:v^a - e^a\:,
	&&\text{for}
	&\gamma &~\neq~ 0\:, \\
	e_1^a &~\equiv~ e^a\:,
	& e_2^a &~\equiv~ \frac{1}{\sqrt{2}} \left( u^a + v^a \right)\:,
	& e_3^a &~\equiv~ \frac{1}{\sqrt{2}} \left( u^a - v^a \right)\:,
	&&\text{for}
	&\gamma &~=~ 0\:,
	\label{eq:ppwave_newbasis_doublenull}
	\end{align}
	whence any KV is reduced to the form $K^a = \omega_2 e_2^a + \omega_3 e_3^a$.
	In either case, the basis satisfies $g^{ab}=e_1^a e_1^b + e_2^a e_2^b - e_3^a e_3^b$ and
	$\tau_2 = \tau_3$ with $\tau_2 \neq 0$.
	Therefore from Figure \ref{fig:caseII_nonnull}
	$\mathrm{rank} {\boldsymbol R}_{\mathrm{cls. 2}}^{\iota~ \#2}$ determines the existence of the 2nd KV.
	
	It is simple to see that the third, fourth and fifth rows of ${\boldsymbol R}_{\mathrm{cls. 2}}^{\iota~ \#2}$ are left nonvanishing,
	yielding the condition to have $\mathrm{rank} {\boldsymbol R}_{\mathrm{cls. 2}}^{\iota~ \#2} = 0$ as $(h_{2,u}/h_2^{3/2})_{,u} = 0$.
	The solution to the ODE categorises into $h_2 = \text{const.}$ or $h_2 \propto u^{-2}$. 
	For $h_2=c_2=\text{const.}$, the 2nd KV arises in the form
\begin{subequations}
	\begin{align}
	K_2~=~\partial_u - \frac{1}{2}\bigl(
	h_0+h_{1,u}^2-2x h_{1,uu}-2c_2 h_1\bigr) \partial_v
	-h_{1,u}\partial_x\:,
	\end{align}
	with the commutator $[K_1, K_2]=0$.
	For $h_2=c_2 u^{-2} (c_2 =\text{const.} \neq 0)$, the 2nd KV reads
	\begin{align}
	K_2~=~&u\partial_u
	-\biggl(v + \frac{u h_0 + \int \D u \:h_0}{2}
	-ux h_{1,uu}
	+\frac{u h_{1,u}^2- \int \D u\: h_{1,u}^2}{2} \notag \\
	&+\biggl(\frac{2}{c_1} -x\biggr) h_{1,u}
	- c_2 \int \frac{\D u}{u^2}\: \biggl(\frac{2}{c_1} + u h_{1,u}\biggr)  \biggr)\partial_v
	-\biggl(\frac{2}{c_1} + u h_{1,u}\biggr) \partial_x\:,
	\end{align}
	with the commutator $[K_1, K_2] = - K_1$.
\end{subequations}

\subsubsection*{\underline{Type B:}}

Here the function $h$ is characterised by the two conditions $(h_{,xxx}/h_{,xx})_{,x} = 0$ and $(h_{,xxx}/h_{,xx})_{,u} \neq 0$,
leading to the form
\begin{align}
h(u,x)~=~h_0(u) + x h_1 (u) + \e^{h_2(u) + x h_3(u)}\:,
\label{eq:ppwave:h_typeB}
\end{align}
where $\{h_0, h_1, h_2, h_3\}$ are arbitrary functions of $u$, but
$h_{3,u}(u) ~\neq~0$ has to be true for the latter condition.
It is required by ${\boldsymbol R}_{[(21)]}^{~ \#2} {\boldsymbol \omega} = 0$ that
\begin{align}
& h_{3,u} \: \omega_v~=~0\:,
&& h_{3,u} \: \omega_e~=~0\:,
\end{align}
concluding that $\omega_v=\omega_e=0$, so there is no possibility of finding the 2nd KV.

\subsubsection*{\underline{Type C:}}
Since the function $h$ is the general solution to $(h_{,xxx}/h_{,xx})_{,u} = 0$, we have
\begin{align}
h(u,x) ~=~ h_0(u) + x h_1 (u) + h_2(u) h_3(x)\:,
\end{align}
where $\{h_0, h_1\}$ are arbitrary functions of $u$, and
$\{h_2, h_3\}$ are respectively nonzero functions of $u$ and of $x$.
As $(h_{,xxx}/h_{,xx})_{,x} = 0$ is satisfied nowhere, it is stipulated that $(h_{3,xxx}/h_{3,xx})_{,x} \neq 0$.
From this and the first row of ${\boldsymbol R}_{[(21)]}^{~ \#2}$, it is inevitable that $\omega_e = 0$.
The leftover components of ${\boldsymbol R}_{[(21)]}^{~ \#2}$ put
the requirements to have $\mathrm{rank} {\boldsymbol R}_{[(21)]}^{~ \#2} = 1$ as
\begin{align}
&\left( h_{2,u}/h_2  \right)_{,u}~=~0\:,
&&h_{1,u}/h_1 ~=~ h_{2,u}/ h_2\:.
\label{eq:ppwave_typeC_conds}
\end{align}
Assuming \eqs \eqref{eq:ppwave_typeC_conds} and using new basis $\{e_1^a, e_2^a, e_3^a\}$ defined by \eq \eqref{eq:ppwave_newbasis_doublenull},
it is easy to see that $\mathrm{rank} {\boldsymbol R}_{\mathrm{cls. 2}}^{\iota~ \#2} = 0$.
For $h_1 = c_1=\text{const.}$ and $h_2 = c_2=\text{const.}$, the 2nd KVs is given by
\begin{subequations}
\begin{align}
K_2~=~\partial_u - \frac{h_0}{2} \partial_v\:,
\end{align}
with the commutator $[K_1, K_2] = 0$.
For $h_1 = c_1 u^{-2} (c_1 =\text{const.})$ and $h_2 = c_2 u^{-2} (c_2 =\text{const.})$, the 2nd KVs is expressed as
\begin{align}
K_2~=~u \partial_u - \frac{1}{2}\left( u h_0 + \int \D u\:h_0 + 2v \right) \partial_v\:,
\end{align}
with the commutator $[K_1, K_2] = -K_1$.
\end{subequations}

\subsubsection*{{\underline{Type D:}}}
	It is immediately seen from the first row of ${\boldsymbol R}_{[(21)]}^{~ \#2}$ that
	\begin{align}
	&\omega_e ~=~ - \frac{\sigma_{,u}}{\sigma_{,x}} \:\omega_v\:,
	&\sigma (u,x) ~\equiv~  \frac{h_{,xxx}}{2h_{,xx}}\:,
	\label{eq:ppwave_rho}
	\end{align}
	where the valuable $\sigma(u,x)$ inherits from the definition \eqref{eq:compatibility_III_21_deg}.
	The remaining entries do not have illuminating expressions to be described here.
		Leaving aside the full implications of ${\boldsymbol R}_{[(21)]}^{~ \#2}$,
	we proceed to the analysis of class 2 to simplify the reasoning.
	By choosing an orthonormal basis $\{e_1^a, e_2^a, e_3^a\}$ as
	\begin{align}
	&e_1^a~\equiv~ \frac{\sigma_{,u}}{\sigma_{,x}}\:u^a + e^a\:,
	&&e_2^a~\equiv~ - \frac{\sigma_{,x}}{\sigma_{,u}}\:v^a + e^a\:,
	&&e_3^a~\equiv~ - \frac{\sigma_{,u}}{\sigma_{,x}}\:u^a + \frac{\sigma_{,x}}{\sigma_{,u}}\:v^a - e^a\:,
	\end{align}
	we have
	\begin{align}
	\kappa_1~=~-\eta_1~=~ \frac{\sigma_{,xx}}{\sigma_{,x}} - \frac{\sigma_{,ux}}{\sigma_{,u}}\:.
	\label{eq:ppwave_D_kappa}
	\end{align}
	From the result of subsection \ref{subsec:caseII_nonnull_branch1},
	\eq \eqref{eq:ppwave_D_kappa} has to be zero so that the 2nd KV can exist.
	Solving a PDE $\sigma_{,xx}/\sigma_{,x}-\sigma_{,ux}/\sigma_{,u} = 0$,  we obtain
	\begin{align}
	\sigma (u,x) ~=~\sigma (x + h_4(u))\:,
	\end{align}
	where $h_4$ is an arbitrary function of $u$.
	Subsequently, the definition of $\sigma$ \eqref{eq:ppwave_rho} gives
	\begin{align}
	h(u,x) ~=~ h_0(u) + x \left( h_1(u) - 2 h_{4,uu} (u)\right) + h_2(u) h_3(x+h_4(u))\:,
	\end{align}
	where $\{h_0, h_1, h_2, h_3\}$ are arbitrary functions of one variable such that
	$(h_{3,xxx}/h_{3,xx})_{,x} \neq 0$ and $h_{4,u} \neq 0$.
	Note that the existence of the 2nd KV is still not clear, so we go on to examining $\mathrm{rank} {\boldsymbol R}_{\mathrm{cls. 2}}^{\iota~ \#2}$.
	It follows from the third and fourth rows of ${\boldsymbol R}_{\mathrm{cls. 2}}^{\iota~ \#2}$ that $\mathrm{rank} {\boldsymbol R}_{\mathrm{cls. 2}}^{\iota~ \#2} = 0$
	if $(h_{2,u}/h_2^{3/2})_{,u} = 0$ and $h_{1,u}/h_1=h_{2,u}/h_2$.
	
	For $h_1 = c_1=\text{const.}$ and $h_2 = c_2=\text{const.}$, the 2nd KV is given by
\begin{align}
K_2~=~\partial_u -\frac{1}{2}\left( h_0 + h_{4,u}^2 - 2x h_{4,uu} -c_1 h_4 \right) \partial_v
-h_{4,u} \partial_x\:,
\end{align}
with the commutator $[K_1, K_2]=0$.
For $h_1 = c_1 u^{-2} (c_1 =\text{const.})$ and $h_2 = c_2 u^{-2} (c_2 =\text{const.})$, the 2nd KVs is expressed as
\begin{align}
K_2~=~&u \partial_u - \frac{1}{2}\biggl(
u h_0 + \int \D u\:h_0 + 2v
-\int \D u \:h_{4,u}^2 -2x (u h_{4,uu} + h_{4,u}) + u h_{4,u}^2  \notag\\
&
-c_1 \biggl( \frac{h_4}{u} + \int \D u \: \frac{h_4}{u^2}\biggr)
\biggr) \partial_v
-u h_{4,u} \partial_x\:,
\end{align}
with the commutator $[K_1, K_2] = -K_1$.

\section{Conclusion}\label{sec:conclusion}

The basic questions we addressed in this paper are
whether there exists a set of invariants associated with the existence of KVs, and if so, how to construct it
for a given Lorentzian manifold.
Our contribution is to give affirmative answers to such questions in dimension $3$,
extending the result for a Riemannian manifold \cite{Kruglikov:2018qcn}.
According to our theorem in Section \ref{sec:introduction},
the number of linearly independent KVs can be counted using
the algorithm described in Figure \ref{fig:main}, even if a given spacetime has a VSI property.
As we have seen in Section \ref{sec:application},
the theorem can classify a given spacetime into a hierarchy based on their level of symmetry.

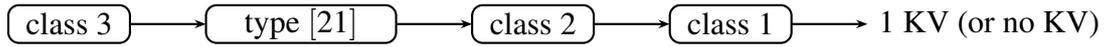
\begin{figure}[h]
	\begin{center}
		\begin{tikzpicture}
		[
		every node/.style={outer sep=0.15cm, inner sep=0cm},
		arrow/.style={-{Stealth[length=0.15cm]},thick},
		rblock/.style={rectangle, rounded corners,draw, minimum height = 0.5cm,
			minimum width=1.6cm, thick, outer sep = 0},
		rgblock/.style={rectangle, rounded corners,draw, minimum height = 0.5cm,
			minimum width=2.5cm, thick, outer sep = 0},
		point/.style={radius=2pt}
		]
		\node [rblock] (first){class 3};
		\node [rgblock, right=1 of first] (second){type $[21]$};
		\node [rblock, right=1 of second] (third){class 2};
		\node [rblock, right=1 of third] (fourth){class 1};
		\node [right=1 of fourth] (fifth){1 KV (or no KV)};
		\draw[arrow] (first) -- (second);
		\draw[arrow] (second) -- (third);
		\draw[arrow] (third) -- (fourth);
		\draw[arrow] (fourth) -- (fifth);
		\end{tikzpicture}
		\caption{A flow which requires the most differential invariants.
			}
		\label{fig:worstcase}
	\end{center}
\end{figure}

It would be instructive to mention the algorithmic efficiency of the Cartan--Karlhede and our formulations.
Given a Lorentzian manifold of dimension $3$,
the Cartan--Karlhede algorithm uses Cartan scalars and requires at most six differentiations of $R_{abc}{}^d$. 
In that case, one must assess  the functionally independent $336$ Cartan scalars, 
whence it reveals the number of KVs in principle.
On the other hand, our algorithm uses the Ricci rotation coefficients, their derivatives and the ratio thereof.
In the worst case, it may be implemented in line with Figure \ref{fig:worstcase}
and $66$ differential invariants are required in total. In conjugation with this, our prescription requires up to the 
3rd derivatives of the curvature in ${\boldsymbol R}^{\#2}_{[(21)]}$.
Thus, our algorithm is more economic than that of the Cartan--Karlhede to count the number of KVs.
The only price to pay for our method to work out is to solve  the eigenvalue problem of the traceless Ricci operator $S^a{}_b$ in the class 3. Fortunately, this is not a demanding task in dimension 3.

As an application, our theorem may enable us to derive the canonical
form of metrics with a high degree of symmetry.
In fact, the calculations we have carried out in \ref{app:4KVs} give the
canonical form of metrics admitting 4 KVs,
which is a reproduction and improvement of the classical result due to Kruchkovich \cite{Kruchkovich}.
As a side remark, we note that our algorithm does not ensure that
local metrics endowed with a certain degree of symmetry exist in all branches of class 1--3.
In other words, it seems that some values of $\mathrm{rank} {\boldsymbol R}_{\mathrm{cls. }\dimcase}$ are prohibited in principle,
resembling Fubini's theorem on the order of the isometry group.
The classification of metrics with 3 KVs makes this fact manifest.
For instance, it can be shown that $\mathrm{rank} {\boldsymbol R}_{[(21)]}^{~\# 1}$ cannot equal to zero.
The detail of this extensive study will be reported in a forthcoming paper.

It is also noteworthy to comment that the local existence of KVs does not immediately give rise to the existence of the global isometry group due to topological restrictions. 
An emblematic sample is the black hole constructed by Ba\~nados, Teitelboim and Zanelli (BTZ) \cite{Banados:1992wn}. The BTZ black hole solves  the vacuum Einstein's equations
with a negative cosmological constant and is obtained by the identification of points in AdS by the discrete isometry. From the curvature points of view, the BTZ metric obviously admits 6 KVs generating $\mathfrak{so}(2,2)$ algebra. In spite of this, only two of them are globally well-defined, since the rest of KVs is not single-valued under identifications~\cite{Banados:1992gq}.
It thus turns out that the global isometry group of the BTZ solution is broken from ${\rm SO}(2,2)$ down to ${\rm SO}(1,1)\times{\rm SO}(2)$.

Our algorithm can be extensible for a (semi-)Riemannian manifold $M$ of higher dimension.
In this case, the Weyl tensor $W_{abc}{}^{d}$ also comes into play. 
In particular, the first obstruction matrix in dimension $\dim M = n$ takes the form
\begin{align}
{\boldsymbol R}_a^{n}~\equiv~
\left[\nabla_a R~~
\nabla_a S^{(2)}~~
\cdots~~
\nabla_a S^{(n)}~~
\nabla_a W^{(2)}~~
\cdots~~
\nabla_a W^{(n(n-1)/2)}
\right]^T\:,
\end{align}
where $W^{(i)}$ are principal traces of the $i$-th powers of the Weyl operator $W^{ab}{}_{cd}$,
considered as an endomorphism of $\Lambda^2T(M)$.
So the invariants associated with the existence of KVs can be specified based on ${\boldsymbol R}_a^{n}$.
This line of extension would be a promising way to improve the past-proposed schemes.  
Further consideration of this shall be done elsewhere.

\section*{Acknowledgement}

KT thanks Boris Kruglikov, Vladimir Matveev, Tohru Morimoto and Kazuhiro Shibuya for their advices and useful comments.
KT is also grateful to Kazuhiro Shibuya, Yoshio Agaoka, Hiroshi Tamaru and other organisers of
The Geometry Conference 2018 held in Hiroshima for their invitation.
The work of MN is partially supported by Grant-in-Aid for Scientific Research from Ministry of Education, 
Science, Sports and Culture of Japan (16H03979 and 17H01091).

\appendix
\section{Relations amongst the Ricci rotation coefficients and their derivatives}\label{app:rels}

In this Appendix,
we collect some relations amongst the Ricci rotation coefficients and their derivatives
which are implicitly used in Sections \ref{sec:caseII} and \ref{sec:caseIII}.

\subsection{For an orthonormal frame}\label{app:rels_ortho}
Given an orthonormal frame $\{e_i^a, i=1,2,3\}$ satisfying
\begin{align}
g^{ab}~=~ \iota e_1^a e_1^b + e_2^a e_2^b -\iota e_3^a e_3^b\:,
\end{align}
where $\iota=\mathrm{sgn} (g_{ab}e_1^a e_1^b)$ and its Ricci rotation coefficients \eqref{eq:ricci_coeffs},
the following relations hold true.

\noindent
{\bf The commutation relations:}
\begin{subequations}
	\begin{align}
	\left[ e_1,e_2 \right]^a &~=~-\iota\:\kappa_1\:e_1^a+\kappa_2\:e_2^a-\iota (\tau_1-\tau_2)e_3^a\:,\\
	\left[ e_2,e_3 \right]^a &~=~-\iota(\tau_2-\tau_3) e_1^a-\eta_2\:e_2^a-\iota\:\eta_3\:e_3^a\:,\\
	\left[ e_3,e_1 \right]^a &~=~
	\iota\:\eta_1\:e_1^a+(\tau_1+\tau_3)e_2^a+\iota\:\kappa_3\:e_3^a\:.
	\end{align}
	\label{eq:app_comm_ortho}
\end{subequations}

\noindent
{\bf The components of the Ricci tensor:}
\begin{subequations}
	\begin{align}
	R_{ab}e_1^a e_1^b ~=~&\eta_1^2 + \iota \eta_1 \eta_2-\kappa_3^2
	+\iota \eta_3 \kappa_1 - \iota \kappa_1^2 -\kappa_2^2
	+2\iota \tau_2 \tau_3
	-\iota \pounds_1 \kappa_3 - \iota\pounds_3 \eta_1 + \pounds_1 \kappa_2 + \pounds_2 \kappa_1\:,\\
	R_{ab}e_2^a e_2^b ~=~& -\eta_3^2 + \eta_1 \eta_2 + \iota \eta_2^2
	-\kappa_1^2 + \kappa_2 \kappa_3 - \iota \kappa_2^2 - 2\tau_3 \tau_1 - \iota \pounds_3 \eta_2
	+\iota \pounds_1 \kappa_2 - \iota \pounds_2 \eta_3 + \iota \pounds_2 \kappa_1\:,\\
	R_{ab} e_3^a e_3^b ~=~&
	\iota \eta_3^2 - \eta_1^2 - \eta_2^2 + \kappa_3^2 - \iota \eta_3 \kappa_1
	-\iota \kappa_3 \kappa_2 - 2\iota \tau_1 \tau_2 + \iota \pounds_3 \eta_1 + \pounds_3 \eta_2
	+\iota \pounds_1 \kappa_3 + \pounds_2 \eta_3 \:,\\
	R_{ab}e_1^a e_2^b ~=~&
	-\eta_3 \kappa_3 - \iota \eta_3 \kappa_2 - \eta_1 \tau_3 + \iota \eta_2 \tau_3 + \eta_1 \tau_2 + \iota \eta_2 \tau_2
	-\iota \pounds_3 \tau_2 - \iota \pounds_2 \kappa_3\:,\\
	~=~&-\eta_3 \kappa_3 - \kappa_3 \kappa_1 - \eta_1 \tau_3 + \iota \eta_2 \tau_3 + \eta_1 \tau_1
	+\iota \eta_2 \tau_1 - \iota \pounds_3 \tau_1 - \iota \pounds_1 \eta_3\:,\\
	R_{ab}e_2^a e_3^b ~=~& -\eta_3 \eta_1 - \eta_1 \kappa_1 -\kappa_3 \tau_3 + \iota \kappa_2 \tau_3 + \kappa_3 \tau_1
	+\iota \kappa_2 \tau_1 + \iota \pounds_3 \kappa_1 - \iota \pounds_1 \tau_3\:,\\
	~=~& -\eta_1 \kappa_1 + \iota \eta_2 \kappa_1 + \kappa_3 \tau_1 + \iota \kappa_2 \tau_1
	-\kappa_3 \tau_2 + \iota \kappa_2 \tau_2 - \iota \pounds_1 \tau_2 + \iota \pounds_2 \eta_1\:,\\
	R_{ab} e_3^a e_1^b ~=~& -\iota \eta_2 \kappa_3 - \eta_2 \kappa_2 + \iota \eta_3 \tau_3 - \iota \kappa_1 \tau_3
	+\iota \eta_3 \tau_2 + \iota \kappa_1 \tau_2 + \pounds_3 \kappa_2 + \pounds_2 \tau_3\:,\\
	~=~& \iota \eta_1 \kappa_2 - \eta_2 \kappa_2 - \iota \eta_3 \tau_1 + \iota \kappa_1 \tau_1
	+\iota \eta_3 \tau_2 + \iota \kappa_1 \tau_2 + \pounds_1 \eta_2 -\pounds_2 \tau_1\:.
	\end{align}
	\label{eq:app_ricci_ortho}
\end{subequations}

\subsection{For a double-null frame}\label{app:rels_null}
Given an orthonormal frame $\{u^a, v^a,e^a\}$ satisfying
\begin{align}
g^{ab}~=~u^a v^b + v^a u^b + e^a e^b\:,
\end{align}
and its Ricci rotation coefficients \eqref{eq:ricci_coeffs_null},
the following relations hold true.

\noindent
{\bf The commutation relations:}
\begin{subequations}
	\begin{align}
	\left[ u,v \right]^a &~=~
	\kappa_v\:u^a -\kappa_u\:v^a + (\tau_u-\tau_v) e^a\:,
	\\
	\left[ v,e \right]^a &~=~
	-\eta_v\:u^a - (\tau_v - \tau_e) v^a + \eta_e\:e^a\:,\\
	\left[ e,u \right]^a &~=~
	(\tau_u+\tau_e)u^a + \eta_u\:v^a -\kappa_e\:e^a\:.
	\end{align}
	\label{eq:app_comm_null}
\end{subequations}

\noindent
{\bf The components of the Ricci tensor:}
\begin{subequations}
	\begin{align}
	R_{ab} u^a u^b ~=~&
	-\kappa_e^2-\kappa_e\kappa_u -2\eta_u \tau_e -\eta_u \tau_u -\eta_u \tau_v
	+\pounds_e \eta_u + \pounds_u\kappa_e\:,\\
	R_{ab} v^a v^b ~=~&
	-\eta_e^2 - \eta_e \kappa_v + 2\eta_v \tau_e - \eta_v \tau_u - \eta_v \tau_v + \pounds_e \eta_v +\pounds_v \eta_e\:,\\
	R_{ab}e^a e^b ~=~&
	-2\eta_u \eta_v -2 \eta_e \kappa_e + \eta_e \kappa_u + \kappa_e \kappa_v
	-\tau_u^2 -\tau_v^2 + \pounds_e \tau_u + \pounds_e \tau_v + \pounds_u \eta_e + \pounds_v \kappa_e\:,\\
	R_{ab}u^a v^b ~=~&
	-\eta_e \kappa_e + \kappa_e \kappa_v - 2 \kappa_u \kappa_v - \tau_e \tau_u + \tau_e \tau_v - \tau_u \tau_v -\tau_v^2
	+\pounds_e \tau_v - \pounds_u \kappa_v + \pounds_v \kappa_e - \pounds_v \kappa_u\:,\\
	~=~&
	-\eta_e \kappa_e + \eta_e \kappa_u - 2 \kappa_u \kappa_v - \tau_e \tau_u - \tau_u^2 + \tau_e \tau_v - \tau_u \tau_v
	+\pounds_e \tau_u + \pounds_u \eta_e - \pounds_u \kappa_v - \pounds_v \kappa_u\:,\\
	R_{ab} v^a e^b ~=~&
	-\eta_v \kappa_e - \eta_v \kappa_u + \eta_e \tau_e - \kappa_v \tau_e + \eta_e \tau_v + \kappa_v \tau_v -\pounds_e \kappa_v -\pounds_v \tau_e\:,\\
	~=~& -2\eta_v \kappa_u -\eta_e \tau_u + \eta_e \tau_v - \pounds_u \eta_v + \pounds_v \tau_u\:,\\
	R_{ab} e^a u^b ~=~&
	-\eta_e \eta_u - \eta_u \kappa_v - \kappa_e \tau_e + \kappa_u \tau_e + \kappa_e \tau_u + \kappa_u \tau_u
	-\pounds_e \kappa_u + \pounds_u \tau_e\:,\\
	~=~& -2\eta_u \kappa_v + \kappa_e \tau_u - \kappa_e \tau_v + \pounds_u \tau_v - \pounds_v \eta_u\:.
	\end{align}
	\label{eq:app_ricci_null}
\end{subequations}

\section{Canonical form of metrics admitting 4 Killing vectors}
\label{app:4KVs}

Using the scheme developed in the present paper, 
we can obtain the canonical form of the metric admitting any number of KVs and the corresponding algebra. 
To make the discussion focused, we investigate in this appendix the case in which 4 KVs exist. 
As described in section  \ref{sec:caseIII}, this occurs only for Segre types
$[1,(11)]$, $[(1,1),1]$ and $[(21)]$. In each case, it turns out that we can actually obtain all the explicit metrics. 
Interestingly,  these spacetimes are all homogeneous, in the sense that local isometry groups possess transitive actions on the manifold.

\subsection{Type $[1,(11)]$}

Let us begin with the case of Segre type $[1,(11)]$. 
Analysis in section \ref{subsec:caseIII_111_21_t} reveals that 4 KVs exist, provided
\begin{align}
\label{}
&\kappa_2 ~=~ \tau_2+\tau_3=0 \,,
&&\tau_2 ~=~{\rm const.} \,,  
\end{align} 
together with \eq \eqref{eq:caseIII_111_21_t_cond}:
\begin{align}
\label{}
&	\kappa_1 ~=~0\,,
&&	\eta_1 ~=~0\,,
&&	\kappa_3~=~ -\kappa_2 \,. 
\end{align}
Here $\pounds _1 \tau_2 =0$ follows from $R_{23}=R_{32}$. 
With these spin connections, the first derivative of $\{e_1, e_2, e_3\}$ reads
\begin{subequations}
\begin{align}
\label{}
\nabla_b e_{1a}~=~&\tau_2 e_{2b}e_{3a}-\tau_2 e_{3b} e_{2a} \,, \\ 
\nabla_b e_{2a}~=~& -\tau_1 e_{1b} e_{3a}+\eta_2 e_{2b} e_{3a}+e_{3b} (-\tau_2 e_{1a}-\eta_3 e_{3a} ) \,, \\ 
\nabla_b e_{3a}~=~& \tau_1 e_{1b} e_{2a}+e_{2b} (\tau_2 e_{1a}-\eta_2 e_{2a})+\eta_3 e_{3b}e_{2a} \,. 
\end{align}
\end{subequations}
It follows that $W_a=e_{2a}+i e_{3a}$ satisfies 
\begin{align}
\label{}
\nabla_b W_a~=~i (\tau_1 e_{1b}-\eta_2 e_{2b}+\eta_3 e_{2b})W_a +i \tau_2 e_{1a} W_b \,,  
\end{align}
hence 
\begin{align}
\label{}
&\nabla_{(a}e_{1b)}~=~0 \,,
&&W_{[a}\nabla_b W_{c]} ~=~ 0 \,. 
\end{align}
Then, there exist real functions $t,  x,  y$ and $\theta , \phi, f, \chi_1, \chi_2$ such that 
\begin{align}
\label{}
&e_{1a}~=~-f(\nabla_a t +\chi_1 \nabla_a x+\chi_2 \nabla_a y)\,,
&&W_a ~=~e ^{i \theta +\phi} (\nabla_a x+i \nabla_a y) \,. 
\end{align}
By the redefinition $t\to \int f^{-1}\D t$,  one can set $f\equiv 1$ without loss of generality.
Exploiting the ${\rm SO}(2) $ gauge freedom which rotates ($e_2, e_3$), $\theta=0$ is always achieved.
The Killing equation $\nabla_{(a}e_{1b)}=0$ then demands that the metric is independent of $t$.
The condition  $\tau_2 (={\rm const.} )$ boils down to
\begin{align}
\label{}
\partial_y \chi_1 -\partial_x \chi_2 ~=~ 2 \tau_2 e^{2\phi} \,. 
\end{align}
Using this relation, the trace-free part of Ricci tensor gives rise to Liouville's equation
\begin{align}
\label{}
&(\partial_x^2+\partial_y^2 )\phi ~=~ -k e^{2\phi} \,,
&&k ~\equiv~ -\frac 12 (3\lambda_1+8\tau_2^2) \,. 
\end{align}
It follows that $\D \Sigma_k^2= e^{2\phi}(\D x^2+\D y^2)$ corresponds to the space $\Sigma_k$ with 
a constant sectional curvature $k$, which can be normalised to be $0$ or $\pm 1$, and 
the scalar curvature is given by $R=2(k+\tau_2^2)$.
The local solution to Liouville's equation can be chosen to be
\begin{align}
\label{}
&\phi~=~- \log \left(1+\frac k4(x^2+y^2)\right) \,,
&&\chi~=~ \frac{\tau_2 }{1+\frac k4(x^2+y^2)}(y \D x-x \D y) \,, 
\end{align}
where $\chi=\chi_1 \D x+\chi_2\D y$. 
Defining $x+i y=\frac 2{\sqrt k}\tan \left(\frac{\sqrt k}2\theta \right)e^{i\phi} $, 
we therefore arrive at
\begin{align}
\label{4KVs_1,(11)}
\D s^2~=~ - \left[\D t -4\tau_2\left( \frac{\sin \left(\frac{\sqrt k}2\theta\right)}{\sqrt k}\right) \D \phi \right]^2 +\D \theta ^2 +\left(
\frac{\sin (\sqrt k\theta )}{\sqrt k}
\right)^2 \D \phi^2  \,, 
\end{align}
If $k=-4\tau_2^2<0$, we have ${\rm AdS}_3$ for which the number of KVs is enhanced to 6. 
Otherwise, we have precisely 4 KVs 
\begin{subequations}
\begin{align}
\label{}
K_1~=~&   \partial_t \,,\\ 
K_2 ~=~&\tfrac{2\tau_2}{\sqrt k}\cos\phi \tan\left(\tfrac{\sqrt k}{2}\theta\right)\partial_t - \sin\phi \partial_\theta -
\tfrac{{\sqrt k}}{2}\cos\phi \left[\cot ^2\left(\tfrac{\sqrt k}{2}\theta\right)-1\right]\tan \left(\tfrac{\sqrt k}{2}\theta\right)\partial_\phi \,,  \\
K_3 ~=~& \tfrac{2\tau_2}{\sqrt k} \sin\phi \tan\left(\tfrac{\sqrt k}{2}\theta\right)\partial_t + \cos\phi \partial_\theta -
\tfrac{{\sqrt k}}{2}\sin\phi \left[\cot ^2\left(\tfrac{\sqrt k}{2}\theta\right)-1\right]\tan \left(\tfrac{\sqrt k}{2}\theta\right)\partial_\phi \,, \\ 
K_4~=~&\partial_\phi \,, 
\end{align}
\end{subequations}
satisfying 
\begin{align}
\label{}
&[K_2, K_3]~=~-\left(2\tau_2 K_1+k K_4 \right)\,,
&&[K_2, K_4]~=~K_ 3 \,,
&&[K_3, K_4]~=~-K_2 \,. 
\end{align}
For $\tau_2=0$, the metric collapses to 
$\mathbb R \times \Sigma_k$, which is locally symmetric. For $\tau_2(k+4\tau_2^2)\ne 0$ with $k=-1$, 
the metric describes the 3-dimensional G\"odel universe, which is sourced by 
a dust with a negative cosmological constant (see e.g. \cite{Klemm:2015mga}).

\subsection{Type $[(1,1)1]$}

In this case, we have $\kappa_1=\kappa_2=\eta_2=\eta_3=0$ and 
$\tau_1=\tau_3={\rm const.}$, for which $e_{2a}$ is Killing and 
$W^\pm_a =e_{1a}\pm e_{3a}$ are hypersurface-orthogonal. 
Since the rest of the derivation is parallel to the $[1,(11)]$ case, we only 
show the final results: 
\begin{align}
\label{4KVs_(1,1)1)}
\D s^2~=~ \left[\D t +4\tau_1\left( \frac{\sin \left(\frac{\sqrt k}2\theta\right)}{\sqrt k}\right) \D \phi \right]^2 +\D \theta ^2 -\left(
\frac{\sin (\sqrt k\theta )}{\sqrt k}
\right)^2 \D \phi^2  \,, 
\end{align}
where $\tau_1 $ is a constant. This is the double Wick-rotated version of \eq \eqref{4KVs_1,(11)}.

\subsection{Type $[(2,1)]$}

A class of metrics with 4 KVs exists also for the type $[(2,1)]$, for which 
\begin{align}
\label{}
&\kappa_e~=~0 \,,
&&\eta_u~=~0 \,,
&&\sigma~=~0 \,.
\end{align}
The second obstruction matrix ${\boldsymbol R}_{[(21)]}$ given by \eq \eqref{eq:compatibility_III_21_def_submaxi}
must vanish identically, yielding 
\begin{align}
\label{}
&\tau_v ~=~{\rm const.} \,,
&&\Sigma~=~{\rm const.} \,, 
\end{align}
where $\pounds _u \tau_v=0$ follows from $S_{ab}e^a u^b=0$. 
These are exhaustive information supplied from the condition for 4 KVs in type $[(2,1)]$. 

The definition of $\Sigma$, $\sigma=0$ and Bianchi identity are combined to give 1st-order system 
for $\varphi=\frac 12 \log (S_{vv})$ as 
\begin{align}
\label{4KVs_dervphi}
&\pounds _u \varphi ~=~-\kappa_u \,,
&&\pounds _v \varphi ~=~\kappa_v -e^{\varphi}\Sigma \,,
&&\pounds _e \varphi ~=~-\tau_e-\tau_v \,.
\end{align}
The compatibility conditions \eq \eqref{eq:app_comm_null} for these equations give 
\begin{align}
\label{}
2e^\varphi \Sigma \tau_v~=~0 \,, 
\end{align}
which branches into (i) $\tau_v=0$ and (ii)  $\Sigma=0$. 

Before proceeding, let us note that the Segre type 
[$(2,1)]$ allows the following gauge freedom for the choice of 
null basis $\{u_a, v_a , e_a\}$: 
\begin{align}
\label{4KVs_gauge1}
&u_a ~\to~ a u_a \,,
&&v_a ~\to~ a^{-1} v_a \,,
&&e_a ~\to~ e_a \,,
\end{align}
and
\begin{align}
\label{4KVs_gauge2}
&u_a ~\to~ u_a \,,
&&v_a ~\to~ v_a -\frac 12 b^2 u_a +b e_a \,,
&&e_a ~\to~ e_a -b u_a \,,
\end{align}
where $a $ and $b$ are arbitrary functions.
By these transformations, $\tau_v \equiv e^a v^b \nabla_b u_a $ and 
$\Sigma\equiv \pounds _v (e^{-\varphi})+\kappa_v e^{-\varphi}$ remain invariant.
In contrast, $\tau_u\equiv -v^au^b\nabla_b e_a$ and $\tau_e\equiv v^a e^b\nabla_b u_a$ vary as 
$\tau_u\to \tau_u$ and $\tau_e \to \tau_e +e^b\nabla_b \log a$ under \eq \eqref{4KVs_gauge1}, which 
permits us to set $\tau_e+\tau_u=0$. Since $[u, e]^a=0$ is now satisfied because the condition $\kappa_e=\eta_u=0$ 
does not change under \eq \eqref{4KVs_gauge1},
one can introduce local coordinates ($x,y,z$) in such a way that $u^a$ and $e^a$ form the coordinate vectors 
\begin{align}
\label{}
&u^a ~=~\left(\partial_y\right)^a\,,
&&v^a ~=~ V_1  \left(\partial_x\right)^a+V_2  \left(\partial_y\right)^a+V_3  \left(\partial_z\right)^a\,,
&&e^a =  \left(\partial_z\right)^a\,,
\end{align}
where $V_i=V_i(x,y,z)$. Lowering indices, we have
\begin{align}
\label{}
&u_a ~=~\frac{\nabla_a x}{V_1} \,,
&&v_a ~=~\nabla_a y -\frac{V_2}{V_1}\nabla_a x \,,
&&e_a ~=~\nabla_a z-\frac{V_3}{V_1}\nabla_a x \,.
\end{align}
The hypersurface orthogonality of $u_a$ is a direct consequence of $\eta_u=0$ and $\kappa_e=0$. In this basis,
$\tau_v$ is computed to be 
\begin{align}
\label{4KVs_tauv}
\tau_v ~=~-\frac 12 V_1 \left[\partial_y \left(\frac{V_3}{V_1}\right)+\partial_z \left(\frac 1{V_1} \right)\right] \,. 
\end{align}
The following discussion will be divided according to $\tau_v=0$ or $\Sigma=0$. 

\subsubsection{$\tau_v=0$ case}

Setting $\tau_v=0$ in \eq \eqref{4KVs_tauv}, one finds a local function $F=F(x,y,z)$ satisfying 
\begin{align}
\label{4KVs_tauv0_F}
&\frac{V_3}{V_1}~=~-\partial_z F \,,
&&\frac 1{V_1}~=~\partial_y F \,.
\end{align}
Inserting this into $S_{ab}e^a =0$, one finds a function $f_1 =f_1 (x)$ such that 
\begin{align}
\label{}
\partial_y V_2 \partial_y F+2 V_2 \partial_y^2 F+\frac{\partial_x \partial_y F-\partial_z F\partial_y \partial_z F}{\partial_y F}~=~f_1 \,,
\end{align}
which is further integrated to give 
\begin{align}
\label{4KVs_tauv0_V2}
V_2 ~=~-\frac{1}{(\partial_y F)^2}\left[\partial_x F-\frac 12 (\partial_z F)^2-f_1(x)F+\frac 12 f_2(x,z)\right]\,, 
\end{align}
where $f_2=f_2(x,z)$ is an arbitrary function of $x$ and $z$. 
Inspecting \eqs \eqref{4KVs_tauv0_F} and \eqref{4KVs_tauv0_V2}, one obtains
\begin{align}
\label{}
&R~=~0 \,,
&&\varphi~=~\frac 12 \log \left(-\frac{\partial_z^2 f_2 }{2(\partial_y F)^2 }\right) \,. 
\end{align}
Substitution of this expression of $\varphi$ into $\sigma\equiv \pounds _e \varphi+\tau_v+\tau_e=0$, 
one gets $\partial_z^3 f_2(x,z)=0$. Upon integration, we find
\begin{align}
\label{}
f_2(x,z)~=~ f_{20}(x)+f_{21}(x)z+f_{22}(x)z^2 \,.
\end{align}
The condition $S_{vv}\ne 0$ asks for $f_{22}(x) \ne 0$. 
$\Sigma\equiv \pounds _v (e^{-\varphi})+\kappa_v e^{-\varphi} $ is now computed  to 
\begin{align}
\label{4KVs_tauv0_Sigma}
\Sigma~=~\frac{-2f_1(x)f_{22}(x)+f_{22}'(x)}{2 (-f_{22}(x))^{3/2}} \,.
\end{align}

We have all ingredients in place to obtain the explicit metric form. 
Defining $\tilde y=F(x,y,z)$, the metric becomes 
\begin{align}
\label{4KVs_tauv0_metric0}
\D s^2~=~2 \D \tilde y \D x+\D z^2+ \D x^2 [f_{20}(x)+f_{21}(x)z+f_{22}(x)z^2+\tilde y f_1 (x) ] \,.
\end{align}
Further change of variable $x=h(\hat x)$, $\tilde y=\hat y/h'(\hat x)$ renders the metric into 
\begin{align}
\label{}
\D s^2~=~2 \D \hat x\D \hat y+\D z^2+\D \hat x^2 \left[\hat f_{20}+\hat f_{21} z+\hat f_{22} z^2+\frac{\hat y}{h'(\hat x)}(\hat f_1+h''(\hat x))  \right]\,,
\end{align}
where $\hat f_1(\hat x)=h'(\hat x)^2 f_1(h(\hat x))$ and $\hat f_{2i}(\hat x)=h'(\hat x)^2 f_{2i}(h(\hat x))$ ($i=0,1,2$). 
By choosing $h_1''(\hat x)=-\hat f_1(\hat x)$ and omitting hats,  one obtains the metric of the following form 
\begin{align}
\label{4KVs_tauv0_metric}
\D s^2~=~2 \D  x\D y+\D z^2+ \D x^2 [ f_{20}(x)+f_{21}(x) z+f_{22}(x) z^2]\,.
\end{align}
This amounts to setting $f_1(x)=0$ in the metric \eqref{4KVs_tauv0_metric0}.
Thus, equation \eqref{4KVs_tauv0_Sigma} is integrated to give 
\begin{align}
\label{4KVs_tauv0_f22}
f_{22}(x)~=~-\frac{1}{(\Sigma x-c_1)^2} \,,
\end{align}
where $c_1$ and $\Sigma $ are constants. This metric describes the pp-wave, whose 4 KVs were 
already obtained in section \ref{subsec:ppwave}. The special case $\Sigma=0$ corresponds to the locally symmetric space 
which admits a covariantly constant Ricci tensor $\nabla_a R_{bc}=0$.

\subsubsection{$\Sigma=0$ case}

We assume $\tau_v \ne 0$ henceforth, since the metric \eqref{4KVs_tauv0_metric} with $\Sigma=0$ in \eq \eqref{4KVs_tauv0_f22} is recovered for the $\tau_v=0$ case.
Equation \eqref{4KVs_tauv} is solved as 
\begin{align}
\label{4KVs_Sigma0_V3y}
\partial_y V_3~=~-2 \tau_v+\frac{\partial_z V_1+V_3 \partial_y V_1}{V_1}\,.
\end{align}
Inserting this into $S_{ab}e^b=0$, one finds a function $k_1(x)$ satisfying 
\begin{align}
\label{4KVs_Sigma0_V2y}
\partial_y V_2~=~-k_1 V_1+V_3 \left(-2\tau_v+\frac{\partial_z V_1}{V_1}\right)
+\frac{2V_2 \partial_y V_1}{V_1}+\partial_x V_1 \,.
\end{align}
It follows that the scalar curvature is a negative constant 
$R=-6 \tau_v^2 <0$. 
The first and third conditions in \eq \eqref{4KVs_dervphi} give rise to 
\begin{align}
\label{4KVs_Sigma0_phi}
\varphi~=~-2\tau_v z+\log (k_2 (x)V_2) \,, 
\end{align}
where $k_2=k_2(x)$ is an arbitrary function. 
Setting $\Sigma=0$ in the second condition of \eq \eqref{4KVs_dervphi}, $k_1$ is subjected to  
\begin{align}
\label{}
k_1(x)~=~-\frac{k_2'(x)}{k_2(x)} \,. 
\end{align}
Comparison of  $S_{vv}=e^{2\varphi}$ in a coordinate basis with 
the one given by \eq \eqref{4KVs_Sigma0_phi} assures the existence of  a function 
$k_3=k_3(x,y)$ such that 
\begin{align}
\label{}
k_3(x,y)~=~& -\frac{\partial_x V_3}{V_1}+\frac{-2V_2+V_3^2}{V_1^3}\partial_z V_1 -\frac{V_3}{V_1^2}\partial_z V_3 
+\frac{V_3}{V_1}\partial_x V_1+\frac{\partial_z V_2}{V_1^2}\notag \\
& +\frac{V_3}{V_1^2}(-k_1 V_1-\tau_v V_3)+\frac{2\tau_v V_2}{V_1^2}+\frac{e^{-4\tau_v z}}{4\tau_v}k_2^2 \,, 
\end{align}
which is arranged into 
\begin{align}
\label{}
\partial_x \left[-\frac{V_3 e^{2\tau_v z}}{V_1k_2}\right] ~=~\partial_z \left[
\frac{e^{2\tau_v z}}{2k_2}\left(\frac{-2V_2+V_3^2}{V_1^2}+\frac{k_3}{\tau_v}+\frac{e^{-4\tau_v z}}{4\tau_v^2}k_2^2 \right)
\right]\,.
\end{align}
This implies the existence of a function $F_1=F_1(x,y,z) $ such that the terms in the square bracket on the left-hand side is $\partial_z F_1$
and the terms in the square bracket on the left-hand side is $\partial_x F_1$. This condition is simplified to 
\begin{subequations}
\label{4KVs_Sigma0_V2V3}
\begin{align}
V_2~=~& \frac{e^{-4\tau_v z}}{8\tau_v^2}V_1^2 \left[k_2^2+4 e^{4\tau_v z}\tau_v k_3 +4\tau_v^2k_2^2(\partial_z F_1)^2-8 e^{2\tau_v z}\tau_v^2 k_2 \partial_x F_1\right]\,, 
\\
V_3~=~& -e^{-2\tau_v z}k_2 V_1 \partial_z F_1 \,. 
\end{align}
\end{subequations} 
The compatibility of \eqs \eqref{4KVs_Sigma0_V3y} and \eqref{4KVs_Sigma0_V2V3} gives 
\begin{align}
\label{4KVs_Sigma0_V1}
V_1~=~-\frac{e^{2\tau_v z}}{k_4(x,y)-k_2(x)\partial_y F_1(x,y,z)} \,, 
\end{align}
where $k_4(x,y)$ is a function independent of $z$. Together with \eq \eqref{4KVs_Sigma0_V1},
the compatibility of \eqs \eqref{4KVs_Sigma0_V2y} and \eqref{4KVs_Sigma0_V2V3} gives 
\begin{align}
\label{}
&k_3~=~k_3(x) \,,
&&k_4(x,y)~=~k_2(x)k_{41}(y) \,. 
\end{align}
We have exhausted the constrains coming from the vanishing of second obstruction matrix in class 3 of Segre $[(21)]$.

Let us now move on to obtaining the metric form.
Defining $\tilde x=\int k_2(x)\D x$, $\tilde y=F_1(x,y,z)-\int k_{41}(y)\D y$, we get
\begin{align}
\label{4KVs_Sigma0_metric0}
\D s^2~=~2 e^{-2\tau_vz} \D \tilde x\D \tilde y+\D z^2+ \D \tilde x^2 \left(-\frac{e^{-4\tau_v z}}{4\tau_v^2}-\frac{\tilde k(\tilde x)}{\tau_v}\right) \,, 
\end{align}
where $\tilde k(\tilde x)=k_{31}(x)/k_2(x)^2$. 
We change coordinates further to 
\begin{align}
\label{}
&\hat x~=~\int e^{-2 \tau_v h(\tilde x)} \D \tilde x \,,
&&\tilde y~=~\hat y-\frac 1{2\tau_v }e^{2\tau_v (\hat z+h(\tilde x))} h'(\tilde x) \,,
&&z~=~\hat z+h(\tilde x) \,, 
\end{align}
and choose $h(\tilde x)$ to satisfy $\tilde k(\tilde x)+\tau_v h'(\tilde x)^2+h''(\tilde x)=0$. Then, the resulting metric is \eq \eqref{4KVs_Sigma0_metric0} replaced by  
 $(\tilde x,\tilde y, z)\to (\hat x,\hat y,\hat z)$ and a vanishing $\tilde k(\tilde x)$. Namely, we obtain 
\begin{align}
\label{4KVs_Sigma0_metric}
\D s^2~=~2 e^{-2\tau_v z}\D x \D y+\D z^2-\frac{1}{4\tau_v^2}e^{-4\tau_v z}\D x^2 \,, 
\end{align}
where we have dropped hats from the variables.
This metric represents the plane-wave (or equivalently the Kundt class), but not the pp-wave. 
The KVs are given by
\begin{align}
\label{}
&K_1~=~\partial_x \,, \qquad K_2=\partial_y \,,
&&K_3~=~y\partial_y +\frac{1}{2\tau_v} \partial_z \,,
&&K_4~=~-\frac{e^{2\tau_vz}}{2\tau_v^2}\partial_x +y^2 \partial_y+\frac{y}{\tau_v}\partial_z \,, 
\end{align}
whose nonvanishing commutation relations are 
\begin{align}
\label{}
&[K_2, K_3]~=~K_2 \,,
&&[K_2,K_4]~=~2 K_3 \,,
&&[K_3, K_4]~=~K_4 \,.
\end{align}
This subalgebra is ${\mathfrak sl}(2,\mathbb R)$. 
One can easily find that the above metric \eqref{4KVs_Sigma0_metric} recovers \eq \eqref{Kruchkovich4} below.

\subsubsection{Remarks}

We conclude that the Segre $[(21)]$ allows two metrics \eqref{4KVs_tauv0_metric}
and \eqref{4KVs_Sigma0_metric} with 4 KVs. Let us compare our results with those 
in the literature. 
The classification of spacetimes admitting 4 KVs for Segre $[(21)]$ has been addressed by \cite{Kruchkovich}. 
In this work, Kruchkovich contrastively obtained the following four classes of metrics 
\begin{subequations}
\begin{align}
\label{}
\D s^2 ~=~& 2(2-c)e^{cx_1}\D x_1 \D x_2+ e^{2x_1} \D x_3^2 \,, \qquad c\ne 1,2 \,, \label{Kruchkovich1}\\ 
\D s^2 ~=~& e^{2 x_1}(2 \D x_1 \D x_2-\D x_3^2 )\,, \label{Kruchkovich2}\\ 
\D s^2 ~=~& e^{-qx_1} \left[ 2 \D x_1 \D x_2 -\frac{4}{\omega^2}\cos^2\left( \frac{\omega x_1}{2} \right)\D x_3^2\right] +k\D x_1^2\,, \quad 
\omega = \sqrt{4-q^2} \,, \quad q^2 < 4 \,, \label{Kruchkovich3}\\
\D s^2 ~=~& e^{2 x_3} \D x_1^2+ 2n e^{x_3} \D x_1 \D x_2 +\varepsilon \D x_3^2 \,, \qquad 
n \ne 0\,, \qquad \varepsilon = \pm 1 \,. \label{Kruchkovich4}
\end{align}
\end{subequations}
However, this classification turns out to be redundant and consistent with our results. 
Indeed, one can  bring the metrics \eqref{Kruchkovich1})-- \eqref{Kruchkovich3} into a universal form: 
\begin{align}
\label{4KVs:universalmetric}
\D s^2~=~2 \D x \D y +\frac {a_0+a_2 z^2}{x^2} \D x^2\pm \D z^2 \,, 
\end{align}
where $a_0$ and $a_2$ are constants. This is nothing but the spacetime \eqref{4KVs_tauv0_metric}
up to the metric signature.
The desired coordinate transformations are: For the metric \eqref{Kruchkovich1}, 
plus sign in \eq \eqref{4KVs:universalmetric} with 
\begin{subequations}
\begin{align}
\label{}
x_1~=~&\frac 1c \log \left(\frac{cx}{2-c}\right)\,, &
x_2~=~& y+\frac{z^2}{2c x} \,, &
x_3~=~& \left(\frac{cx}{2-c}\right)^{-1/c} z\,, \notag  \\
a_2~=~&\frac{1-c}{c^2} \,, & a_0~=~&0\,. &&
\end{align}
For the metric \eqref{Kruchkovich2}, the minus sign in \eq \eqref{4KVs:universalmetric} with 
\begin{align}
\label{}
&x_1~=~\frac 12 \log (2x)  \,,
&&x_2~=~y-\frac{z^2}{4x } \,,
&&x_3~=~\frac{z}{\sqrt{2x} }\,,
&&a_2~=~\frac 14 \,,
&&a_0~=~0 \,.
\end{align}
For the metric \eqref{Kruchkovich3}, 
the minus sign in \eq \eqref{4KVs:universalmetric} with
\begin{align}
\label{}
x_1~=~& -\frac 1q \log (qx) \,, \qquad x_3= \frac{\sqrt{4-q^2} z}{2\sqrt{qx}} \sec  \left(\frac{\sqrt{4-q^2} }{2q}\log (qx)\right)\,, \qquad a_0~=~\frac{k}{q^2}\,, \notag \\
x_2~=~&-y +\frac{z^2}{4qx} \left[q-\sqrt{4-q^2} \tan \left(\frac{\sqrt{4-q^2} }{2q}\log (qx)\right)\right]\,, \qquad 
a_2~=~\frac 1{q^2} \,. 
\end{align}
\end{subequations}
It follows that the metrics admitting 4 KVs in Segre $[(21)]$ type are classified into two: one is the  pp-wave \eqref{4KVs_tauv0_metric}
and the other is the plane-wave \eqref{4KVs_Sigma0_metric}, both of which are homogeneous. This refines the analysis in \cite{Kruchkovich}.

\end{document}